\documentclass[doublespace,a4paper,12pt]{book}
\usepackage[bookmarks]{hyperref}
\usepackage{amsmath,amsthm,amssymb,euscript,mathrsfs,epsf,epsfig}
\usepackage{amscd,fancyhdr,amsfonts,setspace,indentfirst,slashed}
\usepackage[top=0.7in,bottom=1in,left=1.0in,right=1.0in,includehead]{geometry}
\usepackage{array,verbatim,mathtools,tikz}
\usepackage[sort&compress]{natbib}
\usepackage[english]{babel}
\numberwithin{equation}{section}
\usepackage{graphicx, tikz}
\usetikzlibrary{arrows,shapes,positioning}
\tikzstyle{every picture}+=[remember picture]
\tikzstyle{na} = [baseline=-.5ex]
\tikzstyle{format} = [rounded rectangle,
                      thick,
                      minimum size=1cm,
                      draw=red!00!black!80,
                      top color=white,
                      bottom color=red!00!black!00,
                      on grid]

\makeatletter
\@addtoreset{equation}{section}
\makeatother

\newcounter{multieqs}

\def\a{\alpha} \def\b{\beta} \def\g{\gamma} \def\d{\delta} \def\e{\epsilon}
 \def\z{\zeta} \def\h{\eta} \def\q{\theta}
 \def\i{\iota} \def\k{\kappa} \def\l{\lambda} \def\m{\mu}
\def\n{\nu} \def\x{\xi} \def\p{\pi}  \def\r{\rho}
 \def\s{\sigma} \def\t{\tau}  \def\f{\varphi}
\def\ff{\phi} \def\c{\chi}  \def\w{\omega}

\def\G{\Gamma} 
 
   \def\Q{\Theta}
   \def\L{\Lambda} 
 \def\X{\Xi} \def\P{\Pi} 
 \def\S{\Sigma}  
\def\F{\Phi}   \def\W{\Omega}

\def\ba{\bar{a}}\def\bb{{\bar{b}}}\def\bc{\bar{c}}\def\bd{\bar{d}}\def\be{\bar{e}}
\def\bm{\bar{m}}\def\bn{\bar{n}}

\def\bA{\bar{A}}\def\bB{\bar{B}}\def\bC{\bar{C}}\def\bD{\bar{D}}\def\bE{\bar{E}}
\def\bK{\bar{K}}\def\bL{\bar{L}}\def\bM{\bar{M}}\def\bN{\bar{N}}

\def\fr{\frac} \def\dfr{\dfrac} \def\dt{\partial}

\def\ph{\phantom}
\def\mc{\mathcal}
\def\mH{\mathcal{H}}
\def\mL{\mathcal{L}}
\def\tx{\tilde{x}}
\def\ty{\tilde{y}}
\def\tdt{\tilde{\partial}}
\def\mE{\mathcal{E}}

\def\Tr{\mbox{Tr}}

\def\Tr{\mbox{Tr}}

\def\hE{\hat{E}}
\def\tE{\tilde{E}}
\def\XX{\mathbb{X}}
\def\YY{\mathbb{Y}}
\def\RR{\mathbb{R}}
\def\TT{\mathbb{T}}
\def\SS{\mathbb{S}}

\def\bF{{\mathbf F}}
\def\Tr{\mbox{Tr}}

\def\ben{\begin{equation*}}
\def\een{\end{equation*}}
\def\bnn{\begin{eqnarray*}}
\def\enn{\end{eqnarray*}}
\def\bsub{\begin{subequations}}
\def\esub{\end{subequations}}

\def\cf{\mc{F}}
\def\cg{\mc{G}}

\newcommand\bqa {\begin{eqnarray}}
\newcommand\eqa {\end{eqnarray}}

\newcommand{\bear}{\begin{array}}
\newcommand{\enar}{\end{array}}

\newcommand{\br}[2]{\bar{#1}\bar{#2}}

\newcommand{\vW}[4]{W^{\bar{#1}\bar{#2}}_{#3 #4}}
\newcommand{\uW}[4]{W^{#1 #2}_{\bar{#3}\bar{#4}}}

\def\bea{\begin{eqnarray}}
\def\eea{\end{eqnarray}}
\def\pl{\partial}
\def\non{\nonumber}
\def\ab{{\bar{a}}}
\def\bbr{{\bar{b}}}
\def\cb{{\bar{c}}}
\def\db{{\bar{d}}}
\def\eb{{\bar{e}}}
\def\fb{{\bar{f}}}
\def\gb{{\bar{g}}}
\def\hb{{\bar{h}}}

\newcommand{\torus}[4]{
\draw [black, rotate around={#3:(#1,#2)}] (#1,#2) ellipse (2*#4 and 1*#4);
\draw [black, rotate around={#3:(#1,#2)}] (#1-1.15*#4,#2+0.2*#4) arc (-200:20:1.2*#4 and 0.3*#4);
\draw [black, rotate around={#3:(#1,#2)}] (#1-1.15*#4,#2+0.0*#4) arc (170:10:1.15*#4 and 0.3*#4);
\draw [black, dashed,  rotate around={#3:(#1,#2)}] (#1-2.0*#4,#2+0.0*#4) arc (-180:0:2*#4 and 0.8*#4);
}

\begin{document}



\begin{titlepage}

\vfill

\begin{center}
   \baselineskip=16pt
   {\LARGE \bf U-dualities in Type II string theories and \\ M-theory.}
   \vskip 2cm
    {\large\bf  Edvard T. Musaev}\footnote{\tt e.musaev@qmul.ac.uk}
       \vskip 1.6cm
       {\it A thesis submitted for the degree of Doctor of Philosophy}
       \vskip 3cm
             {\it Centre for Research in String Theory, \\
             School of Physics and Astronomy,\\
             University of London, Queen Mary, \\
            Mile End Road, London, E1 4NS, UK} 
\end{center}

\vfill
\setcounter{footnote}{0}
\end{titlepage}

\begin{center} 
\textbf{Abstract}
\end{center} 
\begin{quote}
In this thesis the recently developed duality covariant approach to string and M-theory is investigated. In this formalism the U-duality symmetry of M-theory or T-duality symmetry of Type II string theory becomes manifest upon extending coordinates that describe a background. 

The effective potential of Double Field Theory is formulated only up to a boundary term and thus does not capture possible topological effects that may come from a boundary. By introducing a generalised normal we derive a  manifestly duality covariant boundary term that reproduces the known Gibbons-Hawking action of General Relativity, if the section condition is imposed. It is shown that the full potential can be represented as a sum of the scalar potential of gauged supergravity and a topological term that is a full derivative. The latter is conjectured to capture non-trivial topological information of the corresponding background, such as monodromy around an exotic brane.

Next we show that the Scherk-Schwarz reduction of M-theory extended geometry successfully reproduces known structures of maximal gauged supergravities. Local symmetries of the extended space defined by a generalised Lie derivatives reduce to gauge transformations and lead to the embedding tensor written in terms of twist matrices. The scalar potential  of maximal gauged supergravity that follows from the effective potential is shown to be duality invariant with no need of section condition. Instead, this condition, that assures the closure of the algebra of generalised diffeomorphisms, takes the form of the quadratic constraints on the embedding tensor.

\end{quote}

\newpage

\begin{center} 
\textbf{Acknowledgement}
\end{center} 
\begin{quote}

This Ph.D. thesis has greatly benefited from comments and suggestions of several people. First of all, I would like to thank my supervisor, Dr. David Berman, whose professional knowledge and willingness to help were supportive during the process of making this scientific research. 

I also acknowledge my fellow colleagues at Queen Mary and other institutions for their inspiration and the lively discussions we had, in particular: Ilya Bakhmatov, Andreas Brandhuber, Ilmar Gahramanov, \"Omer Gurdogan, Robert Mooney, Malcolm Perry, Sam Playle, Sanjaye Ramgoolam and Daniel Thompson.

Special thanks due to Emil Akhmedov for his support and guidance during my first steps in theoretical physics and to Evgeniy Patrin for introducing me to the world of mathematics.

Finally, my warmest thanks goes to my beloved mother for her exceptional support and excellent assistance during my whole life.
 \end{quote} 

\newpage

\begin{center} 
\textbf{Declaration}
\end{center} 
\begin{quote}

I hereby declare that this thesis is my own work and effort and is result of collaboration with David Berman, Malcolm Perry and Daniel Thompson. The material presented further describes results of the publications \cite{Berman:2011kg, Berman:2012uy, Musaev:2013rq, Berman:2013ab}. Where other sources of information have been used, they have been acknowledged Additionally, during the course of the preparation of this thesis the author has published another article \cite{Bolmatov:2013kpa} the results of which are not included here.

This thesis has not been submitted previously in whole or in part for a degree
examination at this or any other institution.

 \end{quote}

\newpage
\tableofcontents

\newpage

\chapter{Introduction}

\section{Invitation to the topic}

String theory is arguably the most developed candidate for a theory of everything. It appeared as an attempt to describe strong interactions and dualities in scattering amplitudes. Soon it was rediscovered as a possible theory of quantum gravity \cite{Green:1987sp}. It was realised that the spectrum of a closed string contains excitations of spin 2 which were then identified with gravitons, which caused the significant transition in the understanding of strings from simply tubes between quarks to the most elementary constituents of matter. This resulted in intense studying of fundamental strings and led to discovery of five different consistent superstring theories that live in 10 dimension: Type I, Type IIA and IIB, $SO(32)$ and $E_8$ heterotic strings. These theories differ by gauge symmetries, set of fields, boundary conditions and realisation of supersymmetry.

The situation appeared to be very strange: after years of looking for a theory of everything one eventually ends up with five of them having no way to choose the correct one. The way out of this trouble was tightly connected to the problem of extra dimensions in string theories. Almost one hundred years before these events T. Kaluza and F. Klein suggested one could consider the Maxwell field $A_\m$ as a part of 5-dimensional metric. Assuming, that the fifth dimension is compact with very small radius of compactification they showed that General Relativity on such a background is equivalent to the 4-dimensional theory of electromagnetic field interacting with gravity. The same idea can be used to get rid of extra 6 dimensions of string theories.

For example one can choose a 6-dimensional torus $\mathbb{T}^6$ as an internal space. Since the torus is flat it preserves reparametrisation invariance of the worldsheet and Virasoro algebra, that is local. An amazing feature of Type IIA and Type IIB string theories is that compactified on $\mathbb{T}^1$ they become equivalent on quantum level \cite{Fradkin:1984ai, Kikkawa:1984cp, Sakai:1985cs}. This is a particular case of the so-called T-duality that is the oldest known duality in string theory \cite{Vafa:1997pm, Giveon:1994fu}. It relates two heterotic string theories to each other as well.

T-duality is a perturbative symmetry and can be seen manifestly in the spectrum of a closed string living on a background with compact directions. An example of a non-perturbative symmetry is provided by S-duality of Type IIB string theory in 10 dimensions, that is $SL(2,\mathbb{Z})$. In addition, S-duality relates heterotic $SO(32)$ strings to Type I strings. Finally, type IIA theory in the strong coupling regime behaves as an 11-dimensional theory whose low-energy limit is captured by 11-dimensional supergravity. The same supergravity being compactified on a unit interval $\mathbb{I}=[0,1]$ leads to the low-energy limit of $E_8$ heterotic theory.

The net of dualities that unifies all five string theories gives a hint that there should exist a mother theory that gives all string theories in various limits and lives in 11 dimensions. Such theory is commonly referred to as M-theory and, although it has not been understood in great details, a lot of is already known about its structure.

M-theory describes dynamics of 2- and 5-dimensional membranes (the so-called M2- and M5-branes) and reduces to 11-dimensional supergravity in its low-energy limit. Being compactified on a circle $\mathbb{S}^1$ M-theory is equivalent to Type IIA string theory. A fundamental string then is associated to an M2-brane wrapped around the circle. The other objects of Type IIA string theory like D2, D4 branes for example appear from the fundamental objects of M-theory in a similar way \cite{Obers:1998fb, Schwarz:1998fd, delMoral:2012pr}.

On the other hand M-theory compactified on a torus $\mathbb{T}^2$ gives rise to Type IIB string theory compactified on a circle $\mathbb{S}^1$. S-duality symmetry $SL(2,\mathbb{Z})$ of Type IIB theory becomes transparent in this picture and is just the modular group of the 2-dimensional torus. Together S- and T-dualities are combined into a non-perturbative set of symmetries of M-theory that is called U-duality \cite{Hull:1994ys}.

These dualities provide a powerful instrument for studying string compactifications, moduli stabilization, properties of string backgrounds, and were intensively studied for many years (for review see \cite{Giveon:1994fu, Grana:2008yw, Becker:2003yv, Becker:2003sh}). However, the partition function of a superstring is not manifestly invariant under these  transformations. In \cite{Tseytlin:1990nb, Tseytlin:1990ar,Duff:1989tf} the formulation of the worldsheet action for a string where T-duality of a background is manifest was proposed. The idea was to consider combinations of coordinates of a closed string $X=X_++X_-$ and $\tilde{X}=X_+-X_-$ as independent variables. Then $O(d,d)$ T-duality symmetry becomes manifest if the action is rewritten in terms of $2d$ extended coordinates $\XX=(X,\tilde{X})$. The Buscher procedure, described in details in further sections, gives a well defined algorithm for gauging the isometry, integrating out gauge fields and obtaining the T-dual sigma-model. This leads to the notion of the so-called generalised metric that puts the space-time metric and the gauge fields on an equal footing and allows one to consider diffeomorphisms and gauge transformations as a part of more general transformations of extended space. 

The duality invariant approach on which the thesis is focused, is an incredibly fascinating construction. Among other applications, the most intriguing feature of this approach is that both non-geometric and geometric backgrounds of string theory become geometric in terms of the extended space. Although geometry of the extended space is still a mystery and very little is known about its structure, one already sees useful applications such as gauged supergravities, studying non-geometric fluxes, $SU(3)$ structures, global properties of backgrounds and many others. Good pedagogical reviews of this approach and its applications can be found in \cite{Aldazabal:2013sca,Berman:2013eva,Hohm:2013bwa}.

\section{Structure of thesis}

The focus of this thesis is on the duality invariant approach in the context of string and M-theory. In the next section we start with brief introduction to duality symmetries in string and M-theory. It is explicitly demonstrated how the extended space and the generalised metric follow from Duff's procedure.

In Chapter \ref{reduction} we investigate dimensional reduction of the extended space by U-duality valued Scherk-Schwarz twist matrices. It is shown that this reduction successfully reproduces the known structures of gauged supergravities, such as the embedding tensor, scalar potential and gauge group. A brief introduction to gauged supergravities is presented in the beginning of this chapter. The most laborious calculations of this chapter are contained in the Appendix.

Chapter \ref{bound} is devoted to boundary terms in the duality invariant formalism. The potential for Double Field Theory, commonly written only up to a full derivative term, acquires an extra duality invariant term that reduces to the known Gibbons-Hawking term if the section condition is satisfied. This boundary term is written in terms of a generalised normal. For backgrounds with non-trivial monodromy properties the boundary term does not vanish as is shown explicitly for the example of the $5_2^2$ exotic brane.

\section{Dualities in string and M-theory}

The action for a string on a background defined by metric $G_{\m\n}$ and the Kalb-Ramond 2-form field $B_{\m\n}$ is given by the Howe-Tucker action for the 2-dimensional non-linear sigma model \cite{Brink:1976sc, Deser:1976rb}
\begin{equation}
\label{Pol_action}
S_P=\int d\t d\s \left(\sqrt{-h}h^{ab}G_{\m\n} + \e^{a b}B_{\m\n}\right)\dt_aX^{\m}\dt_bX^\n,
\end{equation}
here $\{\t,\s\}$ are coordinates on the world-sheet of the string and $h_{ab}$ is the world-sheet metric. Embedding of the two-dimensional string world-sheet into the target space is described by $D$ functions $X^\m(\t,\s)$, where the Greek indices run from 1 to $D$.  The symmetries of the  theory include the target space diffeomorphisms $X'^\m=X'^\m(X)$, the world-sheet reparamterizations  $\s'^a=\s'^a(\s^a)$ and the Weyl transformations $h'_{a b}(\s^a)=e^{\w(\s)}h_{a b}(\s^a)$. The quantum corrections respect the Weyl symmetry only for certain choices of the dimension $D$ of the target space (the famous $D=26$ for the bosonic string and $D=10$ for the superstring). For more details the reader is referred to the classical textbooks on string theory  \cite{Green:1987sp, Polchinski:1998rq} and the reviews \cite{Tong:2009np, Akhmedov:2009zz}.

In addition to the symmetries listed above there are number of non-manifest transformations of fields involved in string theories that relate different theories to each other. One example of this kind of dualities in string theory is presented by the target space duality or T-duality \cite{Vafa:1997pm}. 

\subsection{Closed string spectrum}

T-duality is usually better understood in the context of backgrounds consistent with dimensional reduction by compactification. In the case of compactification on a torus T-duality acts along cycles of the torus replacing a cycle with radius $R$ by a cycle with radius $\a'/R$ relating two different theories. 

Consider a closed string and start with flat background with one compact direction $\mathbb{R}^{1,D-2}\times\mathbb{S}^1$ of radius $R$ and set the Kalb-Ramond field to be zero, $B_{\m\n}=0$. Gauge transformations represented by the worldsheet reparametrisations and the Weyl transformation  can be used to further simplify the worldsheet metric and bring it to diagonal form $||h_{ab}||=||\h_{ab}||\equiv\mbox{diag}[1,-1]$. The resulting action is then given by
\begin{equation}
\label{free_string_action}
S=\int d\t d\s \,\h^{ab}\dt_a X^\m\dt_b X_\m.
\end{equation}
Variation of the action with respect to the fields $X^{\m}$ reads
\begin{equation}
\d S=-\int d\t d\s \, \h^{ab} \d X^{\m}\dt_a \dt_b X_{\m} +\int d\t d\s \dt_a\left(\h^{a b}\d X^{\m}\dt_b X_{\m}\right)=0.
\end{equation}
Assuming that the variation $\d X^{\m}(\t,\s)$ is an arbitrary function of $\s$ and $\t$ that vanishes as $\t \rightarrow \pm \infty$, the first term gives rise to the known Klein-Gordon-type equation $\dt^{a}\dt_{a}X^{\m}=0$ while the second leads to boundary conditions. For a closed string the boundary conditions will be
\begin{equation}
\begin{split}
 X^{\hat{\a}}(\t,\s+2\p)&=X^{\hat{\a}}(\t,\s),\quad \mbox{for ${\hat{\a}}=1,\ldots,D-1$}\\
 \theta(\t,\s+2\p)&=\theta(\t,\s)+2\p m R, \quad m\in\mathbb{Z},
\end{split}
\end{equation}
where the compact coordinate of the target space is denoted by $\q$. The integer number ${m}$ shows how many times the closed string is wrapped around the compact direction and is called the \emph{winding number}.

Components of momentum of the string which correspond to the non-compact directions ${\hat{\a}}=1\ldots D-1$ give rise to the mass spectrum, while the remaining component $p_\q$ becomes quantized and leads to the tower of states. This follows from the condition that the string wave-function on the circle $\mathbb{S}^1$ should be uniquely defined. The action of the vertex operator on the ground state of the string gives a general state of the string
\begin{equation}
|\z,p\rangle=\int d\s \P(\z,X^{\m})e^{ip_\m X^{\m}} |0\rangle,
\end{equation}
where $\P(\z,X^{\m})$ is some combination of the polarization of the string $\z_{\m_1\ldots \m_n}$ and the coordinates $X^{\m}$, whose explicit form is irrelevant for the discussion. The quantization of the momentum follows from the phase factor in the exponent and states
\begin{equation}
p_\q=\fr{2\p n}{R}.
\end{equation}
Finally, this leads to the mass spectrum which depends both on the winding number $m$ and the translational mode number $n$
\begin{equation}
\label{mass_spectrum}
M^2=\fr{n^2}{R^2}+\fr{m^2 R^2}{\a'^2} + 2(N+\tilde{N}-2),
\end{equation}
where $N$ and $\tilde{N}$ denote the standard number operator. One can immediately see that the closed string mass spectrum is invariant under change of the radius $R$ to $\a'/R$ with additional replacing the winding modes by the translational modes
\begin{eqnarray}
&&R\longleftrightarrow \fr{\a'}{R},\\
&&m \longleftrightarrow n.
\end{eqnarray}
Hence, the equivalence of small and large circles from the point of view of a closed string is shown in this simple example. 

\subsection{The Buscher rules}

So far, the Kalb-Ramond field was set to be zero and the background metric was taken to be flat for simplicity. Dropping these conditions reveals more complicated structure of T-duality transformations that now involve not only inverting the radius $R$ but also non-trivial transformations of the metric $G_{\m\n}$ and the 2-form field $B_{\m\n}$ that are known as the Buscher rules \cite{Buscher:1985kb, Buscher:1987sk, Buscher:1987qj}. The procedure that derives the Buscher rules may be referred to as a path integral approach, since it is concerned with Lagrange multipliers and integrating out non-dynamical fields. Although in the further description of this procedure the path integral is not mentioned, since it does not change things drastically, few comments on one-loop quantum effects are made in the end.

For further discussion it is useful to write the action \eqref{Pol_action} in the conformal gauge and adopt the light cone world-sheet coordinates $\s_\pm=1/2(\t\pm\s)$ 
\begin{equation}
\begin{aligned}
S_1[\q]&=\int d\s (G+B)_{\m\n}\dt_+X^\m\dt_-X^\n=\\
&=\int d\s \left(G_{\q\q}\dt_+\q\dt_-\q + E_{\hat{\a}\q}\dt_+X^{\hat{\a}}\dt_-\q+E_{\q \hat{\a}}\dt_+\q\dt_-X^{\hat{\a}}+\right.\\
&\ph{=\int d\s (}\left.+E_{\hat{\a}\hat{\b}}\dt_+X^{\hat{\a}}\dt_-X^{\hat{\b}}\right),
\end{aligned}
\end{equation}
where the notation $E_{\m\n}=G_{\m\n}+B_{\m\n}$ was introduced. Since the coordinate $\q$ is a coordinate on the circle $\mathbb{S}^1$ this action is invariant under global $U(1)$ transformations $\q'=\q+\x$, where $e^{i\x}\in U(1)$. The idea is to make this symmetry local by introducing covariant derivatives
\begin{equation}
D\q=d\q+A,
\end{equation}
with the gauge field $A=A_+d\s^+ +A_-d\s^-$. The gauge field $A$ should be fixed to be a pure gauge so not to increase the number of degrees of freedom of the theory. This can be done by making use of a Lagrange multiplier
\begin{equation}
\begin{aligned}
S_2[\q,\l]&=\int d\s (G+B)_{\m\n}\dt_+X^\m\dt_-X^\n=\\
&=\int d\s \left(G_{\q\q}D_+\q D_-\q + E_{{\hat{\a}}\q}\dt_+X^{\hat{\a}}D_-\q+E_{\q {\hat{\a}}}D_+\q\dt_-X^{\hat{\a}}+\right.\\
&\ph{=\int d\s (}\left.+E_{{\hat{\a}}{\hat{\b}}}\dt_+X^{\hat{\a}}\dt_-X^{\hat{\b}}+\l F_{+-}\right).
\end{aligned}
\end{equation}
Integrating over the Lagrange multiplier $\l$ in the string path integral leads to the condition $F_{-+}=0$, whose solutions imply that the gauge field is a pure gauge
\begin{equation}
\begin{split}
&A_+=\dt_+\f\\
&A_-=\dt_-\f.
\end{split}
\end{equation}
This condition reverts the action $S_2[\q,\l]$ back to the initial action $S_1[\q+\f]$ that is equal to $S_1[\q]$ up to a field redefinition.

An alternative way to proceed is to leave the Lagrange multiplier $\l$ but exclude the gauge field $A$. Equations of motion of the gauge field are algebraic and thus can be easily solved providing
\begin{equation}
\begin{split}
&A_+=\fr{1}{G_{\q\q}}\dt_+\l+\fr{1}{G_{\q\q}}E_{\q {\hat{\a}}}\dt_-X^{\hat{\a}} \\
&A_-=-\fr{1}{G_{\q\q}}\dt_-\l+\fr{1}{G_{\q\q}}E_{{\hat{\a}} \q}\dt_+X^{\hat{\a}}.
\end{split}
\end{equation}
The action $S_2[\q,\l]$ with $A_+$ and $A_-$ replaced according to these expressions becomes
\begin{equation}
\begin{aligned}
S_3[\l]&=\int d\s \left(G'_{\l\l}\dt_+\l\dt_-\l + E'_{{\hat{\a}}\l}\dt_+X^{\hat{\a}}\dt_-\l+E'_{\l {\hat{\a}}}\dt_+\l\dt_-X^{\hat{\a}}+\right.\\
&\ph{=\int d\s (}\left.+E'_{{\hat{\a}}{\hat{\b}}}\dt_+X^{\hat{\a}}\dt_-X^{\hat{\b}}\right),
\end{aligned}
\end{equation}
where the gauge was fixed by setting $\q=0$. This action has the same form as $S_1[\q]$ but the background is different
\begin{equation}
\label{Buscher}
\begin{aligned}
G'_{\l\l}&=\fr{1}{G_{\q\q}},\\
E'_{\l {\hat{\a}}}&=\fr{1}{G_{\q\q}}E_{\q {\hat{\a}}},\\
E'_{{\hat{\a}}\l}&=-\fr{1}{G_{\q\q}}E_{{\hat{\a}}\q},\\
E'_{{\hat{\a}}{\hat{\b}}}&=E_{{\hat{\a}}{\hat{\b}}}-E_{{\hat{\a}}\q}\fr{1}{G_{\q\q}}E_{\q {\hat{\b}}}.
\end{aligned}
\end{equation}
These transformations are referred to as the Buscher rules and define the transformation of the background under T-duality. Both the actions $S_1$ and $S_3$ are equivalent to the action $S_2$, thus they are equivalent to each other and describe the same physics. Transformations \eqref{Buscher} are non-linear transformations that mix the metric $G_{\m\n}$ and the Kalb-Ramond field $B_{\m\n}$, thus mixing target space diffeomorphisms with gauge transformations $B'=B+d\L$.

The procedure described above is pure classical and does not take into account contribution from the dilaton measure. A correct one-loop calculation shows that in addition to the T-duality transformations of the metric and the B-field listed above one should consider the transformation of the dilaton
\begin{equation}
\f'-\fr14\ln\det g'=\f-\fr14\ln \det g
\end{equation}
providing the combination $\sqrt{g}e^{-2\f}$ is invariant.

\subsection{Duff's procedure for the F1-string and the M2-brane}

The so-called Duff's procedure reveals another non-trivial feature of T-duality transformations: hidden symmetry between equations of motion and the Bianchi identities \cite{Duff:1989tf, Duff:1990hn}. Starting from this symmetry one can introduce a set of dual coordinates and the so-called dual Lagrangian that governs the dual dynamics. The key point is that equations of motion for the dual coordinates appear to be equivalent to Bianchi identities for the ordinary coordinates and vice versa. A relation between these coordinates leads to the notion of the generalised metric.

Consider a bosonic string on a background given by the metric and the B-field that do not depend on  $X^{\m}$. The reason for this is that we have in mind toroidal compactifications with $X^\m$ the compactified coordinates. The equations of motion for the field $X^{\m}$ that follow from the action \eqref{Pol_action} have the form of the conservation law $\dt_a\tilde{\mc{G}}^a_\m=0$ for some current
\begin{equation}
\label{tilde_G}
\tilde{\mc{G}}^a_\m=\left(\sqrt{-h}h^{ab}G_{\m\n}+\e^{ab}B_{\m\n}\right)\dt_bX^{\n}.
\end{equation}
Locally solutions of this equation can be represented by the Hodge dual of the full derivative $\tilde{\mc{G}}^a_\m:=\e^{ab}\dt_bY_{\m}$ of the would-be dual coordinates $Y_{\m}$, that leads to the following equation
\begin{equation}
\label{EOMX}
\left(\sqrt{-h}h^{ab}G_{\m\n}+\e^{ab}B_{\m\n}\right)\dt_bX^{\n}=\e^{ab}\dt_bY_{\m}.
\end{equation}
Hence, taking the derivative $\dt_a$ of this expression one obtains the equations of motion for $X^{\m}$ on the left hand side and the Bianchi identities $\e^{ab}\dt_a\dt_bY_{\m}=0$ for the field $Y_{\m}$ on the right hand side. 

The equations of motion for the field $X^{\m}$ can be equivalently derived from the first order Lagrangian by introducing an extra independent field $U_a^\m$
\begin{equation}
\mc{L}_x=\fr12(\sqrt{h}h^{ab}G_{\m\n}+\e^{ab} B_{\m\n})U_a^{\m}U_b^{\n}-\tilde{\mc{G}}^a_\m U_a^{\m},
\end{equation}
where the current $\tilde{\mc{G}}^a_\m$ is written in terms of the field $X^\m$. Equations of motion for the fields $U_a^{\m}$ and $X^{\m}$ that follow from this Lagrangian give an algebraic constraint on the auxiliary field  and the equation \eqref{EOMX} respectively.
\begin{equation}
\begin{aligned}
\fr{\dt \mc{L}_x}{\dt U_a^{\m}}=0 & \Longrightarrow  U_a^{\m}-\dt_aX^{\m}=0,  \\
\dt_a \,\fr{\dt \mc{L}_x}{\dt\, \dt_a X^{\m}}=0 & \Longrightarrow \dt_a\left[(\sqrt{-h}h^{ab}G_{\m\n} +\e^{ab}B_{\m\n})U_b^{\n}\right]=0.
\end{aligned}
\end{equation}
Solution of the first line, given by $U_a^{\m}=\dt_a X^{\m}$, implies that the second line is exactly \eqref{EOMX}.

The dual Lagrangian for the field $Y_{\m}$ has exactly the same form as $\mc{L}_x$ but with $\tilde{\mc{G}}_\m^a$ expressed in terms of $Y_\m$
\begin{equation}
\mc{L}_y=\fr{1}{2}\left(\sqrt{-h}h^{ab}G_{\m\n}+\e^{ab}B_{\m\n}\right)U_a^{\m}U_b^{\n}-\e^{ab}\dt_a Y_{\m} U_b^\m.
\end{equation}
The variation of the corresponding action with respect to $Y_{\m}$ gives the Bianchi identities $\e^{ab}\dt_a U_b^\m\equiv \e^{ab}\dt_a\dt_b X^{\m} =0$ for the field $X^{\m}$ while variation with respect to $U_a^{\m}$ implies
\begin{equation}
\left(\sqrt{-h}h^{ab}G_{\m\n}+\e^{ab}B_{\m\n}\right)U_b^{\n}=\e^{ab}\dt_bY_{\m}.
\end{equation}
It is straightforward to solve this equation with respect to $U_a^{\m}$ and to write the solution as
\begin{equation}
\label{EOMY}
\e^{ab}U_b^M=\left(\sqrt{-h}h^{ab}p^{\m\n}+\e^{ab}q^{\m\n}\right)\dt_b Y_\n,
\end{equation}
where $p_{\m\n}\equiv G_{\m\n}+B_{\m\a}B^\a{}_{\n}$ and $p_{\m\a}q^{\a\n}=B_\m{}^\n$. This expression has exactly the same form as \eqref{EOMX}, but with $X^{\m}$ replaced by $Y_{\m}$ and the corresponding background transformation, that is actually a T-duality rotation. Hence, the intermediate result is that doubling of coordinates reveals hidden symmetry of equations of motion for a bosonic string and the Bianchi identities. 

To make this symmetry manifest it is useful to make the following definitions
\begin{equation}
\label{idents}
\begin{aligned}
&\mc{G}^a_\m=\sqrt{-h}h^{ab}\dt_bY_{\m}, & \mc{F}^{a\m}&=\sqrt{-h}h^{ab}\dt_b X^{\m},\\
&\tilde{\mc{G}}^a_\m=\e^{ab}\dt_b Y_\m, & \tilde{\mc{F}}^{a\m}&=\e^{ab}\dt_b X^{\m},
\end{aligned}
\end{equation}
and to rewrite two sets of equations \eqref{EOMX} and \eqref{EOMY} as
\begin{equation}
\label{FG_eqns}
\begin{split}
\tilde{\mc{G}}^a_\m=&G_{\m\n}\mc{F}^{a\n}+B_{\m\n}\tilde{\mc{F}}^{a\n} \\
\tilde{\mc{F}}^{a\m}=&p^{\m\n}\mc{G}^a_{\n}+q^{\m\n}\tilde{\mc{G}}^a_\n .
\end{split}
\end{equation}
The first equation here is just the equation $\eqref{tilde_G}$, while the second one can be reduced to \eqref{EOMY} multiplying by $p_{\a\m}$. Structure of these equations suggests to combine indices into one set introducing matrix notations
\begin{equation}
\label{EOM}
\h_{MN}\tilde{\F}^{iN} =\mc{H}_{MN}\F^{iN},
\end{equation}
where the capital Latin indices $M,N=1\ldots 2n$. Here the objects $\mc{G}$ and $\mc{F}$ were collected into two $2n$-rows
\begin{eqnarray}
\tilde{\F}^{iM}=\begin{bmatrix}
\tilde{\mc{F}}^{a\m}\\
\tilde{\mc{G}}^a_\m
\end{bmatrix}, &   \F^{iM}= \begin{bmatrix}
{\mc{F}}^{a\n}\\
{\mc{G}}^a_{\n}
 \end{bmatrix}
\end{eqnarray}
and $2n\times 2n$ matrices $\mc{H}$ and $\h$ are defined as
\begin{eqnarray}
\mc{H}_{MN}=
\begin{bmatrix}
G_{\a\b}-B_{\a\r}B^{\r}{}_\b & -B_\a{}^{\n}\\
 B^{\m}{}_{\b}& G^{\m\n} 
\end{bmatrix}, &
 \h_{MN}=\begin{bmatrix}
0 &\d^\m_\n\\
\d^\a_\b & 0
\end{bmatrix}.
\end{eqnarray}
Now the $SO(n,n)$  symmetry of \eqref{EOM} becomes apparent. The objects $\tilde{\F}$ and $\F$ transform in the fundamental representation of $SO(n,n)$, the matrix $\mc{H}$ transforms as a 2-tensor and the matrix $\h_{MN}$ is an $SO(n,n)$ invariant tensor:
\begin{equation}
\label{Oddtransf}
\begin{aligned}
&\F'^{iM}=\mc{O}^M{}_N\F'^{iN},  &  \mc{H}'_{MN} &= \mc{O}_M{}^K\mc{H}_{KL}O_N{}^L,\\
&\tilde{\F}'^{iM}=\mc{O}^M{}_N\tilde{\F}'^{iN},& \h_{MN} &= \mc{O}_M{}^K\h_{KL}O_N{}^L.
\end{aligned}
\end{equation}
Note that the last equation together with the explicit form of $\n_{MN}$ implies that $\mc{O}\in SO(n,n)$. 

The matrix $\mc{H}_{MN}$, that is the so-called generalised metric, allows to consider the ordinary metric $G$ and the 2-form field $B$ on an equal footing. Moreover, while T-duality is realised by the non-linear transformations of the supergravity fields \eqref{Buscher} the generalised metric transforms linearly \eqref{Oddtransf}. One can check that the linear $SO(d,d)$ transformations of $\mc{H}_{MN}$ are precisely equivalent to the Buscher rules.

This procedure is not a unique property of string theory and can be applied to dynamics of extended objects of other dimensions such as D-branes \cite{Bakhmatov:2011ab,Bakhmatov:2011be} or M2- and M5-branes of M-theory \cite{Berman:2010is}. The bosonic part of the action for the theory involving M2-branes interacting with the 3-form field $C_{\m\n\a}$ can be written as
\begin{equation}
\label{2.0}
 \mc{S}=\int d^3{\x}\sqrt{- h}\left[\fr12h^{ab}G_{\m\n}\dt_a X^\m\dt_b X^\n+\fr16\e^{abc}C_{\a\m\n}\dt_a X^\a \dt_b X^\m \dt_c X^\n-\fr12\right],
\end{equation}
where the integration is performing over the M2-brane world-volume $\sqrt h d^3\x$ that lives in the bulk with the metric $G_{\m\n}$. The
3-form matter field couples to the brane in the way that is a natural generalization of electromagnetic and Kalb-Ramond coupling:
\begin{equation}
 \begin{split}
    & A_\m \dt_\t X^\m;\\
    & B_{\m\n} \e^{ab} \dt_a X^\m \dt_b X^\n,
 \end{split}
\end{equation}
corresponding to zero- and one-dimensional fundamental objects respectively. The M5-branes are carriers for magnetic charge associated with the field strength $F[C]=dC_3$.

Consider the specific case of $SL(5)$ duality group that arises in $\mathbb{T}^4$ spatial reductions of M-theory so that there are 4 commuting Killing vectors. The metric and 3-form are still
independent of the four coordinates $X^\m$ associated with these Killing vectors. Suppose in addition that there are no other directions in space-time in
which the M2--brane is moving. Under such simplifications the equations of motion for the field $X^\m$ that follow from \eqref{2.0} have again the form of the conservation law  $\dt_a \tilde{\mc{G}}^a_\m=0$ with the current defined as
\begin{equation}
\label{2.1.1}
\tilde{\mc{G}}^a_\m(X)=\sqrt{-h}G_{\m\n}\dt_a X^\n+\fr{1}{2} \e^{abc}C_{\m\n\r}\dt_b X^\n \dt_c X^\r.
\end{equation}
At least locally solutions of this equation can be written in the following form
\begin{equation}
\tilde{\mc{G}}_\m^a(Y)=\e^{abc}\dt_b X^\n \dt_c Y_{\m\n},
\end{equation}
where the dual coordinate $Y_{\m\n}$ was naturally introduced. As in the case of the F1-string the equations of motion for the dual coordinate that follow from the dual first order Lagrangian are exactly the Bianchi identities for the field $X^\m$. We can write the first order Lagrangian $\mc{L}_x$ and its dual $\mc{L}_y$ as
\begin{equation}
 \begin{aligned}
\mc{L}_x =&-\fr12\sqrt{-h} h^{ij} U_i^\m U_j^\n G_{\m\n} -\fr13\e^{ijk} U_i^\m U_j^\n U_k^\a  C_{\m\n\a}+U_i^\m\mc{G}^i_\m(X),\\
&\mbox{and}\\
\mc{L}_y =&-\fr12\sqrt{-h} h^{ij} U_i^\m U_j^\n G_{\m\n} -\fr13\e^{ijk} U_i^\m U_j^\n U_k^\a  C_{\m\n\a}+U_i^\m\mc{G}^i_\m(Y),
\end{aligned}
\end{equation}
where the auxiliary field $U_i^\m$ was introduced as before. Equations of motion for the auxiliary field that follow from the Lagrangian $\mc{L}_x$ imply the algebraic constraint $U_i^\m=\dt_iX^\m$, while the variation of the corresponding action with respect to the field $X^\m$ gives $\dt_a\tilde{\mc{G}}^a_\m=0$. The dual Lagrangian $\mc{L}_y$ gives the following equations of motion for the auxiliary field 
\begin{equation}
\sqrt{-h} h^{ij}U_j^\n G_{\m\n} +\e^{ijk}  U_j^\n U_k^\a  C_{\m\n\a}=\mc{G}^i_\m(Y)=\e^{ijk}\dt_j X^\n \dt_k Y_{\m\n}.
\end{equation}
Variation of the dual action with respect to $Y_{\m\n}$ leads to the Bianchi identities $\e^{ijk}\dt_j\dt_k X^\m=0$ on the filed $X^\m$.

In analogy with Duff's procedure for the F1-string one introduces the following variables
\begin{equation}
\label{identsM2}
\begin{aligned}
&\mc{G}^a_{\m\n}=\sqrt{-h}h^{ab}\dt_bY_{\m\n}, & & \mc{F}^{a\m}=\sqrt{-h}h^{ab}\dt_b X^{\m},\\
&\tilde{\mc{G}}^a_\m=\e^{abc}\dt_{b}X^{\n}\dt_c Y_{\m\n}, & & \tilde{\mc{F}}^{a\m\n}=\e^{abc}\dt_b X^{\m}\dt_c X^{\n}.
\end{aligned}
\end{equation}
The fields $\cg^a_{\m\n}$ and $\cf^{a\m}$ are straight derivatives of the coordinates and are therefore
rather like displacements, whereas $\tilde{\cg}^a_{\mu}$ and $\tilde{\cf}^{a\m\n}$ 
are rather like field strengths. These allow to write equations that follow from the Lagrangians $\mc{L}_x$ and $\mc{L}_y$ in a simple form
\begin{equation}
\label{FG_eqnsM2}
\begin{split}
\tilde{\mc{G}}^a_\m=&G_{\m\n}\mc{F}^{a\n}+C_{\m\n\r}\tilde{\mc{F}}^{a\n\r} \\
\tilde{\mc{F}}^{a\m\n}=&p^{\m\n,\a\b}\mc{G}^a_{\a\b}+q^{\m\n\r}\tilde{\mc{G}}^a_\a .
\end{split}
\end{equation}
The fields $p_{\m\n,\r\s}$ and $q^{\m\n\r}$ that naturally appear here are defined by the following relations
\begin{equation}
\begin{aligned}
p_{\a\b,\m\n}q^{\m\n\r}&=-C_{\a\b\g}G^{\m\g},\\
p_{\a\b,\l\q}p^{\l\q,\m\n}&=\d_\a^{[\m}\d_\a^{\n]},\\
p_{\a\b,\m\n}&=G_{\a[\m}G_{\n]\b}-C_{\a\b\r}C^\r{}_{\m\n},
\end{aligned}
\end{equation}

 Introducing the generalised index $M=\{\m,\a\b\}$, that runs from 1 to 10 labelling the $\bf{10}$ representation of $SL(5)$, 
we can write the above equations in the following compact form
\begin{equation}
{\F}^{a M} = M^{MN} \tilde{\F}_N^a,
\end{equation}
where the matrix $M^{MN}$ is the desired generalised metric
\begin{equation}
\label{2.3}
	M^{MN}
= 	
	\begin{bmatrix}
	      G^{\m\n}+\fr12 C^{\m\d\g}C^\n{}_{\d\g} &	& \fr{1}{\sqrt{2}}C^{\m}{}_{\r\s} \\
	      &&\\
	      \fr{1}{\sqrt{2}}C^{\n}{}_{\a\b}	&	& G_{\a\b,\r\s}
        \end{bmatrix}
\end{equation}
and the variables \eqref{identsM2} were collected into the objects
\begin{equation}
\begin{aligned}
\F^{a M} = 
\begin{bmatrix}
	      \cf^{a\m}\\ \\
	      {\cg}^a_{\a\b}
\end{bmatrix},&&
\tilde{\F}^a_{M} = 
\begin{bmatrix}
	      \tilde{\cg}^a_{\n}\\
	      \\
	      \tilde{\cf}^{b\r\s}
	\end{bmatrix}.
\end{aligned}
\end{equation}
The tensor $G_{\m\n,\r\s}=\fr12(G_{\m\r}G_{\n\s}-G_{\m\s}G_{\n\r})$ is used to lower and raise an antisymmetric pair of indices. Finally, the Bianchi identities and the equations of motion can be unified as $\dt_a\tilde{\F}^a_M=0$.

\section{Extended geometry and generalised metric}
\label{GG_metric}

As it was shown in the previous section the generalised metric $\mH_{MN}$ naturally appears when considering theory of closed string on toroidal backgrounds. This metric appeared in the early works on T-duality and defined the first quantized Hamiltonian of a closed string on toroidal backgrounds \cite{Giveon:1988tt, Duff:1989tf}. The Duff's procedure reveals the hidden $O(n,n)$ symmetry of equations of motion for a closed string and the Bianchi identities leading to the generalised metric transforming in a linear representation of the duality group. The matrix $\mH_{MN}$ parametrizes the coset $O(n,n)/O(n)\times O(n)$ that appears in toroidal reductions of supergravity. 

In mathematical literature the concept of a generalised metric appears in generalisations of Calabi-Yau and symplectic manifolds and is usually referred to as the generalised geometry \cite{Hitchin:2004ut, Hitchin:2005in, Gualtieri:2003dx, Cavalcanti:2011wu}. This formalism is based on two ideas: the first is to replace the tangent bundle $T$ of a manifold $M$ by direct sum of the tangent and the cotangent bundles $T\oplus T^*$, the second is to replace the Lie bracket on sections of $T$ by the Courant bracket on sections of $T\oplus T^*$ \cite{courant1990dm}. 

Denoting tangent and cotangent vectors by $Y$ and $\x$ respectively, elements of a fibre $T_p\oplus T_p^*$ at the point $p$ can be represented as a formal sum $v=Y+\x$ with the natural inner product 
\begin{equation}
(v,v)=(Y+\x,Y+\x):=i_Y \x.
\end{equation}
Here $i_Y\xi$ is the interior product or the evaluation $\x(Y)$ that is just index contraction if written in components
\begin{equation}
\begin{aligned}[l]
\i_Y\xi=Y^a\xi_a.
\end{aligned}
\end{equation}
The Courant bracket is defined as a generalisation of the ordinary Lie bracket of vector fields $[Y_1,Y_2]$
\begin{equation}
[Y_1+\x_1,Y_2+\x_2]=[Y_1,Y_2]+\mc{L}_{Y_1}\x_2+\mc{L}_{Y_2}\x_1-\fr12d(i_{Y_1}\x_2-i_{Y_2}\x_1).
\end{equation} 
Failure of the Courant bracket to satisfy the Jacobi identity is an important feature of the algebra it defines. Indeed, for any sections $u, v$ and $w$  of the generalised tangent bundle $T\oplus T^*$ the following is true
\begin{equation}
[[u,v],w]+[[w,u],v]+[[v,w],u]=\fr13 d\left(([u,v],w)+([w,u],v)+([v,w],u)\right),
\end{equation}
which implies that the Courant bracket is not a bracket of any Lie algebra. Application of the formalism of generalised geometry to string theory translates this aspect to the so-called section condition that restricts dynamics of the system. 

Although the generalised tangent space $T\oplus T^*$ has dimension doubled compared to the conventional tangent space $T$, Hitchin's generalised geometry does not introduce extra coordinates. In other words the space $M$ still remains the ordinary manifold. On the contrary, in string or M-theory extended geometry dual coordinates enter the game. The generalised metric introduced by Gualtieri \cite{Gualtieri:2003dx} becomes now a conventional metric defined on the extended space, that still does not admit the structure of a Riemann manifold \cite{Jeon:2010rw,Jeon:2011cn,Jeon:2011sq,Hohm:2011si,Hohm:2012gk,Park:2013gaj}.

String moduli that enter the matrix $\mH_{M N}$ are the metric $G_{ij}$ and the NS-NS gauge field $G_{ij}$. It is known that the low energy effective action for a closed string is that of the supergravity whose bosonic part is
\begin{equation}
\label{sugra_action1}
S_{eff}=\int d x \sqrt{-G}e^{-2\ff}\left(R[G]+4(\dt \ff)^2-\fr{1}{12}H^2\right).
\end{equation}
Here $R[G]$ is the Riemann curvature of the metric $G_{ij}$, $H=dB$ is the field strength of the gauge field $B=B_{ij}dx^i\wedge dx^j$ and the dilaton is denoted as $\ff$. Although, the partition function for string theory is invariant under the T-duality transformations \eqref{Buscher}, it is very non-trivial to show this explicitly. Thus, the symmetry is not manifest or hidden. 

The notion of generalised metric, that puts the metric and the 2-form field on equal footing and transforms linearly under the action of T-duality, allows to rewrite the low energy effective action in a duality covariant 
form. Moreover, the so-called extended space that unifies translational and winding modes has to be introduced.

 In the work by Kugo and Zwiebach \cite{Kugo:1992md} it was shown that a closed string on a toroidal background considers translational and winding modes equally. E.g. they both contribute to the mass spectrum of the closed string on the background given by the torus $\mathbb{S}^1$ of radius $R$ \eqref{mass_spectrum}
\begin{equation}
M=\fr{n^2}{R^2}+\fr{m^2 R^2}{\a'}+(N+\tilde{N}-2),
\end{equation}
where $n$ and $m$ are the (discrete) translational momentum and the winding number. The $\mathbb{Z}_2$ action of T-duality exchanges $n$ and $m$ and replaces $R$ by its inverse.

The winding number can be thought of as a discrete momentum that is dual to the ordinary momentum under the action of T-duality. The inverse Fourier transformation turns the translational mode into the ordinary coordinate $x^i$ and the winding mode into the so-called dual coordinate, that is denoted as $\tx_i$. The theory is now considered as living on a doubled torus $\mathbb{T}^n\times \tilde{\mathbb{T}}^n$ with coordinates $(x^i,\tx_i)$. It is convenient to double not only the compact coordinates but other $d=D-n$ coordinates as well introducing a theory that has manifest $O(D,D)$ invariance and lives on the extended space with $2D$ dimensions.

The construction of the double field theory was developed by Hull, Hohm and Zwiebach in \cite{Hull:2009zb, Hull:2009mi, Hohm:2010jy, Hohm:2010pp}. The main feature of this formalism, in addition to the doubling of coordinates, is a condition that restricts the extended space to a $D$-dimensional space if satisfied. It originates from the Virasoro algebra constraint $L_0-\bar{L}_0=0$ of the closed string theory and states that all fields and all their products must be annihilated by the operator $\dt_i\tdt^i$ (sum over $i$ is understood). 

The strong constraint can be written in an $O(D,D)$ covariant form $\dt_M\dt^M=0$ upon collecting ordinary and dual coordinates into one object 
\begin{equation}
\XX^M=
\begin{bmatrix}
\tx_i\\
x_i
\end{bmatrix},
\end{equation}
that represents coordinates on the extended space labelled by $M=1..2D$. Using this notation it is straightforward to show that the strong constraint effectively reduces the number of dimensions to $D$. Indeed, in the momentum representation section condition reads
\begin{equation}
P_MP^M=0,
\end{equation}
where $P^M=(\tilde{p}^i,p_i)$ is the momentum corresponding to the coordinate $\XX^M$. This implies that momenta $P$ and $P'$ associated with Fourier components of any two fields should be mutually orthogonal and isotropic
\begin{equation}
\begin{aligned}[l]
P\cdot P'=0, & & P\cdot P=0,& & P'\cdot P'=0.
\end{aligned}
\end{equation}
The maximal dimension of such isotropic subspace in the space of signature $(D,D)$ is $D$. Indeed, the equation $P\cdot P=0$ can be written as 
\begin{equation}
p_a\tilde{p}^a=\fr14(k_a+q_a)(k^a-q^a)=0,
\end{equation}
 that has two solutions $k_a=\pm q_a$. Here the new variables $k=p+\tilde{p}$ and $q=p-\tilde{p}$ were introduced. Different choices of this subspace correspond to picking a particular T-duality frame.

Recall the explicit form of the generalised metric and the Buscher rules written in the duality invariant formalism
\begin{equation}
\label{f1_metr}
\begin{aligned}
\mc{H}^{MN}=
\begin{bmatrix}
   G_{ij}-B_i^{\ph{i}a}B_{aj}   & B_i^{\ph{i}k} \\
	&\\
   -B^l_{\ph{l}j}		& G^{kl}
\end{bmatrix}, && \mH'_{MN}=\mc{O}_M{}^K\mH_{KL}\mc{O}^L{}_N,
\end{aligned}
\end{equation}
where $\mc{O}$ is an element of the group $O(D,D)$. The effective action can be expressed in terms of the generalised metric in the duality covariant form \cite{Hohm:2010pp}
\begin{equation}
\label{Odd_action}
\begin{aligned}
    S=\int dx d\tilde x e^{-2d}&\left(\fr18\mc{H}^{MN}\dt_M\mc{H}^{KL}\dt_{N}\mc{H}_{KL} -\fr12
\mc{H}^{KL}\dt_L\mc{H}^{MN}\dt_N\mc{H}_{KM} - \right.\\
      & \left.- 2\dt_M d\dt_N\mc{H}^{MN}+4\mc{H}^{MN}\dt_Md\dt_Nd\lefteqn{\ph{\fr12}}\right) \, .
      \end{aligned}
\end{equation}
Here the dilaton $\ff$ and the determinant of the metric $G=\det{||G_{mn}||}$ are combined into a single object called the duality invariant dilaton
\begin{equation}
d=\ff-\fr14\log G.
\end{equation}
The capital Latin indices are raised and lowered by the $O(D,D)$ invariant $2D\times 2D$ constant metric
\begin{equation}
\h_{MN}=
\begin{bmatrix}
0 & \mathbf{1}_{D\times D} \\
\mathbf{1}_{D\times D} & 0
\end{bmatrix}.
\end{equation}
Structure of the generalised metric implies that application of this rule to $\mH_{MN}$ gives the inverse matrix
\begin{equation}
\begin{aligned}
\mH^{MN}=\h^{MK}\mH_{KL}\h^{LN} && \mH^{MN}\mH_{NK}=\d^M_K.
\end{aligned}
\end{equation}
This suggests to define a matrix $S^M{}_N=\mH^M{}_N=\h^{MK}\mH_{KN}$, that satisfies $S^{t}\h S=\h$ and thus is an element of $O(D,D)$. The matrix $S$ is what was initially defined as the generalised metric in the mathematical literature.

In addition to the global invariance with respect to T-duality transformations, supergravity action is invariant under local symmetries given by the diffeomorphisms and gauge transformations of the Kalb-Ramond field. The formalism of the double field theory allows to unify these transformations and write them in a T-duality (or equivalently an $O(D,D)$) covariant form. 

Consider a generalised gauge parameter $\S^M$ that combines the vector field $\x^a$, which defines the diffeomorphisms, and $\tilde{\x}_a$ that is the parameter of the gauge transformations
\begin{equation}
\S^M=
\begin{bmatrix}
\tilde{\x}_a\\
\x^a
\end{bmatrix}.
\end{equation}
 Then the duality covariant local transformation of an arbitrary generalised vector $V^{M}$ consistent with diffeomorphisms and gauge transformations reads
\begin{equation}
\label{gen_Lie_1}
\begin{aligned}
\d_\S V^M&=\S^N\dt_N V^M -V^N\dt_N \S^M + \dt^M \S_S V^S\\
&=L_\S V^{M}+Y^{MK}{}_{RS}\dt_K \S^R V^S.
\end{aligned}
\end{equation}
It is convenient here to introduce the $O(D,D)$ invariant tensor $Y^{MK}{}_{RS}\equiv \h^{MK}\h_{RS}$ to emphasise that $\d_\S V^M$ is a deformation of the ordinary Lie derivative $L_\S V^M$. This suggests to view the transformation \eqref{gen_Lie_1} as the generalised Lie derivative and write
\begin{equation}
\mc{L}_\S  V^M\equiv \d_\S V^M =L_\S V^M + Y^{MK}{}_{RS}\dt_K \S^R V^S.
\end{equation}
The action of $\mL_\S$ can be defined on any tensor $T^{A_1\ldots A_n}{}_{B_1\ldots B_M}$ by processing each index in the same pattern. In addition one consistently defines the transformation of the dilation to be
\begin{equation}
\d_\S d=\S^M\dt_M d-\fr12\dt_M\S^M =L_\S d - \fr12\dt_M\S^M.
\end{equation}
The generalised Lie derivatives of the $O(D,D)$ metric $\h_{MN}$ and the Kronecker symbol $\d^M{}_N$ vanish explicitly:
\begin{equation}
\begin{aligned}
\mL_\S \h_{MN}\equiv 0, && \mL_\S \d^M{}_N\equiv 0.
\end{aligned}
\end{equation}

Algebra of the local transformations $\d_\S$ is closed only up to the section condition \\ $\dt^M\bullet\,\dt_M\bullet=0$. Indeed, consider commutator of two generalised Lie derivatives calculated on a generalised vector $V_M$
\begin{equation}
[\mL_{\S_1},\mL_{\S_2}]V_M=-\mL_{[\S_1,\S_2]_C}V_M+F_M(\S_1,\S_2,V),
\end{equation}
where $[\S_1,\S_2]_C$ that naturally appears here is the C-bracket which was introduced by Siegel in \cite{Siegel:1993th} and is defined as \cite{Hull:2009zb}
\begin{equation}
\label{C_brckt}
[\S_1,\S_2]_C{}^M=\S^N_{[1}\dt_N\S_{2]}^M-\fr12\S^P_{[1}\dt^M\S_{2]P}.
\end{equation}
The extra term $F_M$ has the following form
\begin{equation}
\label{F_Odd}
F_M(\S_1,\S_2,V)=-\fr12\S_{[1N}\dt^Q\S_{2]}{}^N\dt_Q V_M+\dt^Q\S_{[1M}\dt_Q\S_{2]}{}^P V_P
\end{equation}
and is zero if the strong constraint $\dt_M\bullet\dt^M\bullet=0$ is satisfied.

Although the generalised effective potential \eqref{Odd_action} is written in terms of the fields $\mH_{MN}$ and $d$ living on the extended space of dimension $2D$, the strong constraint effectively reduces the number of dimensions to $D$. As it will be explained further, in the Scherk-Schwarz reduction of the extended space formalism the section condition can be relaxed and turned into conditions on the so-called embedding tensor (or structure constants). This relation between the extended space geometry and the deformations of supergravities means that the extended space is more than a mathematical trick.

It is straightforward to investigate different solutions of the section condition. For this purpose it is convenient to expand the duality invariant action \eqref{Odd_action} and write it in terms of $\dt_i$ and $\tilde{\dt}^i$ that are derivatives with respect to the ordinary and the dual coordinates. The natural form of this expansion suggested by the structure of the effective action itself reads \cite{Hohm:2010jy}
\begin{equation}
\begin{aligned}
S=S^{(0)}+S^{(1)}+S^{(2)}=\int dxd\tx\left(\mL^{(0)}+\mL^{(1)}+\mL^{(2)}\right),
\end{aligned}
\end{equation}
where the number in the superscript denotes the order of the dual derivative $\tdt^i$ in the corresponding expression. Hence, the first term contains no dual derivatives and thus has the form of the conventional supergravity action (up to boundary terms)
\begin{equation}
\label{S0}
\begin{aligned}
&&\mL^{(0)}=& e^{-2d}\left[-\fr14 G^{ij}\dt_i G^{kl}\dt_j G_{kl}+\fr12 G^{ij}\dt_i G^{kl}\dt_j G_{kl} +2\dt_i d\dt_jG^{ij} +4G^{ij}\dt_i d\dt_j d -\fr{1}{12}H^2 \right]\\
&&=& e^{-2\ff}\sqrt{-g}\left(R[G]+4(\dt \ff)^2-\fr{1}{12}H^2\right)+\mbox{boundary terms}.
\end{aligned}
\end{equation}
Here $H=dB$ is the field strength for the Kalb-Ramond field, $R[G]$ is the Riemann curvature for the metric $G_{kl}$ and in the second line the explicit form of the invariant dilaton $d=\ff -\fr14 \log G$ was used.

Since the whole formalism is duality invariant and the term $\mL^{(2)}$ contains only dual derivatives $\tdt^i$ it has to be very similar to $\mL^{(0)}$ and T-dual to it. Introducing the field $\mE_{ij}=g_{ij}+b_{ij}$ this term can be written as
\begin{equation}
\label{S2}
\begin{aligned}
&&\mL^{(2)}=e^{-2d}&\left[ -\fr14 g^{ik}g^{jl}g^{pq}\left(\mE_{pr}\mE_{qs}\tdt^r \mE_{kl}\tdt^s \mE_{ij}-\mE_{ir}\mE_{js}\tdt^r \mE_{lp}\tdt^s \mE_{kq}-\mE_{ri}\mE_{sj}\tdt^r \mE_{pl}\tdt^s \mE_{qk}\right) \right.\\
&&&\left.\ph{\fr14}-g^{ik}g^{jl}\left(\mE_{ip}\mE_{qj}\tdt^p d\tdt^q \mE_{kl}+\mE_{pi}\mE_{jq}\tdt^p d\tdt^q \mE_{lk}\right)+
4g^{ij}\mE_{ik}\mE_{jl}\tdt^kd\tdt^ld\right].
\end{aligned}
\end{equation}
Starting from the $O(D,D)$ transformations of the generalised metric $\mH_{MN}$ it can be shown that the field $\mE_{ij}$ transforms as
\begin{equation}
\mE'(\XX') = (a\mE(\XX)+b)(c\mE(\XX)+d)^{-1},
\end{equation}
where $a,b,c$ and $d$ are $D\times D $ blocks of an $O(D,D)$ matrix
\begin{equation}
\mc{O}_M{}^K=
\begin{bmatrix}
a & b \\
c & d
\end{bmatrix}
\end{equation}
with straightforward constraints on them following from $\mc{O}^{t}\h\mc{O}=\h$. In the special case when T-duality acts in all directions, i.e. $a=0, b=1, c=1$ and $d=0$ these relations imply
\begin{equation}
\mE'=\mE^{-1},
\end{equation}
and the corresponding dual metric is $g'_{kl}=\mE_{ki}g^{ij}\mE_{lj}$. Hence, the Lagrangian $\mL^{(2)}$ can be obtained from $\mL^{(0)}$ using the following rules
\begin{equation}
\begin{aligned}
& \mE_{ij} \rightarrow \mE'^{ij}, && g^{ij} \rightarrow g'_{ij}, && \dt_i\rightarrow \tdt^i.
\end{aligned}
\end{equation}
Verification of T-duality between these two terms is straightforward and is provided in details in \cite{Hohm:2010jy}. Thus, the terms $\mL^{(0)}$ and $\mL^{(2)}$ give the same supergravity action written in different T-duality frames. Namely, the first one corresponds to $\tdt^i\bullet=0$ solution of the strong constraint while the second survives if nothing depends on $x^i$ (alternatively $\dt_i\bullet =0$).

Finally, the term $\mL^{(1)}$ has the following form
\begin{equation}
\label{S1}
  \begin{aligned}
    {\cal L}^{(1)} = e^{-2d}&\left[\frac{1}{2}{g}^{ik}{g}^{jl}{g}^{pq}
    \left({\cal E}_{pr}\,\tilde{\partial}^{r}{\cal E}_{kl}\,\partial_{q}{\cal E}_{ij}
    -{\cal E}_{lr}\,\tilde{\partial}^{r}{\cal E}_{ip}\,
    \partial_{k}{\cal E}_{jq}+{\cal E}_{rl}\,
    \tilde{\partial}^{r}{\cal E}_{pi}\,\partial_{k}{\cal E}_{qj}\right) \right.\\
    &+{g}^{ip}{g}^{jq}\left({\cal E}_{rq}\,\partial_{p}d\,\tilde{\partial}^{r}
    {\cal E}_{ij}-{\cal E}_{pr}\,\tilde{\partial}^{r}d\,\partial_{q}{\cal E}_{ij}
    +{\cal E}_{rp}\,\tilde{\partial}^{r}d\,\partial_{q}{\cal E}_{ij}
    -{\cal E}_{qr}\,\partial_{p}d\,\tilde{\partial}^{r}{\cal E}_{ji}\right)\\
    &\left. \ph{fr14}-8{g}^{ij}\,{\cal E}_{ik}\,\tilde{\partial}^{k}d\,\partial_{j}d\, \right] \;.
  \end{aligned}
 \end{equation}
and contributes to the action for such choices of T-duality frames that include both dual and ordinary coordinates, for example $\dt_1=0$ and $\tdt^2,\ldots \tdt^D=0$. 

This decomposition of the effective action has very close relation to fluxes in type II string theory. One can find reviews on fluxes in string theory \cite{Grana:2005jc, Shelton:2005cf,Wecht:2007wu}, in supergravity \cite{Samtleben:2008pe} and in application to extended geometry \cite{Aldazabal:2010ef, Grana:2006is}. In addition, recently some progress has been made in this direction in \cite{Andriot:2012an,Andriot:2012wx, Dibitetto:2012ia, Dibitetto:2012rk}. The first and the most intuitive example of a flux that is called the H-flux is given by the integration of $H=dB$ over a 3-torus $\mathbb{T}^3$
\begin{equation}
\int_{\mathbb{T}^3} H.
\end{equation}
Starting with the Kalb-Ramond field with the only non-zero component $B_{xy}=Nz$, where $N\in \mathbb{Z}$ the H-flux is given by the integer $N$. Using the Buscher rules one can show that T-duality in the direction $x$ leads to the following metric
\begin{equation}
ds^2=(dx+f^x{}_{yz}z dy)^2 +dy^2 +dz^2.
\end{equation}
Here $f^x{}_{yz}=N$  is the so-called f-flux that is T-dual to the H-flux and the metric is that of the twisted torus. Indeed, one can consider the torus $\mathbb{T}^3$ as a $\mathbb{T}^2$ fibration over a circle $\mathbb{S}^1$ parametrized by the coordinate $z$. Then, going around the circle $z \sim z+2\p$ one has to shift the coordinate $x$ as $x \sim x+ 2\p f^x{}_{yz} y$ in order to have well defined metric. 
Finally, T-dualities in the directions $y$ and $z$ turn f-flux into Q-flux and R-flux respectively providing the following chain
\begin{equation}
\begin{aligned}
H_{xyz} & &\overset{T_x}{\longrightarrow}& & f^x{}_{yz}& & \overset{T_y}{\longrightarrow} & & Q^{xy}{}_{z} & &  \overset{T_z}{\longrightarrow} & & R^{xyz}.
\end{aligned}
\end{equation}
Q and R fluxes are non-geometric in the sense that the first one leads to non-com\-mu\-ta\-ti\-vity of the string coordinates and the second implies non-associativity \cite{Andriot:2012an}
\begin{equation}
\begin{aligned}
[x^a,x^b]&\sim Q^{ab}{}_{c}x^c,\\
[x^a,x^b,x^c]&\sim R^{abc}.
\end{aligned}
\end{equation}

These ideas naturally fit in the picture of the extended geometry. The solution $\tdt^i=0$ of the section condition leaves only the term \eqref{S0} that contains contribution from the H-flux. T-duality in the direction $x$ in this language means that the only non-zero derivatives are now $(\tdt^1,\dt_2,\dt_3)$ and one has to include both $\mL^{(1)}$ and $\mL^{(2)}$ into the consideration. After some algebra the effective action takes exactly the same form 
\begin{equation}
S_f=\int d\tx_1 dx^2dx^3 \left(R[g'] -\fr{1}{12}H'^2\right)
\end{equation}
but the metric $g'$ now contains a contribution from the f-flux. Following the same simple pattern it is straightforward to show that the Q and R fluxes correspond to the coordinates $(\tx_1,\tx_2,x^3)$ and $(\tx_1,\tx_2,\tx_3)$  respectively.

\section{Extended geometry for M-theory}
\label{M}

Apart from fundamental strings that are one-dimensional, string theories contain various excitations represented by extended objects. These are $Dp$-branes that appear as $p$-dimensional subspace where strings endpoints can travel and $D$ stands for Dirichlet boundary conditions. Type IIA string theory contains even dimensional $Dp$-branes that interact with $p+1$ odd forms $C_{(1)}, C_{(3)},\ldots$, while even dimensional branes appear in type IIB strings coupled to odd forms $C_{(0)}, C_{(2)},\ldots$. All these excitations along with KK monopoles appear naturally from compactifications of an 11 dimensional quantum theory whose fundamental objects are 2 dimensional M2 branes and their duals 5-dimensional M5 branes. Lacking any better name this conjecture was called M-theory. For a review see \cite{Obers:1998fb, Schwarz:1998fd} and \cite{Vafa:1997pm}.
\subsection{M-theory and U-duality}

M-theory that is formulated in 11 dimensions firstly appeared as a theory which describes non-perturbative strong coupling limit of Type IIA string theory. The extra compact dimension is generated dynamically in string theory and has radius $R=l_p g_s^{2/3}$, where $g_s$ is the string coupling constant and $l_p$ denotes the 11d Planck length \cite{Witten:1995zh, Townsend:1995gp}. In the limit when the string coupling is large, the extra dimension becomes uncompactified. Clearly, the relation between Type IIA string theory and M-theory is non-perturbative and cannot be derived from analysis of the string spectrum. The proper tool to investigate this correspondence is perturbative duality symmetries of string theories $O(d,d,\mathbb{Z})$, non-perturbative $SL(2,\mathbb{Z})$ Schwarz and West symmetries of type IIB string theory \cite{Schwarz:1983wa} and exceptional $E_{d(d)}(\mathbb{Z})$ Cremmer-Julia symmetries of supergravity \cite{Cremmer:1978km, Cremmer:1978ds, Cremmer:1979up}. 

Relations between Type IIA string theories and 11-dimensional supergravities were known long time ago. Namely, upon compactification of 11-dimensional supergravity on a circle $\mathbb{S}^1$ of radius $R$ one obtains Type IIA 10-dimensional supergravity that appears to be the low-energy limit of the corresponding string theory \cite{Campbell:1984zc, Giani:1984wc}. The dimensional reduction is carried by splitting the fundamental (bosonic) fields of 11d SUGRA, the metric and the 3-form field $C_{mnk}$, into 10 dimensional metric, gauge fields and the dilaton. The metric anzats is then given by
\begin{equation}
ds_{11}^2=e^{4\ff/3}\left(dx^{11}+A_\m dx^\m\right)^2+e^{-2\ff/3}ds_{10}^2
\end{equation}
where the index $\m=1\ldots 10$ labels 10 dimensions of the resulting theory, $\ff$ denotes the dilaton and $ds_{10}^2$ is the 10-dimensional interval. The vector field $A_\m$ is the RR 1-form gauge potential of the 10-dimensional theory. The 3-form field $C_{mnk}$ gives rise to the 10-dimensional RR 3-form potential $C_{\m\n\r}$ and the NS-NS 2-form Kalb-Ramond field $B_{\m\n}$ thus completing the bosonic sector of the theory:
\begin{equation}
\begin{aligned}
\mbox{NS-NS} &&: && g_{\m\n}, && B_{\m\n}, && \ff\\
\mbox{RR} &&: &&A_{\m}, && C_{\m\n\r}.&&
\end{aligned}
\end{equation}
Since, the 11-dimensional theory does not have dimensionless couplings the string coupling $g_s$ is generated dynamically reading $g_s^2=e^{2\ff}$. Hence, in order to relate M-theory to Type IIA string theory by compactification on a circle one has to consider the 11-dimensional supergravity as a low-energy limit of M-theory.

As it was discussed in the previous sections, Type IIB string theory compactified on a circle of radius $R$ is T-dual to Type IIA string theory compactified on a circle of radius $\a'/R$. The bosonic NS-NS sector of these two theories is the same and transformation of the fields $g_{\m\n}, B_{\m\n}$ and $\ff$ under T-duality is given by the Buscher rules \eqref{Buscher}. These $\mathbb{Z}_{2}$ transformation is a part of the full T-duality group $O(d,d,\mathbb{Z})$. This allows us to relate M-theory to Type IIB string theory.

In its turn, Type IIB string theory possesses a global $SL(2,\mathbb{Z})$ symmetry that is called S-duality \cite{Hull:1994ys}. It is instructive to consider the bosonic sector of Type IIB supergravity and its transformation properties. Two fields, the dilaton $\ff$ and the axion $\c$, are naturally combined in a complex field
\begin{equation}
\r=\c+i e^{-\f}
\end{equation}
under the action of S-duality that is given by
\begin{equation}
\r \longrightarrow \fr{a \r + b}{c \r + d}.
\end{equation}
Here the integer numbers $a,b,c$ and $d$ compose the corresponding $SL(2,\mathbb{Z})$ matrix with $ad-bc=1$. A pair of 2-form potentials that come from NS-NS and RR sector transform as a doublet. The remained bosonic fields that are the graviton and the 4-form potential are invariant under S-duality. 

The non-perturbative S-duality symmetry of Type IIB string theory  becomes manifest in the approach of M-theory  being a non-trivial remnant of 11-dimensional diffeomorphism invariance and U-duality. The duality group $SL(2)$ is now a modular group of the compact torus $\mathbb{T}^2$ and the complex field $\r$ defines modular parameter \cite{Schwarz:1995dk, Aspinwall:1995fw}. On the other hand, Type IIB theory is T-dual to Type IIA theory. This symmetry together with symmetries of a d-torus form the U-duality symmetry. These are given by the known exceptional Cremmer-Julia symmetry groups $E_{d(d)}$. 

Although the S-duality part is manifest in M-theory and is originated from diffeomorphisms, the whole exceptional symmetry does not have such simple explanation. As it was shown in the previous sections the extended geometry approach allows to write the effective potential of Type II string theory in T-duality covariant variables, i.e. the generalised metric $\mH_{MN}$ and the dilaton $d$. This section is a brief review of the same approach to U-duality.

\subsection{Duality invariant actions}

In string theory extended geometry one introduces an $O(d,d)$ covariant object, generalised metric $\mH_{MN}$, that parametrizes coset $O(d,d)/O(d)\times O(d)$ and is written in terms of the metric $G_{mn}$ and the Kalb-Ramond 2-form field $B_{mn}$. In this formalism $2d$ coordinates of the extended space are associated to every string charge and to every field. The usual space-time coordinates $x^a$ are associated to the metric, while the dual coordinates $\tx_a$ are associated to the 2-form. Mathematically this is realised by exploiting the Hitchin's concept of generalised tangent bundle that is a direct sum of the tangent and the cotangent bundles of the space-time $M$
\begin{equation}
\label{bundle_Odd}
TM\oplus T^*M.
\end{equation}
It is important that the base of this bundle is still a $d$-dimensional space $M$. The non-trivial transition to the extended space occurs when one starts to think of fibres of the generalised tangent bundle as tangent spaces to a $2d$ dimensional extended space $\mc{M}$. Although, a lot is known about covariant derivatives, curvature and infinitesimal tensor transformations on this space its geometry is still unclear \cite{Hohm:2012mf,Jeon:2011cn,Jeon:2010rw,Hohm:2010xe,Hohm:2011si,Hohm:2012gk}.

In the generalization of this formalism to M-theory one considers U-duality group, that is $E_{d}$ for duality acting in $d$ directions. Since fundamental objects of M-theory are represented by M2 and M5 branes, the corresponding extended space becomes slightly more involved than in the case of T-duality where only winding modes of the F1 string contribute. This gives rise to the ordinary coordinates $x^a$, dual coordinates $y_{ab}$ for the M2 brane, $z_{abcde}$ for the M5 brane and so on. The generalised tangent bundle \eqref{bundle_Odd} is replaced then by the following construction
\begin{equation}
TM\oplus \L^2T^*M \oplus \cdots
\end{equation}
As before, a typical fibre of this bundle over the ordinary space $M$ is understood as a generalised tangent space to the extended space $\mc{M}$. The generalised metric, that unifies the metric $G_{mn}$ and the gauge fields, parametrises the coset
\begin{equation}
M_{MN}\in \fr{E_d}{H_d},
\end{equation}
where $H_d$ is a maximal compact subgroup of the U-duality group $E_d$.

To describe the extended space of M-theory one exploits the idea of non-linear realisation of space-time symmetries that was known long ago \cite{Isham:1971dv, Ogievetsky:1973ik, Borisov:1974bn}. Borisov and Ogievetsky showed that the theory of general relativity in four dimensions can be described in terms of a non-linear realisation of the groups $G=GL(4,\mathbb{R})\ltimes \mathbb{R}^4$ and $H=SO(3,1)$. Here the group $G$ is the semi-direct product of the structure group $GL(4,\mathbb{R})$ and the group of space-time translations $\mathbb{R}^4$. The semi-direct product implies that generators of the latter transform under the fundamental representation of the group $GL(4,\mathbb{R})$. The coset $G/H$ is identified with space-time. 

The same structure appears when one constructs a supergravity theory. The coset of the super-Poincar\'e group with respect to the Lorentz group leads to the notion of superspace \cite{West:1990in}. However, this formalism does not include all symmetries of supergravity: generators of U-duality transformations obviously are not in the super-Poincar\'e group. It is known \cite{West:2001as,Riccioni:2007ni} that eleven dimensional supergravity can be naturally formulated in terms of a non-linear realisation of very extended algebra, that is commonly denoted as $E_{11}$. The suggested $E_{11}$ covariant way to include both space-time generators $P_a$ and duality symmetry generators, was to collect them into the first fundamental representation of $E_{11}$ denoted by $l_1$, that is infinite dimensional.

The theory in $D$ dimensions with U-duality acting in $d$ dimensions then can be obtained by deleting a certain node in the Dynkin diagram of $E_{11}$ . This corresponds to taking a subalgebra $GL(D)\oplus E_{d}$ of the algebra $E_{11}$
\begin{figure}[http]
$$
\begin{array}{ccccccccccccccccc}
     & &       & &         & &       &  &\bullet&11&       & &        \\
     & &       & &       & &       &  &   |   &  &          &
&         \\
\bullet &-&\ldots&-&\otimes &-&\ldots&- &\bullet&- &\bullet&-&
\bullet \\
    1    & &         & &    D     & &      &  &   8   &  &   9   &
&    10    \\
\end{array}
$$
\caption{Dynkin diagram of the algebra $E_{11}$ with node $D$ deleted.}
\label{Dynkin}
\end{figure}
 The factor $GL(D)$ together with space-time translations that are contained in $l_1$ gives rise to gravity in $D$ dimensions as it should be. The remained factor $E_{d}$ is the known Cremmer-Julia duality group of maximal supergravity in $D$ dimensions \cite{Cremmer:1978km, Cremmer:1979up, Julia:1980gr}. Thus these symmetries are naturally reproduced in the non-linear realisation of $E_{11}$.

The representation $l_1$ that contains an infinite number of generators
\begin{equation}
\begin{aligned}
P_a,& & Z_{a_1 a_2}, & & Z_{a_1a_2a_3a_4a_5}, & & \ldots
\end{aligned}
\end{equation}
is decomposed into representations of $GL(D)\oplus E_{d}$. In addition to ordinary space-time coordinates one finds an infinite number of coordinates that correspond to higher level fields. Coordinates that are scalars with respect to $GL(D)$ but transform under $E_d$ are in the $10, 16_s,\overline{27}, 56$ and $248\oplus 1$ of $SL(5),SO(5,5),E_6,E_7$ and $E_8$ for $d=4,5,6,7$ and 8 respectively \cite{Riccioni:2009xr, West:2004iz}.
\begin{table}[http]
\begin{center}
 \begin{tabular}{|c|c|c|c|c|}
\hline
 d&  Global duality group & Local duality group & $\mc{R}_V$\\
\hline 
  1 & $SO(1,1)$ & $1$ & $\bf{1}$ \\
  2 & $SL(2)$ & $SO(2)$ & $\bf{3}$ \\
  3 & $SL(3)\times SL(2)$ & $SO(3)\times SO(2)$ & \bf{6}\\
  4 & $SL(5)$ 	& $SO(5)$			& $\bf{10}$\\
  5 & $SO(5,5)$	& $SO(5)\times SO(5)$ & $\bf{16}_s$	\\
  6 & $E_6$	& $USp(8)$		& $\overline{\bf{27}}$\\
  7 & $E_7$	& $Sp(8)$	& $\bf{56}$		\\
  8 & $E_8$	& $SO(16)$	& $\bf{248\oplus1}$	\\
\hline  
 \end{tabular}
 \end{center} 
 \caption{Global and local duality groups and the representation $\mc{R}_V$.}
 \label{Duality_table}
\end{table}
 A set of ordinary coordinates together with a certain number of these scalar dual coordinates parametrizes the extended space of M-theory. The corresponding set of generators transforms in the representation $\mc{R}_V$ of $E_d$ that is listed in the table above..

The non-linear realisation leads not only to the extended space but allows one to construct a generalised vielbein using the conventional vielbein and form fields. The dynamics of strings and branes in the presence of background fields can be formulated in terms of the non-linear realisation as well. The coordinates of the extended space correspond to brane charges and momentum. It is instructive to go through a case with a certain number of dimensions in more details. For a full analysis the reader is referred to \cite{Hull:2007zu} that describes the geometry of the extended space in terms of brane charges in the spirit of Hitchin's' generalised geometry, and to the papers \cite{Berman:2010is} and \cite{Berman:2011jh}, that construct duality invariant actions for M-theory using the Duff's procedure and the non-linear realisation briefly described above.

Consider the case where U-duality group is $SL(5)$ that acts in 4 space dimensions. Winding modes of 5-branes do not appear in the formalism and thus one has to include only wrappings of 2-branes. This results in the generalised tangent space
\begin{equation}
TM\oplus \L^2T^*M,
\end{equation}
whose fibres are understood as tangent spaces to the 10-dimensional extended space parametrized by the coordinates $(x^\a,y_{\a\b})$ with $\a,\b=1,\ldots,4$. The coordinates $x^\a$ are associated with the metric and represent space-time coordinates, while $y_{\a\b}=-y_{\b\a}$ are associated with the M2-brane charge and represent dual coordinates. Already this simple example shows that in contrast to the case of T-duality, where each space-time coordinate has its dual, here numbers of space-time and dual coordinates are not equal. All ten coordinates of the extended space are combined in an object that transforms in the representation $\mathbf{10}$ of the U-duality group $SL(5)$
\begin{equation}
\XX^M=
\begin{bmatrix}
x^\a\\
y_{\a\b}
\end{bmatrix}.
\end{equation}
Capital Latin indices here and in all expressions in this thesis label the representation $\mc{R}_V$ of the corresponding U-duality group. It is convenient to represent the 10-dimensional index $M$ as an antisymmetric pair of two indices in $\mathbf{5}$ using the following identifications \cite{Berman:2011cg}
\begin{equation}
\XX^{ab}=\left\{
\begin{aligned}
&\XX^{5\a}=x^\a,\\
&\XX^{\a 5}=-x^\a,\\
&\XX^{\a\b}=\fr12\e^{\a\b\m\n}y_{\m\n},
\end{aligned}\right.
\end{equation}
where small Latin indices run from 1 to 5 and $\e^{\a\b\m\n}$ is the 4 dimensional alternating symbol ($\e^{1234}=1$). Generalised vectors $V^{ab}$ then carry indices labelling the representation $\mathbf{10}$ of $SL(5)$. Tensors of other ranks may carry any number of small Latin indices, even or odd. 

As in the case of the $O(d,d)$ geometry one constructs a generalised tangent bundle whose fibre are the formal sums $V=v+\r$, where $v=v^\a\dt_\a$ is a vector and $\r=\r_{\a\b}dx^\a\wedge dx^\b$ is a 2-form. Structure group  of this fibre bundle is $SL(5,\mathbb{R})$, that can be reduced to $SO(5,\mathbb{R})$ upon introducing a generalised vielbein globally. The coset $SL(5)/SO(5)$ is parametrised by the metric $g_{\m\n}$ and the 3-form field $C_{\m\n\r}$, that are collected into the generalised metric
\begin{equation}
M_{MN}=
\begin{bmatrix}
g_{\m\n}+\fr12 C_{\m\a\b}C_\n{}^{\a\b} & \fr{1}{\sqrt{2}}C_{\n{}^{\r\s}} \\
&\\
\fr{1}{\sqrt{2}}C_{\m}^{\g\d} & g^{\g\d,\r\s}
\end{bmatrix},
\end{equation}
where the object $g^{\m\n,\a\b}=g^{\m[\a}g^{\b]\n}$ is used to raise and lower antisymmetric pairs of indices.

The generalised Lie derivative of a generalised vector $V^M$ in the U-duality invariant formalism has exactly the same form as in the Double Field Theory \eqref{gen_Lie_1} 
\begin{equation}
\label{gen_Lie_2}
\begin{aligned}
\d_\S V^M&=\S^N\dt_N V^M -V^N\dt_N \S^M + \dt^M \S_S V^S\\
&=L_\S V^{M}+Y^{MK}{}_{RS}\dt_K \S^R V^S\\
&=\mc{L}_{\S} V,
\end{aligned}
\end{equation}
but the invariant tensor $Y^{MN}_{KL}$ is defined in a different way. Its exact form follows from the condition that algebra of transformations \eqref{gen_Lie_2} is closed upon the section condition
\begin{equation}
\label{closure}
\begin{aligned}
[\mc{L}_{V_1},\mc{L}_{V_2}] & =\mc{L}_{[V_1,V_2]_C}+F_0,\\
Y^{MN}_{KL}\dt_M \bullet \dt_N \bullet &=0 \Longrightarrow F_0=0.
\end{aligned}
\end{equation}
Substituting \eqref{gen_Lie_2} into the closure condition one finds the following expressions for the invariant tensor \cite{Coimbra:2011ky,Berman:2012vc}:
\begin{equation}
\label{Y}
\begin{array}{rcl}
O(d,d)_{strings}: & \quad & Y^{MN}_{PQ}  = \eta^{MN} \eta_{PQ},, \vspace{0.2cm} \\
SL(5):  & \quad & Y^{MN}_{PQ}= \e^{\a MN}\e_{\a PQ},   \\[0.2cm]
SO(5,5): &\quad & Y^{MN}_{PQ}  = \frac{1}{2} (\G^i)^{MN} (\G_i)_{PQ} \ ,  \\[0.2cm]
E_{6(6)}: &\quad & Y^{MN}_{PQ}  = 10 d^{MN R} d_{PQR} \ ,  \\ [0.2cm]
E_{7(7)}: &\quad & Y^{MN}_{PQ}  = 12 c^{MN}{}_{PQ} + \delta^{(M}_P \delta^{N)}_Q + \frac{1}{2} \e^{MN} \e_{PQ } \ . 
\end{array}
\end{equation}
Here the index $\a$ runs from 1 to 5 labelling the representation $\bf{5}$ of $SL(5)$ and the index $i$ labels the 10-dimensional vector representation of $SO(5,5)$. The invariant metric on $O(d,d)$ is denoted by $\h_{MN}$, $\e_{\a MN}=\e_{\a,\b\g,\d\e}$ is the $SL(5)$ alternating tensor, $SO(5,5)$ gamma-matrices $\G^{iMN}$ are $16\times16$ gamma-matrices in Majorana-Weyl representation, the tensors $d_{MNK}$ and $c^{MN}{}_{KL}$ are symmetric invariant tensors of $E_6$ and $E_7$ respectively.

The invariant tensor $Y_{KL}^{MN}$ is subject to various important relations that will be used later \cite{Berman:2012vc}
\begin{equation}
\begin{split}
\label{rel}
&Y^{(MN}_{KL}Y^{L)R}_{PQ}-Y^{(MN}_{PQ}\d^{R)}_{K}=0 \mbox{ , for $d\leq5$},\\
&Y^{MN}_{KL}=-\a_d P_{K}{}^{M}{}_{L}{}^{N}+\b_d\d^M_K\d^N_L+\d^M_L\d^N_K,\\
&Y^{MA}_{KB}Y^{BN}_{AL}=(2-\a_d)Y^{MN}_{KL}+(n\b_d+\a_d)\b_d\d^M_K\d^N_L+(\a_d-1)\d^M_L\d^N_K.
\end{split}
\end{equation}
Here $d=11-D$ is the number of compact directions and $P_A{}^B{}_C{}^D$ is the projector on the adjoint representation of the corresponding duality group. It is defined as $P_A{}^B{}_C{}^DP_D{}^C{}_K{}^L=P_A{}^B{}_K{}^L$ and $P_A{}^B{}_B{}^A=\mbox{dim}(adj)$. The coefficients $\a_d$ and $\b_d$ depend on the duality group and for the cases in question take numerical values 
$(\a_4,\b_4)=(3,\fr{1}{3})$, $(\a_5,\b_5)=(4,\fr{1}{4})$, $(\a_6,\b_6)=(6,\fr{1}{3})$. The last line in \eqref{rel} with $n=\d^A_A$ is a direct consequence of the second relation and properties of the projector $P_A{}^B{}_C{}^D$. The first line is true only for $d\leq5$ and the relevant identity for $E_{6(6)}$ duality group reads
\begin{equation}
10P_Q{}^{(M}{}_T{}^NP_R{}^{P)}{}_S{}^{T}-P_R{}^{(M}{}_S{}^N\d^{P)}_Q-\fr13d^{MNP}d_{QRS}=0.
\end{equation}

The generalised metric is a dynamical field of the theory and along with its derivatives contributes to the effective potential. The explicit form of the potential for the $SL(5)$ case was found by D. Berman and M. Perry in \cite{Berman:2010is} and has the following form
\begin{equation}
 \begin{aligned}
\label{V_sl5}
    V_{SL(5)}&=\sqrt{g}\left[\fr{1}{12}M^{MN}(\dt_M{}M_{KL})(\dt_N{M^{KL}})-\fr12{}M^{MN}(\dt_N{}M^{KL})(\dt_L M_{MK})\right.+\\
      &\left.+\fr{1}{12}M^{MN}(M^{KL}\dt_M M_{KL})(M^{RS}\dt_N M_{RS})-\fr14 (M^{RS}\dt_K M_{RS})(\dt_L M^{KL})\right],
 \end{aligned}
\end{equation}
where $g=\det ||g_{\m\n}||$ is the determinant of the four dimensional metric. It is easy to show that $g$ is always proportional to a certain power of $\det ||M_{MN}||$.

The potential \eqref{V_sl5} is invariant under the transformations \eqref{gen_Lie_2} up to the section condition. Taking a special solution of the section condition $\dt_y=0$ that effectively removes all dependence of the dual coordinate $y_{\m\n}$, turns the effective potential to that of the supergravity theory (the bosonic part) up to A boundary term:
\begin{equation}
\label{sgr}
V_{SL(5)} \rightarrow \sqrt{g}\left(R[g]-\fr{1}{48}F[C]^2\right),
\end{equation}
where $R[g]$ is the curvature of the metric $g_{\m\n}$ and $F=dC$ is the field strength of the 3-form field $C_{\m\n\r}$.

When considering duality transformations acting in more than 4 dimensions one has to include coordinates associated with the M5-brane and the KK6-brane. The corresponding bundle is then
\begin{equation}
TM\oplus \L^2T^*M\oplus \L^5 T^*M \oplus \L^6TM.
\end{equation}
The expression for the generalised Lie derivative \eqref{gen_Lie_2} remains the same up to choosing an appropriate $Y$-tensor \eqref{Y}. In four dimensions the last two terms do not contribute since elements of say $\L^5T^*M$ are  5-forms that can't be defined in four dimensions. The same is true for 5 dimensions and the term $\L^6TM$ whose elements are 6-vectors. Finally, in dimensions $d\geq6$ one considers the full bundle \cite{Hull:2007zu, Berman:2011pe, Berman:2011jh}.

In higher dimensions it is more convenient to introduce a generalised vielbein rather giving an explicit expression for the generalised metric 
\begin{equation}
M_{MN}=E^{\bA}_ME^{\bB}_{N} M_{\bA\bB},
\end{equation}
where the barred indices run from 1 to $n$ labelling flat directions and $M_{\bA \bB}$ is diagonal. Explicit expressions for the generalised vielbein in dimensions $d=5,6,7$ and the effective potential were found in \cite{Berman:2011jh} starting from the non-linear realisation of $E_{11}$. These are given in the next sections in application to Scherk-Schwarz reductions.

Finally, it is necessary to mention the work \cite{Thompson:2011uw} that considers the reduction from M-theory to type II string theory in the duality invariant formalism. It is shown that one successfully reproduces the structures of $O(3,3)$ geometry starting from $SL(5)$ invariant theory. The DFT section condition naturally emerges from the $SL(5)$-covariant section condition.

\chapter{Dimensional reductions}
\label{reduction}
\section{Introduction}

The extended geometry formalism of string and M-theory encodes the low energy limit of the theories in a manifestly T(U)-duality invariant way. The resulting theory is formulated on a space that is parametrised by an enhanced set of coordinates both ordinary and dual that correspond to momentum and winding modes. In this chapter we show that Scherk-Schwarz reduction \cite{Scherk:1979zr} of the extended space consistently reproduces structures of gauged supergravities such as the scalar potential, the embedding tensor and the gauge group.

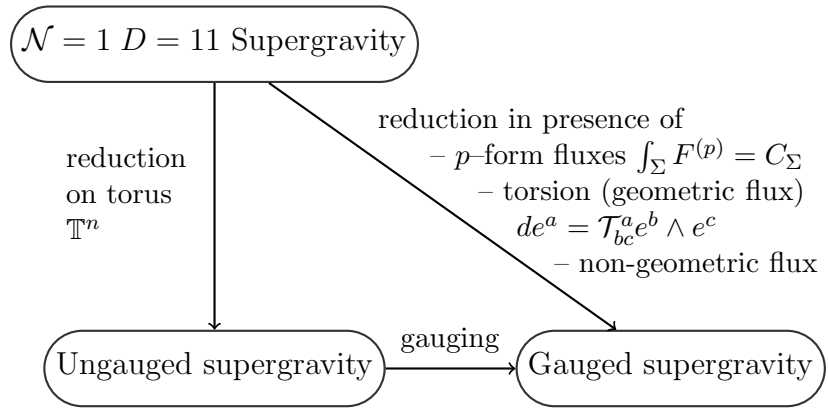
\begin{figure}[ht]
\label{gauged}
\centering
\begin{tikzpicture}[thick,
                    text height=1ex,
                    text depth=.15ex
                    ]
\node at (0,-3) [format] (N11) {$\mc{N}=1\;D=11 $ Supergravity};
\node at (0,-7.3) [format] (Ung) {Ungauged supergravity};
\node at  (6,-7.3) [format] (Ga)  {Gauged  supergravity};
\node at (3.1,-7.3) [above=0.1] {\small gauging};
\node at (0.3,-5) [left=0.1] {\parbox{2cm}{ \small reduction \\ on torus \\ $\mathbb{T}^n$}};
\node at (2,-5) [right=0.0]  {\parbox{6cm}{\small reduction in presence of \\\ph{1}\hspace{0.5cm}-- $p$--form fluxes 
								{\color{black}$\int_\S F^{(p)}=C_\S$} 
\\\ph{1}\hspace{1cm} -- torsion (geometric flux) \\ \ph{1}\hspace{1.5cm} {\color{black}$de^a=\mc{T}^a_{bc}e^b\wedge e^c$} \\\ph{1}\hspace{2cm} -- non-geometric flux}};

\path[<-] (Ung) edge   (N11);
\path[->] (N11) edge  (Ga);
\path[->] (Ung)  edge  (Ga);
\end{tikzpicture}
{\caption{\small This diagram demonstrates relations between toroidal reductions of $\mc{N}=1$ $D=11$ supergravity, gaugings and more complicated dimensional reductions.}}
\end{figure}

Gauged supergravities appear as consistent supersymmetric deformations of toroidal compactifications of 11-dimensional $\mc{N}=1$ supergravity (for review see \cite{Samtleben:2008pe}). These are represented by the horizontal line on the Figure \ref{gauged} where the general picture is sketched.

Gauged supergravities were first constructed in \cite{deWit:1981eq} by incorporating the structure of Yang-Mills theories to the maximal supergravity and then generalised to higher dimensions in \cite{Gunaydin:1985cu, Pernici:1984xx} and to other non-compact groups \cite{Hull:1984vg, Hull:1984qz}. The diagonal line on the picture demonstrates the relation of gauged supergravities to flux compactifications that was realised recently (for review see \cite{Grana:2006is, Blumenhagen:2006ci,Geissbuhler:2013uka}.

Starting from the 11-dimensional supergravity one performs the Kaluza-Klein reduction on a simple $n$-dimensional torus and ends up with ungauged supergravity where none of the matter fields is charged under gauge group $U(1)^{n_V}$. Here $n_V$ is the number of vector multiplets that come from certain components of metric and 3-form field. 

More complicated reductions on non-trivial $n$-dimensional manifolds like a sphere, reductions in presence of torsion or non-zero fluxes of $p$-form fields (geometric fluxes), or reductions with non-geometric fluxes lead to gauged supergravity. In these theories the matter fields transform under a gauge group that is a  subgroup of the global $E_{d(d)}$ Cremmer-Julia duality group. The non-trivial geometry of the internal space typically allows one to introduce a scalar potential that supports an effective cosmological constant and provides terms for moduli stabilization, leads to spontaneous symmetry breaking etc. (see \cite{Becker:2003yv,Becker:2003sh,Grana:2005jc,Shelton:2005cf} for review).  A  universal approach to gauged supergravities is the embedding tensor  which describes how gauge group generators are embedded into the global symmetry group.  Treated as a spurionic object the embedding tensor provides a manifestly duality covariant description of gauged supergravities.  
 
  In addition to the global $E_{d(d)}$ symmetry the toroidally reduced theories also posses a global $\mathbb{R}^+$ scaling symmetry  known as the trombone symmetry (this is an on-shell symmetry for $D\neq2$).   This gives rise to a more general class of gaugings whereby a subgroup of the full global duality group $E_{d(d)}\times \mathbb{R}^+$  is promoted to a local symmetry.  The embedding tensor approach was extended to incorporate such trombone gaugings in \cite{LeDiffon:2008sh}. The embedding tensor $\widehat{\Theta}_M^{\underline{\a}}$ projects generators  $t_{\underline{\a}}$ of the global duality group $E_{n(n)}\otimes\mathbb{R}^+$ to some subset $X_{M}=\widehat{\Q}_{M}{}^{\underline{\a}} t_{\underline{\a}}$ which generate the gauge group and enter into covariant derivatives:
  \begin{equation}
  \label{cov}
  D = \nabla - g A^{M} X_{M} \ .
    \end{equation}
The index $\underline{\a}$ of the embedding tensor is a multiindex which labels the adjoint representation of the duality group. According to its index structure the embedding tensor is in the $\mc{R}_V\times \mc{R}_{adj}$ representation. Here $\mc{R}_{adj}$ is the adjoint representation of the global duality group and $\mc{R}_V$ is an $n_V$-dimensional representation in which the vector fields transform. In general the embedding tensor decomposes as
 \begin{equation}
 \widehat{\Q}_M{}^{\underline{\a}} \in \mc{R}_V\otimes\mc{R}_{adj}=\mc{R}_V\oplus\ldots
 \end{equation}
The preservation of supersymmetry gives a linear constraint restricting the embedding tensor only to some representations, e.g. for the cases considered in further sections we have
\begin{equation}
\begin{split}
&\widehat{\Q}_M{}^{\underline{\a}} \in {\bf 10\oplus 15 \oplus  \overline{40}}\mbox{, for $D=7$} \\
&\widehat{\Q}_M{}^{\underline{\a}} \in {\bf 16}_s\oplus {\bf 144}_c\mbox{, for $D=6$} \\
&\widehat{\Q}_M{}^{\underline{\a}} \in \bf{27\oplus 351}\mbox{, for $D=5$}.
\end{split}
\end{equation}
The trombone gauging that is always in the representation $\mc{R}_V$ corresponds to the on-shell symmetry and does not appear in the action. Hence, the scalar potentials written further below do not include this gauging. 

In this chapter Scherk-Schwarz compactifications of the extended geometry formalism are considered. In what follows the extended space parametrized by the coordinates $\XX^M$ represents the internal space. In other words coordinates of the $d$-dimensional internal space are extended while the external non-compact space is parametrised by ordinary the coordinates $x_{(D)}$ with $D=11-d$. 

Dependence of any covariant object defined on the extended space is given by the so-called Scherk-Schwarz twist matrices that act like a vielbein. With this anzats the generalised diffeomorphism \eqref{gen_Lie_2} turns into a gauge transformation generated by the same algebra that one encounters in gauged supergravities. The corresponding group appears to be a subgroup of the global duality group and the embedding tensor becomes naturally written in terms of the twisting matrices. Hence, one connects geometric properties of the extended space to the algebraic properties of the theory.

An important feature of the extended space formalism is that one needs a constraint for closure of the algebra of generalised diffeomorphisms and to have an invariant effective potential (see \eqref{closure}). In the generalised Scherk-Schwarz reduction this constraint is promoted into the so-called quadratic constraint on the embedding tensor that is basically the condition of closure of the algebra
\begin{equation}
\label{clos}
[X_M,X_N]=X_{MN}{}^KX_K,
\end{equation}
where $X_M$ is a generator of the algebra and the structure constants $X_{MN}{}^K$ are written in terms of the twist matrices and their first derivatives.

Therefore in this chapter we do not impose section condition. Rather we require closure of the algebra of the generators $X_M$. In general the functions $X_{MN}{}^K$ are not necessarily constants, since one is free to choose the twist matrices almost in an arbitrarily way. The condition that $X_{MN}{}^K$ are indeed structure constants of an algebra is promoted to a constraint on the twist matrices. In other words, since, the whole extended space is considered to be internal, instead of projecting on a subspace by section condition we choose it to be of  a particular shape, defined by the twist matrices. In what follows we assume that there exist non-trivial twist matrices for each gauging, however this has not been proved yet.

It is worth mentioning, that to ensure invariance of the effective action written in the Scherk-Schwarz anzats one has to introduce an extra term of the form
\begin{equation}
\int d\XX \;Y^{MN}{}_{KL}\dt_M E^{\hat{A}K}\dt_N E^L_{\hat{A}}
\end{equation}
that is zero up to section condition.  Here the generalised vielbein is denoted as $E^{\hat{A}}_A$ with hatted indices parametrising generalised tangent space (fiber indices). Although, this term introduces extra degrees of freedom as it is not invariant under the local H transformations this appears to be not an issue of the Scherk-Schwarz reduced action. In this case the twisted vielbeins do not depend on $\XX^M$ and the derivatives in the extra term act only on the twist matrices $W^A_{\bB}$. 

The extra term allows to organise all terms in the twisted effective potential in expressions that involve only the generators $X_{MN}{}^K$ and the twisted generalised metric. The details of this procedure for the cases of $SL(5)$, $SO(5,5)$ and $E_{6(6)}$ duality groups is given in further sections.

\section{Scalar sector of maximal gauged supergravity}

In this section we briefly review the structure of the scalar sector of the maximal gauged supergravities in $D=5,6$ and $7$ and introduce expressions that we will need further. Sections devoted to different dimensions $D$ exploit their own conventions for indices and fields, that should not be confused. Since the review is very brief and does not cover all the details we refer the reader to the relevant papers \cite{deWit:2004nw, Bergshoeff:2007ef, Samtleben:2005bp,LeDiffon:2008sh} and \cite{Samtleben:2008pe}.

\subsection{$D=7$ supergravity}     

The global symmetry group of the ungauged $D=7$ maximal supergravity is $E_{4(4)}=SL(5)$ whose generators can be expressed as 
\begin{equation}
\begin{aligned}
(T_a^b)^i_j & = \delta_a^i \delta_j^b - \frac{1}{5} \delta_a^b \delta^i_j \\
(T_a^b)^{ij}_{kl} & = 2 (T_a^b)^{[i}_{[k} \delta_{l]}^{j]},
\end{aligned}
\end{equation}
in the representations $\bf 5$ and $\bf 10$ respectively. These generators are traceless, $\Tr(T_a^b)= 0 $ and obey a relation $T_a^a = 0 $. Here small Latin indices run from 1 to 5 labelling the fundamental representation. Commutation relations that define the algebra of $SL(5)$ read
\begin{equation}
[T_a^b, T_c^d ] = \delta_c^b T_a^d-  \delta_a^d T_c^b \  . 
\end{equation}  
Also we have 
\begin{equation}
Tr_{\bf{5}}( T_a^b T_c^d) = - \delta_a^d \delta_c^b + \frac{1}{5} \delta_a^b \delta_c^d \ . 
\end{equation}

The embedding tensor for $D=7$ gauged supergravity, with no trombone gauging, is given by  \cite{Samtleben:2005bp}:
\begin{equation}
\Theta_{mn, p}{}^q = \delta^q_{[m} Y_{n]p} - 2 \epsilon_{mn p rs}Z^{rs ,q} 
\end{equation}
with $Y_{mn}= Y_{(mn)}$ in the $\bf{15}$ and $Z^{rs, q} = Z^{[rs],q}$ in the $\bf{\overline{40}}$ so that $Z^{[rs, q] }  = 0$.  
It is traceless  $\Theta_{mn, p}{}^p = 0 $ and hence the gauge group generators in the $\bf{5}$ and $\bf{10}$ are given by
\begin{equation}
X_{mn , p }{}^q = \Theta_{mn, p}{}^q \ , \quad 
X_{mn,  pq }{}^{rs} = 2 \Theta_{mn, [p}{}^{[r}\delta^{s]}_{q]} \ .  
\end{equation}

To incorporate the trombone gauging an extra generator $(T_0)_p^q = \delta_p^q $ corresponding to the $\mathbb{R}^{+}$ is introduced and an ansatz is proposed (we follow exactly  \cite{LeDiffon:2008sh} where the procedure is carried out for all the other exceptional groups)
\bea
\hat{\Theta}_{mn, 0} &=& \theta_{mn} \ , \non  \\
\hat{\Theta}_{mn, p}{}^q &=& \delta^q_{[m} Y_{n]p} - 2 \epsilon_{mn p rs}Z^{rs ,q}  + \zeta \theta_{ij} (T_p^q)^{ij}_{mn}  \ ,
\eea
where $\theta_{mn} = \theta_{[mn]}$ is in the $\bf{10}$. 
Then the gauge generators in the fundamental are given by 
\bea
\hat{X}_{mn, p}{}^q &=&  \hat{\Theta}_{mn, 0} (T_0)^q_p + \hat{\Theta}_{mn , r}{}^s (T_r^s)_p^q   \non  \\ 
&=& \delta^q_{[m} (Y_{n]p} -2 \zeta \theta_{n] p}  ) - 2 \e_{mn p rs}Z^{rs ,q}  +\frac{1}{5} (5-2 \zeta) \theta_{mn} \delta_p^q,
\eea
and in the ${\bf 10}$ by 
\bea
\hat{X}_{mn, pq}{}^{rs} &=&  \hat{\Theta}_{mn, 0} (T_0)^{rs}_{pq} + \hat{\Theta}_{mn , a }{}^b (T_b^a)^{rs}_{pq}   \non \\ 
&=& 2 \Theta_{mn, [p}{}^{[r}\delta^{s]}_{q]} +  2\theta_{mn} \delta^{[r}_{[p} \delta^{s]}_{q]}   + \zeta  \theta_{ij} (T_a^b)^{ij}_{mn} (T^b_a)^{rs}_{pq} \non \\
&=& 2 \Theta_{mn, [p}{}^{[r}\delta^{s]}_{q]} +  \left(2 + \frac{\zeta}{5}  \right)\theta_{mn} \delta^{[r}_{[p} \delta^{s]}_{q]}  + \zeta \theta_{pq} \delta^{[r}_{[m} \delta^{s]}_{n]}   -\frac{1}{4} \zeta  \theta_{ij}\epsilon^{ij rs a} \epsilon_{a mn pq}  \non \\
\eea
One now calculates the symmetric part of the gauging 
\begin{equation}
\begin{aligned}
 \hat{X}_{mn, pq}{}^{rs}  + \hat{X}_{pq, mn}{}^{rs}  &  =2 \epsilon_{a mn pq}\left( Z^{rs ,a}  - \frac{\zeta}{4} \epsilon^{rs a i j }\theta_{ij} \right)  +
 \\&+    \left( 2 + \frac{6 \zeta}{5} \right) \left(\theta_{mn} \delta^{[r}_{[p} \delta^{s]}_{q]} +\theta_{pq} \delta^{[r}_{[m} \delta^{s]}_{n]}  \right) \  .
 \end{aligned}
\end{equation}
The requirements of supersymmetry as explained in \cite{LeDiffon:2008sh} are that this falls in  the same representation as without the trombone gauging hence we fix   $\zeta = - \frac{5}{3}$.    Although the symmetric part of the gauging does not depend on $Y$ note that the  that the antisymmetric part of the gauging  depends on $\theta$ $Z$ and $Y$.

The abelian vector fields $A^{\hat{\a}}_{ab}=A^{\hat{\a}}_{[ab]}$ of ungauged supergravity transform in the representation $\overline{\bf 10}$ of  $SL(5)$. These are turned into non-abelian fields by introducing a deformation given by the embedding tensor $\Q_{mn,p}{}^q$  that acts as structure constants.

The scalar fields of the theory are elements of the coset $SL(5)/SO(5)$ and are most conveniently described by an $SL(5)$ valued matrix $\mc{V}_a{}^{\dot{m}\dot{n}}$. It satisfies $\mc{V}_a{}^{\dot{m}\dot{n}}\W_{\dot{m}\dot{n}}$ and transforms as \cite{Samtleben:2005bp}
\begin{equation}
\begin{aligned}
\mc{V} \to G\mc{V} H, & & G\in SL(5), & &H\in SO(5).
\end{aligned}
\end{equation}
  Here the dotted Latin indices run from 1 to 4 labelling the fundamental representation of $USp(4)\simeq SO(5)$ and $\W_{\dot{m}\dot{p}}=\W_{[\dot{m}\dot{p}]}$ is the invariant symplectic form.
 
A coset representative is fixed by imposing a gauge condition with respect to the local $SO(5)$ invariance that result in a minimal parametrization of the coset space in terms of the $24-10=14$ physical scalars
 \begin{equation}
 \label{m}
 m_{ab}=\mc{V}_a{}^{\dot{m}\dot{n}}\mc{V}_b{}^{\dot{p}\dot{q}}\W_{\dot{m}\dot{p}}\W_{\dot{n}\dot{q}}.
 \end{equation}
The scalar potential of maximal gauged  supergravity can be expressed totally in terms of the unimodular $USp(4)$ invariant matrix $m_{ab}$ and the gaugings (except the on-shell trombone gauging)
 \begin{equation}
 \begin{split}
 V_{scalar}=&\fr{1}{64}\left(3X_{mn,r}{}^sX_{pq,s}{}^rm^{mp}m^{nq}-X_{mp,q}^{n}X_{nr,s}^{m}m^{pr}m^{qs}   \right)+\\
 &+\fr{1}{96}\left(X_{mn,r}{}^sX_{pq,t}{}^um^{mp}m^{nq}m^{rt}m_{su} + X_{mp,q}{}^nX_{nr,s}{}^mm^{pq}m^{rs}    \right).
 \end{split}
 \end{equation}
 In further sections it will be shown that the generalised metric acts as the coset representative \eqref{m} in Scherk-Schwarz reduction as it has precisely 14 components and provides exactly the scalar potential of the maximal gauged supergravity.

\subsection{$D=6$ supergravity}     

Maximal supergravity in six dimensions is invariant under the global duality group $SO(5,5)$. The representation $\mc{R}_V$ is now the spinorial representation ${\bf 16}_s$ of $SO(5,5)$. We let the capital Latin indices run from 1 to 16 labelling this representation and the small Latin indices run from 1 to 10 labelling the $\bf{10}$ representation of $SO(5,5)$. Then the components of the projected generators $X_M$ can be written in the spinorial representation as
\begin{equation}
X_{MN}{}^K=(X_M)_N{}^K=\widehat{\Q}_M{}^{ij} t_{ij}=\widehat{\Q}_M{}^{ij}(\G_{ij})_N{}^K,
\end{equation}
where $\G_{ij}=\G_{[i}\G_{j]}$ are the generators $t_{ij}$ in the spinorial representation while $\G_i$ are $16\times16$ gamma matrices in the Majorana representation. This means that they are real and symmetric
\begin{equation}
\G_i{}^{MN}=\G_i{}^{NM}.
\end{equation}
As it was shown in \cite{Bergshoeff:2007ef} and \cite{LeDiffon:2008sh} the gauge group generators are given by 
\begin{equation}
X_{MN}{}^K=-\q^{iL}\G^j{}_{LM}(\G_{ij})_{N}{}^K-\fr{1}{10}(\G^{ij})_M{}^L(\G_{ij})_N{}^K\q_L-\q_M\d_N{}^K.
\end{equation}
The generators are only written in terms of the gauging $\q^{iM}\in{\bf 144}$ and the trombone gauging $\q_M\in{\bf 16}$. The symmetric part $Z_{MN}{}^K=X_{(MN)}{}^K$ then reads
\begin{equation}
Z_{MN}{}^K=\G_{iMN}\hat{Z}^{iM},\quad Z^{iM}=-\q^{iM}-\fr25\G^{iMN}\q_N.
\end{equation}
Since the gauging $\q^{iM}$ is in the $\bf 144$ representation it satisfies the linear constraint \\ $\q^{iM}\G_{iMN}~=~0$.

Scalar fields of the theory are elements of the coset space $SO(5,5)/SO(5)\times SO(5)$ that can be conveniently parametrised by $SO(5,5)$ valued $16\times 16$ matrices $V_M{}^{\a\dot{\a}}$ \cite{Tanii:1984zk}. its inverse is defined by
\begin{equation}
V_{M}{}^{\a\dot{\a}}V^N{}_{\a\dot{\a}}=\d_M{}^N,\quad V_{M}{}^{\a\dot{\a}}V^M{}_{\b\dot{\b}}=\d^\a_\b\d^{\dot{\a}}_{\dot{\b}}.
\end{equation}
Here the dotted and the undotted small Greek indices run from 1 to 4 and label the spinor representation $\bf 4$ of each $SO(5)$ in the coset. 

In the absence of the trombone gauging the scalar potential can be written as
\begin{equation}
V_{scalar}=g^2\Tr\left(T^{\hat{a}}\widetilde{T}^{\hat{a}}-\fr12T\widetilde{T}\right),
\end{equation}
where tilde denotes transposition and the $T$-tensors are given by \cite{Bergshoeff:2007ef}
\begin{equation}
\label{T_tensor}
\begin{split}
(T^{\hat{a}})^{\a\dot{\a}}&=\mc{V}_i{}^{\hat{a}}\q^{iM}V_M{}^{\a\dot{\a}}\\
(T^{\hat{\dot{a}}})^{\a\dot{\a}}&=-\mc{V}_i{}^{\hat{\dot{a}}}\q^{iM}V_M{}^{\a\dot{\a}}\\
T&=T^{\hat{a}}\g^{\hat{a}}=-T^{\hat{\dot{a}}}\g^{\hat{\dot{a}}}.
\end{split}
\end{equation}
Here the hatted small Latin indices label the vector representation $\bf{5}$ of $SO(5)$ and dots again distinguish between two $SO(5)$'s in the coset. The gamma matrices $\gamma^{\hat{a}}$  and $\gamma^{\hat{\dot{\a}}}$ are $4\times4$ chiral gamma matrices whose vector indices are contracted without raising and lowering. The $10\times 5$ matrices $\mc{V}$ are defined as
\begin{equation}
\begin{split}
&\mc{V}_i{}^{\hat{a}}=\fr{1}{16}V_M{}^{\a\dot{\a}}(\g^{\hat{a}})_{\a}{}^{\b}\G_i{}^{MN}V_N{}_{\b\dot{\a}},\\ &\mc{V}_i{}^{\hat{\dot{a}}}=-\fr{1}{16}V_M{}^{\a\dot{\a}}(\g^{\hat{\dot{a}}})_{\dot{\a}}{}^{\dot{\b}}\G_i{}^{MN}V_N{}_{\a\dot{\b}}.
\end{split}
\end{equation}
According to the quadratic constraint the dotted and the undotted $T$ tensors are not independent and satisfy
\begin{equation}
\label{TT}
T^{\hat{a}}{}_{\a\dot{\a}}T^{\hat{a}}{}_{\b\dot{\b}}=T^{\hat{\dot{a}}}{}_{\a\dot{\a}}T^{\hat{\dot{a}}}{}_{\b\dot{\b}}.
\end{equation}

\subsection{$D=5$ supergravity}     
     
In five dimensions the global duality group of the maximal supergravity is $E_{6(6)}$ that is the maximal real subgroup of the complexified $E_6$ group. The representation $\mc{R}_V$ in this case is given by the $\bf 27$ representation of $E_{6(6)}$ and the capital Latin indices run from 1 to 27. The corresponding invariant tensor is a fully symmetric tensor $d_{MNK}$ that satisfies the following identities 
\begin{equation}
\label{d_ident0}
\begin{split}
d_{MPQ}d^{NPQ}&=\d_M^N,\\
d_{MRS}d^{SPT}d_{TNU}d^{URQ}&=\fr{1}{10}\d_{(M}^{P}\d_{N)}^{Q}-\fr25d_{MNR}d^{RQP},\\
d_{MPS}d^{SQT}d_{TRU}d^{UPV}d_{VQW}d^{WRN}&=-\fr{3}{10}\d_M^N.
\end{split}
\end{equation}

The linear constraint implied by supersymmetry restricts the full embedding tensor $\widehat{\Q}_M{}^{\underline{\a}}$ to the $\bf 27\oplus 351$ representation of $E_{6(6)}$. In the absence of the trombone gauging the embedding tensor reads
\begin{equation}
\Q_M{}^{\underline{\a}}=Z^{PQ}(t^{\underline{\a}})_R{}^Sd^{RKL}d_{MNK}d_{SQL}.
\end{equation}
The symmetric part of the gauge group generators $Z_{MN}{}^K=X_{(MN)}{}^K$ is then given by 
\begin{equation}
Z_{MN}{}^K=d_{MNL}\hat{Z}^{KL}, \quad \hat{Z}^{KL}=Z^{KL}-\fr{15}{2}d^{KLM}\q_{M}.
\end{equation}
A non-trivial relation among the generators of $E_{6(6)}$ that follows from the last line in \eqref{rel} is
\begin{equation}
(t^{\underline{\a}})_M{}^K(t_{\underline{\a}})_N{}^L=\fr{1}{18}\d_M^K\d_N^L+\fr16\d_M^L\d_N^K-\fr53d_{MNR}d^{RKL}.
\end{equation}

Scalar fields of the theory live in the coset space $E_{6(6)}/USp(8)$ and can be pa\-ra\-met\-rised by the scalar matrix $\mc{V}_M{}^{ij}$ with small Latin indices labelling the $\bf 8$ representation of $USp(8)$. The scalar matrix $\mc{V}_M{}^{ij}$ is antisymmetric in $ij$ and satisfies $\mc{V}_M{}^{ij}\W_{ij}=0$, where $\W_{ij}=-\W_{ji}$ is the symplectic invariant of $USp(8)$. Thus, the scalar matrix has $27\times 27$ components and its inverse is defined as
\begin{equation}
\begin{split}
&\mc{V}_M{}^{ij}\mc{V}_{ij}{}^N=\d_M^N\\
&\mc{V}_{ij}{}^M\mc{V}_M{}^{kl}=\d_{ij}{}^{kl}-\fr18\W_{ij}\W^{kl}.
\end{split}
\end{equation}

The matrix $\mc{V}$ can be used to elevate the embedding tensor to the so-called $T$-tensor that is $USp(8)$ covariant field dependent tensor. We need this tensor since it appears in the scalar potential. The convenient relation to be exploited  below is \cite{deWit:2004nw}
\begin{equation}
\label{XT}
X_{MN}{}^P=\mc{V}_{M}{}^{mn}\mc{V}_N{}^{kl}\mc{V}_{ij}{}^P\left[2\d_k{}^iT^j{}_{lmn} + T^{ijpq}{}_{mn}\W_{pk}\W_{ql}\lefteqn{\ph{\fr12}}\right]
\end{equation}
The tensor $T^{klmn}{}_{ij}$ belongs to the $\bf 315$ representation while $T^i{}_{jlm}$ is in the $\bf 36\oplus 315$. It is possible to write these two tensors in terms of two pseudoreal, symplectic traceless, tensors $A_1{}^{ij}\in\bf 35$ and $A_2{}^{i,jkl}\in \bf 315$ as
\begin{equation}
\label{TA}
\begin{split}
T^{klmn}{}_{ij}&=4A_2{}^{q,[klm}\d^{n]}{}_{[i}\W_{j]q}+3A_2{}^{p,q[kl}\W^{mn]}\W_{p[i}\W_{j]q},\\
T_i{}^{jkl}&= -\W_{im}A_2{}^{(m,j)kl}-\W_{im}\left(\W^{m[k}A_1{}^{l]j}+\W^{j[k}A_1{}^{l]j}+\fr14\W^{kl}A_1{}^{mj}\right).
\end{split}
\end{equation}
Tensors $A_1$  and $A_2$ satisfy $A_1{}^{[ij]}=0$, $A_2{}^{i,jkl}=A_2{}^{i,[jkl]}$ and $A_2{}^{[i,jkl]}=0$. The scalar potential then can be written as
\begin{equation}
V_{scalar}=g^2\left[3|A_1{}^{ij}|^2-\fr13|A_2{}^{i,jkl}|^2\right],
\end{equation}
where $|\;|^2$ stands for the contraction of all indices.

\section{Scherk-Schwarz reduction}

In contrast to the Kaluza-Klein reduction here the dependence on internal coordinates is hidden in so-called twist matrices $W^{A}{}_{\bar{B}}(X)$ that are subject to various constraints. For the case at hand we consider the whole extended space as an internal space and let the remained $D$-dimensional space to be whatever it wants to be \cite{Aldazabal:2011nj, Grana:2012rr}:
\begin{equation}
\label{twist}
V^{A}(\XX^M,x_{(D)})=W^{A}{}_{\bB}(\XX) V^{\bB}(x_{(D)}),
\end{equation}
where $V^A$ is a generalised vector on extended space defined by its transformation \eqref{gen_Lie_2} and $W^A_{\bB}$ is the Scherk-Schwarz generalised twisting matrix. The anzats for tensor of higher rank is introduced in a similar way.

From now on we will not include the dependence on $x_{(D)}$ since it does not affect the extended geometry formalism. The barred indices are the twisted ones (flat) and the unbarred are the untwisted ones (curved). To simplify notation we will use the unbarred indices for the flat space in cases where this does not cause confusion.

The important feature of the Scherk-Schwarz reduction is that it allows non-abelian gauge groups. Substituting the anzatz \eqref{twist} into the local transformations of the initial theory that are given by the generalised Lie derivative \eqref{gen_Lie_2} we obtain the following transformation of the vector $Q^A$
\begin{equation}
\label{Lie_twisted}
\d_\S Q^{A}=(\mc{L}_\S Q)^A=W^{A}{}_{\bar{B}}X_{\br{K}{L}}{}^{\bar{B}}\S^{\bar{K}}Q^{\bar{L}}.
\end{equation}
Here the coefficients $X_{MN}{}^K$ are defined as
\begin{equation}
\label{Xgen}
X_{\br{A}{B}}{}^{\bar{C}}\equiv2W_C{}^{\bar{C}}\dt_{[\bar{A}}W_{\bar{B}]}{}^C+Y^{\br{C}{D}}_{\br{M}{B}}W_C{}^{\bar{M}}\dt_{\bar{D}}W_{\bar{A}}{}^{C}
\end{equation}
with the antisymmetrisation factor of $1/2$, and are assumed to be constants. One should note that in the case of extended geometry these ``structure constants'' are not antisymmetric.

We recall the closure constraint \eqref{closure}
\begin{equation}
\mc{L}_{[X_1,X_2]_C} Q^M -  [\mc{L}_{X_1},\mc{L}_{X_2}]Q^{M } = -  F_0^{M}.
\end{equation}
Assuming that $X_{MN}{}^K$ is constant and substituting the twist anzats \eqref{twist} and the explicit from of $F_0$ \cite{Berman:2011jh, Berman:2011cg} this implies
\begin{equation}
\frac{1}{2} \left( X_{\bar{A} \bar{B} }{}^{\bar{C}} -    X_{\bar{A} \bar{B} }{}^{\bar{C} }\right)X_{\bar{C} \bar{E}}{}^{\bar{G}} - X_{\bar{B} \bar{E}}{}^{\bar{C}}X_{\bar{A} \bar{C}}{}^{\bar{G}} +   X_{\bar{A} \bar{E}}{}^{\bar{C}}X_{\bar{B} \bar{C}}{}^{\bar{G}}  = 0 \  
\end{equation}
for any $X_1$ and $X_2$. If we define $X_{MN}{}^K=(X_M)_N{}^K$ this can be written in the suggestive form
 \begin{equation}
  \label{closure1}
 [ X_{\bar{A}}, X_{\bar{B}}]  = - X_{[\bar{A} \bar{B}]}{}^{\bar{C}} X_{\bar{C}}  \ . 
 \end{equation}
This  allows one to interpret the structure constants as the components of the generators $X_M$ of the algebra of transformations
\begin{equation}
\label{G_transf}
\d_{\S}Q^{\bar{A}}=X_{\br{K}{L}}{}^{\bar{A}}\S^{\bar{K}}Q^{\bar{L}}
\end{equation}
in adjoin representation. By making use of the closure constraint \eqref{closure1} we find the Jacobiator
\begin{equation}
\begin{split}
\label{Jac}
&[\delta_{\S_1}, [\delta_{\S_2}, \delta_{\S_3} ] ]V^{\bar{F}} + c.p. = \\ & \left( X_{[\bar{A}\bar{B}]}{}^{\bar{E}}X_{[\bar{E} \bar{C}] }{}^{\bar{G}}   +   X_{[\bar{C}\bar{A}]}{}^{\bar{E}}X_{[\bar{E} \bar{B}] }{}^{\bar{G}} +X_{[\bar{B}\bar{C}]}{}^{\bar{E}}X_{[\bar{E} \bar{A}] }{}^{\bar{G}}   \right)X_{\bar{G} \bar{D}}{}^{\bar{F} } \S_1^{\bar{A}}\S_2^{\bar{B}} \S_3^{\bar{C}} V^{\bar{D}},
\end{split}
\end{equation}
where {\it c.p.} denotes cyclic permutations. The right hand side of this equation is the Jacobi identity of the antisymmetric part $X_{[MN]}{}^K$ projected into the algebra generator. For the consistency of the algebra of transformations the right hand side should vanish. We emphasise that the Jacobi identity for $X_{[MN]}{}^K$ needs only to hold after the projection.

We need $X_{MN}{}^K$ to be not only constants but also invariant objects under the local symmetry transformations. As it will be shown later it is necessary so that the reduced action does not depend on the internal coordinates and transforms as a scalar. As it follows from the definition \eqref{Lie_twisted} the structure constants $X_{MN}{}^K$ should transform as a generalised tensor
\begin{equation}
\delta_{\S} X_{\bar{A} \bar{B}}{}^{\bar{C}}  =   \S^{\bar{E}} \left(  [ X_{\bar{E}}, X_{\bar{A}}] _{\bar{B}}{}^{\bar{C}} +  X_{\bar{E} \bar{A}}{}^{\bar{D}} (X_{\bar{D}})_{\bar{B}}{}^{\bar{C}} \right).
\end{equation}
This leads to the final quadratic constraint on the structure constants
 \begin{equation}
  \label{closure2}
 [ X_{\bar{A}}, X_{\bar{B}}]  = - X_{\bar{A} \bar{B}}{}^{\bar{C}} X_{\bar{C}}  \ . 
 \end{equation}
We conclude from this constraint that the symmetric part $Z_{MN}{}^K=X_{(MN)}{}^K$ should vanish when projected into a generator
\begin{equation}
Z_{\bar{A} \bar{B}}{}^{\bar{C}} X_{\bar{C}}  = 0 \ .
\end{equation}

The quadratic constraint \eqref{closure2} on its own is enough to ensure that the Jacobiator \eqref{Jac} vanishes and the algebra is closed. This can be seen by considering the Jacobi identity for the commutator appearing in \eqref{closure2}. Hence the closure condition can be relaxed from the section condition that restricts fields and their products to a condition on the structure constants $X_{MN}{}^K$ that define the algebra of gauge transformations. 

\begin{figure}
\centering
\begin{tikzpicture}[thick,
                    text height=1ex,
                    text depth=.15ex
                    ]
\node at (0,-1) [format] (ST) { M-theory};                   
\node at (6,0) [format] (GG) {Extended geometry};                   
\node at (0,-3) [format] (N11) {$\mc{N}=1\;D=11 $ Supegravity};
\node at (0,-7.3) [format] (Ung) {Ungauged supergravity};
\node at  (6,-7.3) [format] (Ga)  {Gauged  supergravity};
\node at (2.8,-1.7) [rotate=26, above=0.1] {\small section};
\node at (3.2,-1.3) [rotate=26, below=0.1] {\small condition};
\node at (1.4,-2.0) [left=0.1] {\small low energy};
\node at (3.1,-7.3) [above=0.1] {\small gauging};
\node at (6.7,-2.3) [above=0.1] {\parbox{3cm}{\small \begin{tabular}{r}Generalised \\ Scherk-Schwarz \\ reduction \end{tabular}}};
\node at (0.3,-5) [left=0.1] {\parbox{2cm}{ \small reduction \\ on torus \\ $\mathbb{T}^n$}};
\node at (2,-5) [right=0.0]  {\parbox{6cm}{\small reduction in presence of \\\ph{1}\hspace{0.5cm}-- $p$--form fluxes 
								{\color{black}$\int_\S F^{(p)}=C_\S$} 
\\\ph{1}\hspace{1cm} -- torsion (geometric flux) \\ \ph{1}\hspace{1.5cm} {\color{black}$de^a=\mc{T}^a_{bc}e^b\wedge e^c$} \\\ph{1}\hspace{2cm} -- non-geometric flux}};

\path[<-] (Ung) edge   (N11);
\path[->] (N11) edge  (Ga);
\path[->] (Ung)  edge  (Ga);
\path[->] (ST) edge (N11);
\path[->] (GG) edge (N11);
\path[->] (GG) edge [out=-25, in=25] (Ga);
\end{tikzpicture}
{\caption{\small This diagram demonstrates relations between reductions of $\mc{N}=11$ supergravity and Scherk-Schwarz reductions in the extended geometry formalism.}}
\end{figure}
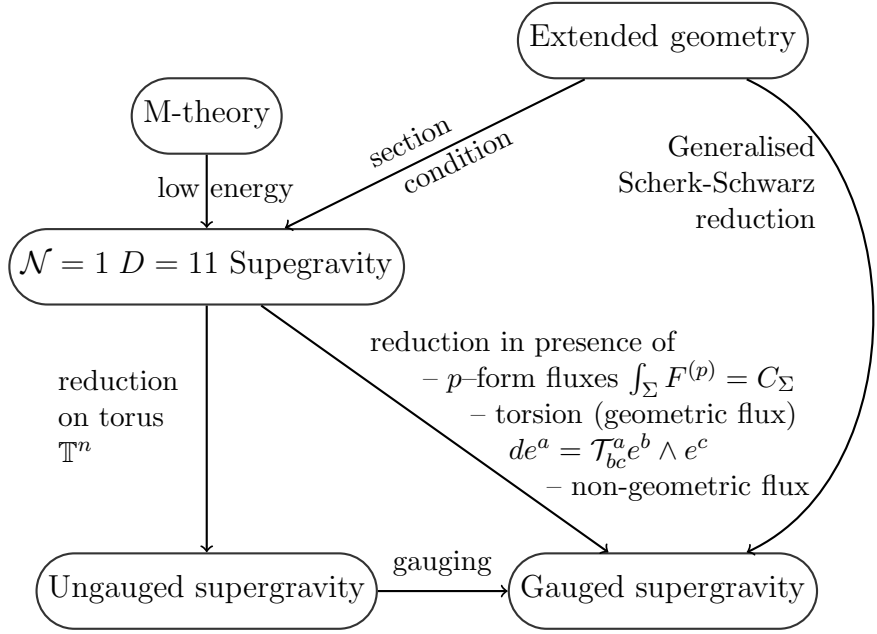

\section{Algebraic structure}

The general form of the structure constants $X_{AB}{}^{C}$ is always the same and is given by \eqref{Xgen}. Under a particular U-duality group these split into certain representations that depend on the duality group and are identified with gaugings. In this section we give an explicit derivation of the embedding tensor and all gaugings starting from $X_{AB}{}^C$ in its general form.

\subsection{$SL(5)$: reduction to 7 dimensions}

The extended space formalism introduced in \cite{Berman:2010is} starts with 4 compact dimensions and rewrites the low energy action in an $SL(5)$ invariant form. The generalised metric parametrises the coset $SL(5)/SO(5)$ and has 14 components, given by the metric and the 3-form field in 4 dimensions. In Section \ref{M} it was shown in details that the ordinary coordinates $x^\m$ and the dual ones $y_{\a\b}$ can be collected into an $SL(5)$-covariant extended coordinate
\begin{equation}
\XX^{ab}=
\begin{bmatrix}
\XX^{5\m}\\
\XX^{\m\n}
\end{bmatrix}=
\begin{bmatrix}
x^{\m}\\
\fr12\e^{\m\n\a\b}y_{\a\b}
\end{bmatrix},
\end{equation}
where $\e^{\m\n\a\b}$ ($\e^{1234}=1$) is the 4-dimensional alternating symbol. Then the generalised Scherk-Schwarz twisting \eqref{twist} takes the following form
\begin{equation}
Q^{ab}(\XX^{mn},x_{(D)})=W^{ab}_{\br{c}{d}}(X)Q^{\br{c}{d}}(x_{(D)}).
\end{equation}
The twisting matrix $\uW{a}{b}{c}{d}$ is written in the representation $\bf{10}$ of $SL(5)$ and can be decomposed into a product of two matrices $V^a_{\bc}$ in the fundamental representation
\begin{equation}
\label{WtoV}
\uW{a}{b}{c}{d}=\fr12 \left(V^a_{\bc}V^{b}_{\bd}-V^a_{\bd}V^{b}_{\bc}\right).
\end{equation}

Recall the explicit form of the invariant tensor $Y^{MN}{}_{KL}$ from the table \eqref{Y} where all relevant cases are collected
\begin{equation}
Y^{MN}_{PQ}= \e^{a MN}\e_{a PQ} \Longrightarrow \fr{1}{8}\e^{amnkl}\e_{apqrs},
\end{equation}
where each antisymmetric pair of indices $M=[ab]$ that will be a dummy index in $X_{MN}{}^K$ carries a factor of one half. Then the would-be structure constants written in terms of the twist matrices read
\begin{equation}
\label{F}
\begin{aligned}
X_{\br{c}{d},\br{e}{f}}{}^{\br{a}{b}}&= \frac{1}{2} \left(   \vW{a}{b}{m}{n} \partial_{\bar{c} \bar{d}} \uW{m}{n}{e}{f} -   \vW{a}{b}{m}{n} \partial_{\bar{e} \bar{f}} \uW{m}{n}{c}{d}   +\frac{1}{4} \epsilon^{ \bar{a} \bar{b} \bar{i} \bar{j} \bar{k}}\epsilon_{\bar{k} \bar{p} \bar{q} \bar{c} \bar{d} }  \vW{p}{q}{m}{n} \partial_{\bar{i} \bar{j}} \uW{m}{n}{a}{b}  \right) \nonumber  \\ 
&= \fr12\vW{a}{b}{m}{n}\uW{p}{q}{c}{d}\dt_{p q}\uW{m}{n}{e}{f}+\fr12\d^{\bar{a}\bar{b}}_{\be\bar{f}}\dt_{mn}\uW{m}{n}{c}{d} +2\vW{a}{b}{m}{n}\uW{m}{p}{e}{f}\dt_{pq}\uW{q}{n}{c}{d} \ . 
\end{aligned}
\end{equation}
We encounter our first constraint on the Scherk--Schwarz twist element which is that these objects are constant. However, an immediate difference to the $O(d,d)$ case is that these ``structure constants" are not anti-symmetric in their lower indices -- to correct this misnomer we shall refer to them as gaugings rather than structure constants.  By making use of the invariance of the epsilon tensor and the decomposition   \eqref{WtoV}, the symmetric part of the gaugings can be extracted as
\begin{equation}
X_{\br{c}{d},\br{e}{f}}{}^{\br{a}{b}}+X_{\br{e}{f},\br{c}{d}}{}^{\br{a}{b}} = \frac{1}{8} \epsilon_{\bar{i} \bar{c} \bar{d} \bar{e} \bar{f} } \epsilon^{\bar{j} \bar{m} \bar{n} \bar{a}\bar{b}} V^{\bar{i}}_{p} \partial_{\bar{m} \bar{n}} V^p_{\bar{j}} \  .
\end{equation}
To see the full content of the gauging it is in fact  helpful to decompose according  \eqref{WtoV}.  One finds that 
\begin{equation}
X_{\br{c}{d},\br{e}{f}}{}^{\br{a}{b}}  = 2 X_{\bar{c}\bar{d} , [\bar{e}}{}^{[\bar{a}}\delta_{\bar{f}]}^{\bar{b}]}  \ ,
\end{equation}
with
\begin{equation}
X_{\bar{c}\bar{d} , \bar{e}}{}^{\bar{a}} = \frac{1}{2} V^{\bar{a}}_{m} \partial_{\bar{c} \bar{d} } V_{\bar{e}}^m + (T^{\bar{p}}_{\bar{q}})^{\bar{m} \bar{n}}_{\bar{r} [\bar{c}} \left( V^{\bar{r}}_t \partial_{\bar{m} \bar{n}} V^t_{\bar{d}]}\right) (T^{\bar{q}}_{\bar{p}})^{\bar{a}}_{\bar{e}} - \frac{1}{10} \delta^{\bar{a}}_{\bar{e}} V^{\bar{m}}_t \partial_{\bar{m} [\bar{c}} V_{\bar{d}]}^t \ ,
\end{equation}
in which  $(T^{\bar{q}}_{\bar{p}})^{\bar{a}}_{\bar{e}}$ and $(T^{\bar{p}}_{\bar{q}})^{\bar{m} \bar{n}}_{\bar{r} \bar{c}} $ are the $SL(5)$ generators in the $\bf{5}$ and $\bf{10}$ respectively (see appendix).  This result can be expressed as 
\begin{equation}
X_{\bar{c}\bar{d} , \bar{e}}{}^{\bar{a}} = \delta^{\bar{a} }_{[\bar{c} } Y_{\bar{d}] \bar{e} }  - \frac{10}{3} \delta^{\bar{a} }_{[\bar{c} }  \theta_{\bar{d}] \bar{e} }   - 2 \epsilon_{\bar{c}\bar{d} \bar{e} \bar{m} \bar{n} } Z^{\bar{m} \bar{n}, \bar{a} }  + \frac{1}{3} \theta_{\bar{c}\bar{d}} \delta_{\bar{e}}^{\bar{a} } \ ,
\end{equation}
where $Y_{\bar{c} \bar{d}}= Y_{\bar{d} \bar{c}}$ is in the $\bf{15}$ and is given by 
\begin{equation}
Y_{\bar{c} \bar{d} } =  V^{\bar{m}}_t \partial_{\bar{m} (\bar{c} } V^t_{\bar{d})}  \ , 
\end{equation}
and $ Z^{\bar{m} \bar{n}, \bar{p} } = -  Z^{\bar{n} \bar{m}, \bar{p} } $ is in the $\overline{\bf{40}}$ such that  $ Z^{[\bar{m} \bar{n}, \bar{p}] } = 0 $  is given by 
\begin{equation}
Z^{\bar{m} \bar{n}, \bar{p} } = - \frac{1}{24} \left( \epsilon^{\bar{m} \bar{n} \bar{i} \bar{j} \bar{k} } V_{t}^{\bar{p}} \partial_{\bar{i}\bar{j}} V_{\bar{k}}^t    + V_{t}^{[\bar{m}} \partial_{\bar{i}\bar{j}} V^{|t|}_{\bar{k}}  \epsilon^{\bar{n}]  \bar{i} \bar{j} \bar{k}  \bar{p}}  \right)  \ ,
\end{equation}
and $\theta_{\bar{c} \bar{d}}= - \theta_{\bar{d} \bar{c}}$ is in the $\bf{10}$ and is given by 
\begin{equation}
\theta_{\bar{c} \bar{d}}  =  \frac{1}{10 } \left( V^{\bar{m}}_t \partial_{\bar{c} \bar{d}} V_{\bar{m}}^t  - V^{\bar{m}}_t \partial_{\bar{m} [\bar{c} } V^t_{\bar{d}]} \right)  \ . 
\end{equation}
It is note worthy that although $\bf{10 \otimes 24}=  \bf{10 \oplus 15 \oplus \overline{40} \oplus  175}$ the $\bf{175}$ makes no appearance in the gaugings produced by Scherk--Schwarz reduction.  

\subsection{$SO(5,5)$: reduction to 6 dimensions }

Maximal supergravity in 6 dimensions possesses a global duality group $E_{5(5)}=SO(5,5)$. The local group of the theory is $SO(5)\times SO(5)$. Thus the target space of scalar fields of the theory is given by the coset
\begin{equation}
\frac{SO(5,5)}{SO(5)\times SO(5)}.
\end{equation}
The corresponding extended space of the Berman-Perry formalism has 16 dimensions and the representation $\mc{R}_V$ appears to be the spinorial representation of $SO(5,5)$.

The invariant tensor of the duality group is given by the contraction of two gamma matrices in the Majorana representation 
\begin{equation}
Y^{MN}_{KL}=\fr{1}{2}\G^{iMN}\G_{iKL},
\end{equation}
that are thus symmetric and real. Here capital Latin indices run from 1 to 16 and small Latin indices run from 1 to 10 labelling the vector representation of $SO(5,5)$. Since the generators $t^{\underline{\a}}$ of the duality group  in the spinorial representation are given by $\G^{ij}$, where the multiindex $\underline{\a}$ is represented by the antisymmetric pair of vector indices, the projector with correct normalisation is defined as
\begin{equation} 
P_N{}^M{}_L{}^K=-\fr{1}{32}(\G^{ij})_N{}^M(\G_{ij})_L{}^K.
\end{equation}

All gaugings of the maximal supergravity appear as components of the structure constants (or the embedding tensor). Start with the trace part of the structure constants \eqref{Xgen}
\begin{equation}
\label{XtraceSO}
X_{\br{M}{N}}{}^{\bar{N}}=4\dt_{C}W^C_{\bar{M}} +W^{\bar{C}}_C\dt_{\bar{M}}W^C_{\bar{C}} =:-16\q_{\bar{M}}.
\end{equation}
By making use of the algebra of gamma matrices the symmetric part of the gaugings can be extracted as
\begin{equation}
\label{ZSO}
\begin{split}
X_{(AB)}{}^C&=\G_{iAB}Z^{iC},\\
Z^{\br{i}{C}}&=\fr14 \G^{\bar{j}\br{C}{D}}G_{\bar{j}}{}^i\dt_{\bar{D}}G_{i}{}^{\bar{i}},
\end{split}
\end{equation}
where the twist matrices in the vector representation $G_{i}{}^{\bar{j}}$ are defined as
\begin{eqnarray}
\G^{iAB}G_{i}{}^{\bar{j}}=\G^{\bar{j}\br{C}{D}}W_{\bar{C}}{}^AW_{\bar{D}}{}^B.
\end{eqnarray}
According to its indices the gauging $Z^{iB}$ is in the $\bf{16\otimes10=16\oplus144}$ representation of $SO(5,5)$. Separating the $\bf{16}$ part of the gauging we obtain the trombone gauging $\q_M$
\begin{equation}
Z^{iM}\G_{iMN}=-4\q_N.
\end{equation}
What is left lives in the $\bf{144}$ representation and is defined as
\begin{equation}
\q^{iM}=-Z^{iM}-\fr25\G^{iMN}\q_N.
\end{equation}
After some algebra (see Appendix A) the structure constants can be rewritten in terms of only these objects 
\begin{equation}
\label{XSO}
X_{MN}{}^K= -\q^{iL}\G^j_{LM}(\G_{ij})_N{}^K - \fr{1}{10}(\G^{ij})_M{}^L(\G_{ij})_N{}^K\q_L - \d_N^K\q_M.
\end{equation}
This has the same structure as the embedding tensor of the maximal supergravity in 6 dimensions
\begin{equation}
X_{MN}{}^K \in \bf{16\oplus 144}.
\end{equation}
with gaugings explicitly written in terms of the twist matrices as
\begin{equation}
\label{gaugingsSO}
\begin{split}
\q^{\br{i}{M}}&=-\fr14 \G^{\bar{j}\br{M}{D}}G_{\bar{j}}{}^i\dt_{\bar{D}}G_{i}{}^{\bar{i}}-\fr25\G^{\br{i}{M}\bar{N}}\q_{N},\\
\q_{\bar{N}}&=-\fr{1}{16}\G^{\br{A}{D}\bar{i}}\G_{\br{j}{A}\bar{N}}G_{\bar{i}}{}^{i}\dt_{\bar{D}}G_{i}{}^{\bar{i}}.
\end{split}
\end{equation}
It is straightforward to check that the second line here is the same as \eqref{XtraceSO} using the definition of $G_{i}{}^{\bar{i}}$ and the relation
\begin{equation}
G_{\bar{j}}{}^j\dt_{\bar{A}}G_j{}^{\bar{i}}=\fr18\G^{\br{B}{K}\bar{i}}\G_{\bar{j}\br{C}{K}}W_C{}^{\bar{C}}\dt_{\bar{A}}W_{\bar{B}}{}^C.
\end{equation}

\subsection{$E_{6(6)}$: reduction to 5 dimensions }

In five dimensions vector fields of maximal supergravity transform in the $\mathbf 27$ representation of the global duality group $E_{6(6)}$. The scalar fields transform non-linearly and are parametrised by elements of the coset
\begin{equation}
\frac{E_{6(6)}}{USp(8)}.
\end{equation}
The group $USp(8)$ is the R-symmetry group of the theory. 

The U-duality invariant formalism of the extended geometry provides the extended space to be 27 dimensional and the generalised vector indices $A, B \ldots$ label the $\mathbf{27}$ representation of $E_{6(6)}$. The invariant tensor is given by the $E_{6(6)}$ symmetric invariant tensor $d_{MNK}$
\begin{equation}
Y^{MN}_{RS}=10d^{MNK}d_{KRS},
\end{equation}
that is subject to the following useful identities
\begin{equation}
\label{d_indent}
\begin{split}
d_{MPQ}d^{NPQ}&=\d_M^N,\\
d_{MRS}d^{SPT}d_{TNU}d^{URQ}&=\fr{1}{10}\d_{(M}^{P}\d_{N)}^{Q}-\fr25d_{MNR}d^{RQP},\\
d_{MPS}d^{SQT}d_{TRU}d^{UPV}d_{VQW}d^{WRN}&=-\fr{3}{10}\d_M^N.
\end{split}
\end{equation}
The trace part of the structure constant is identified with the trombone gauging $\q_{M}$ and reads
\begin{equation}
\label{XtraceE}
X_{\br{M}{N}}{}^{\bar{N}}=9\dt_{C}W^C_{\bar{M}}+W^{\bar{C}}_C\dt_{\bar{M}}W^C_{\bar{C}}=-27\q_{\bar{M}}.
\end{equation}
The intertwining tensor is given by the symmetric part of the structure constant $Z_{MN}{}^K=X_{(MN)}{}^K$ and is parametrised by the tensor $\hat{Z}^{MN}$ in the $\mathbf{27\oplus351}$
\begin{equation}
\begin{split}
&Z_{MN}{}^K=d_{MNR}\hat{Z}^{RK}
\end{split}
\end{equation}
Taking the symmetric part of \eqref{Xgen} and by making use of the identities \eqref{d_indent} we have for the symmetric part
\begin{equation}
\label{ZE}
\hat{Z}^{\br{M}{N}}=5d^{\br{M}{K}\bar{L}}W^{C}_{\bar{L}}\dt_{\bar{K}}W^{\bar{N}}_C
\end{equation}
that has the same structure as \eqref{ZSO} if one notices that
\begin{eqnarray}
W^{M}_{\bar{M}}d^{\bar{M}\br{K}{L}}=d^{MKL}W_{K}^{\bar{K}}W_{L}^{\bar{L}}
\end{eqnarray}
since the twist matrices interpolate between the barred and the unbarred indices. 

Subtracting the part of the tensor \eqref{ZE} that is symmetric in $MN$ we are left with the gauging in the $\mathbf{351}$ and the trombone gauging
\begin{equation}
\begin{split}
Z^{MN}&=\hat{Z}^{MN}+\fr{15}{2}d^{MNK}\q_{K},\\
\q_{\bar{N}}&=5\,d^{\br{M}{B}\bar{L}}d_{\br{M}{N}\bar{K}}W^L_{\bar{L}}\dt_{\bar{B}}W^{\bar{N}}_L.
\end{split}
\end{equation}
Thus the structure constant $X_{MN}{}^K$ is in the $\mathbf{27\oplus351}$ representation of $E_{6(6)}$.


\section{Scalar potential}

The effective potential $V=V(M_{AB}, \dt_KM_{AB})$ that depends on the generalised metric $M_{AB}$ and its derivatives after twisting should become the scalar potential for the appropriate gauged SUGRA. It appears that one must add an extra term of type
\begin{equation}
Y^{AB}_{MN}\dt_{A}E_{\X}{}^M\dt_BE_{\Q}{}^N\d^{\X\Q},
\end{equation}
where $E_\X{}^M$ is a generalised vielbein and $\d^{\X\Q}$ is the Kronecker delta. It can be always added to the action since it is zero up to the section condition. This term is necessary for two major reasons. Firstly, this term allows to write the action in terms of the structure constants $X_{MN}{}^K$. Secondly, this terms provides the action that is invariant under gauge transformations \eqref{G_transf}. Not to be confused, one should think of this term as a term that \emph{has always been in the action} but has usually been dropped because of the section condition. After the section condition is relaxed it is important to add this term since it provides the invariance of the action.

\subsection{$D=7$ supergravity}

The effective potential that defines the $SL(5)$ covariant dynamics has the following form
\begin{equation}
\label{lag}
V =  \sqrt{g}    \left(  \fr12 V_1 - \fr12 V_2 + \fr14 V_3 +\fr{1}{12} V_4 +V_5 \right)
\end{equation}
where
\begin{equation}
\begin{aligned}
V_{1} & = M^{MN} \pl_M M^{KL} \pl_N M_{KL} , & V_2 & =  M^{MN} \pl_M M^{KL} \pl_K M_{NL}  \ ,  \\ 
 V_{3} & = -\pl_M M^{MP} \left( M^{RS} \pl_P M_{RS} \right)  , & V_{4} & = M^{MN}  \left( M^{RS} \pl_M M_{RS} \right) \left( M^{KL} \pl_N M_{KL} \right) \\
 V_{5}& = \epsilon^{a MN} \epsilon_{a PQ}  E^{\hat{A}}_{R} M^{RS} E^{\hat{B}}_{S}  \partial_{M} E^{P}_{\hat{A}} \partial_N E^Q_{\hat{B}} \ , 
\end{aligned}
\end{equation}
with the generalised metric (see Section \ref{M}) and the vielbein $E^{\hat{A}}_M$ where the hatted Latin indices label flat coordinates
\begin{equation}
\begin{aligned}
M_{MN}&=
\begin{bmatrix}
g_{\m\n}+\fr12 C_{\m\a\b}C_\n{}^{\a\b} & \fr{1}{\sqrt{2}}C_{\n{}^{\r\s}} \\
&\\
\fr{1}{\sqrt{2}}C_{\m}^{\g\d} & g^{\g\d,\r\s}
\end{bmatrix},&
M_{MN} = E^{\hat{A}}_M \delta_{\hat{A}\hat{B}} E^{\hat{B}}_N.
\end{aligned}
\end{equation}

We now apply the Scherk--Schwarz ansatz   to the terms in the action to find the reduced theory.   We will find it convenient to work not with the $10 \times 10$ matrix {\it{big}}  $M_{MN}$ but instead with the $5 \times 5$ little $m_{mn}$ defined by 
\begin{equation}
\begin{aligned}
M_{MN}&=M_{mn,pq}= m_{mp} m_{nq} - m_{mq} m_{pq}\ ,\\
M^{MN}&=M^{mn,pq}= m^{mp} m^{nq} - m^{mq} m^{pq}\ .
\end{aligned}
\end{equation}
The metric in the fundamental representation is given by 
\begin{equation}
m_{mn}=  \left( \begin{array}{cc} g^{-1/2}  g_{\mu \nu}  & V_\nu \\
V_\m & \det{g}^{1/2}( 1+V_\m V_\n g^{\m \n}   )  \end{array}\right)
\end{equation}
where $V^\m = \frac{1}{6} \e^{\m \n \rho \sigma} C_{\nu \rho \sigma}$ and $ \e^{\m \n \rho \sigma} $ is the alternating {\it tensor}. This object has determinant $\det m_{mn} = \det{g}^{-\frac{1}{2} }$.  In terms of little $m$ the terms in the potential reads\footnote{We found the computer algebra package {\tt{Cadabra}} \cite{Peeters:2007wn,Cadabra} a useful tool for verifying some of the more laborious manipulations in this section} 
\bea
V_{1} &=& \frac{3}{2} m^{pr} m^{qs} \pl_{pq} m^{mn} \pl_{rs} m_{mn}   - \frac{1}{2}  m^{pr} m^{qs} Tr(m^{-1}\pl_{pq} m)  Tr(m^{-1}\pl_{mn} m) \ , \nonumber\\ 
V_{2} &=&   m^{pr} m^{qs} \pl_{pq} m^{kl} \pl_{ks} m_{rl} - \pl_{pq} m^{pk} \pl_{kl} m^{lq} \ , \nonumber\\ 
V_{3} &=&4m^{mq}m^{ij} \pl_{pq}m_{ij} \pl_{mk} m^{kp} ,  \nonumber \\ 
V_{4} &=&    8 m^{pr} m^{qs} Tr(m^{-1}\pl_{pq} m)  Tr(m^{-1}\pl_{mn} m)  \  . 
\eea
For little m the Scherk--Schwarz ansatz is then  
\begin{equation}
\label{ansatz}
m_{mn} =  V_m^\ab m_{\ab \bbr} V_n^\bbr     \ . 
\end{equation}

Let us introduce some notation:
\begin{equation}
\Lambda^{\ab}{}_{\bbr \cb \db} = V^{\ab}_m \partial_{\bbr \cb} V^m_{\db} \ ,  \quad 
\Lambda^{\ab}{}_{\bbr \cb \ab} = \chi_{\bbr \cb} \ ,  \quad \Lambda^{\ab}{}_{\ab \bbr \cb} = \psi_{\bbr \cb} \ . 
\end{equation}
Assuming that $\partial_{mn } m_{\ab\b} = 0$ we obtain
\bea
 V_{1}  &=& - m^{\ab \bbr}  m^{\cb \db} \left[3\Lambda^{\eb}{}_{\ab\cb \fb} \Lambda^{\fb}{}_{\bbr \db \eb}   + 2    \chi_{\ab \cb} \chi_{\bbr \db}  +3   m_{\eb \fb} m^{\gb \hb} \Lambda^{\eb}{}_{\ab \cb \gb} \Lambda^{\fb}{}_{\bbr\db\hb} \right]  \nonumber \\
V_{2} &=&-  m^{\ab \bbr} m^{\cb \db}\left[ 2\Lambda^{\eb}{}_{\fb \ab \cb} \Lambda^{\fb}{}_{\db \bbr \eb} +  \Lambda^{\eb}{}_{\fb \ab \cb} \Lambda^{\fb}{}_{\eb \bbr \db}  -   \Lambda^{\eb}{}_{\fb \ab \bbr} \Lambda^{\fb}{}_{\eb \cb \db} + 2 \psi_{\eb \cb} \Lambda^{\eb}{}_{\db \ab \bbr} -\psi_{\ab\cb} \psi_{\bbr \db}    \right]  \nonumber \\
&& \quad -  m^{\ab \bbr} m^{\cb \db}\left[  m_{\eb \fb} m^{\gb \hb} \Lambda^{\eb}{}_{\ab \cb \gb} \Lambda^{\fb}{}_{\hb\db\bbr}  \right] \nonumber \\
V_{3} &=& - 8 m^{\ab \bbr} m^{\cb \db} \left[ \chi_{\eb \ab} \Lambda^{\eb}{}_{\bbr \cb \db}   + \chi_{\bbr \cb} \psi_{\ab \db} \right]\nonumber \\
V_{4} &=& 32 m^{\ab \bbr} m^{\cb \db} \chi_{\ab \cb} \chi_{\bbr \db} 
\eea
for the original terms in the action.  For the extra term (which vanishes upon the strong constraint) we find 
\begin{equation}
\begin{split}
V_{5} &=  - \e^{\ab \bbr \cb \db\eb} \e_{\ab  \fb \gb \hb \bar{i} } \left(  m^{\bar{p}  \bar{i} } m^{\bar{q} \gb}    \right) \Lambda^\fb{}_{\bbr \cb \bar{p} } \Lambda^\hb{}_{\db \eb \bar{q} }      \\   
 &=    -4 (\psi_{\ab \bbr} m^{\ab \bbr})^2   + 4  m^{\ab \bbr} m^{\cb \db}\left[  \psi_{\ab \cb} \psi_{\bbr \db} +2 \psi_{\eb \cb} \Lambda^{\eb}{}_{\db \ab \bbr}  +  \Lambda^{\eb}{}_{\fb \ab \bbr}    \Lambda^{\fb}{}_{\eb \cb \db}  -    \Lambda^{\eb}{}_{\fb \ab \db}    \Lambda^{\fb}{}_{\eb \cb \bbr}   \right]
  .  \end{split}
\end{equation}

To proceed we shall simplify matters by assuming
\begin{equation}
\det g = 1 \ , \quad   \det m = 1 \ ,  \quad   \det V = 1 \ , 
\end{equation}
 and further that the trombone gauging vanishes.  Then we have the following identifications:
 \begin{equation}
 \chi_{\ab \bbr} = 0 \  , \quad \psi_{[\ab \bbr]} = 0 \ ,  \quad  ,  Y_{\ab \bbr}  =  \psi_{(\ab \bbr)}  \ , \quad   Z^{\bbr \cb , \ab}    = - \frac{1}{16}  \Lambda^{\ab}{}_{\db \eb \fb}\e^{\db \eb \fb \bbr \cb}   \ . \end{equation}

Using the invariance of the $\epsilon$-tensor we then find the following relations: 
\begin{equation}
\begin{split}
 64  Z^{\ab \bbr, \cb} Z^{\db \eb \fb} m_{\ab \db} m_{\bbr \eb} m_{\cb \fb} =& \Lambda^{\ab}{}_{\bbr \cb \db} \Lambda^{\eb}{}_{ \fb \gb \hb} m_{\ab \eb } m^{\bbr \fb} m^{\cb \gb} m^{\db \hb} - 2 \Lambda^{\ab}{}_{\bbr \cb \db} \Lambda^{\eb}_{\fb \gb \hb} m_{\ab \eb} m^{ \bbr \fb} m^{ \cb \hb } m^{ \db \gb}  \\
   64   Z^{\ab \bbr, \cb} Z^{\db \eb \fb}  m_{\ab \db} m_{\bbr \cb} m_{\eb \fb} =& \Lambda^{\ab}{}_{\bbr \cb \db} \Lambda^{\bbr}{}_{\ab \eb \fb} m^{ \cb \fb} m^{ \db \eb } - \frac{1}{2}\Lambda^{\ab}{}_{\bbr \cb \db} \Lambda^{\db}{}_{\eb \fb \ab} m^{\bbr \eb} m^{ \cb \fb}  \\
   &   - \Lambda^{\ab}{}_{\bbr \cb \db} \Lambda^{\bbr}_{\ab \eb \fb} m^{\cb \eb} m^{ \db \fb} - 2 \Lambda^{\ab}{}_{\bbr \cb \db} \Lambda^{\bbr}{}_{\eb \fb \ab } m^{ \cb \eb } m^{ \db \fb }   \\ 
   & +  \frac{1}{2} \Lambda^{\ab}{}_{\bbr \cb \db} \Lambda^{\eb}{}_{ \fb \gb \hb} m_{\ab \eb } m^{\bbr \fb} m^{\cb \gb} m^{\db \hb} -  \Lambda^{\ab}{}_{\bbr \cb \db} \Lambda^{\eb}_{\fb \gb \hb} m_{\ab \eb} m^{ \bbr \fb} m^{ \cb \hb } m^{ \db \gb}       \ .
\end{split}
\end{equation}

Putting things together we then find that 
\begin{equation}
\frac{1}{12} V_1 - \frac{1}{2} V_2 - \frac{1}{8} V_5  =- 32 V_{gauged}  + \Lambda^{\ab}{}_{\bbr \cb \db} \Lambda^{\bbr}{}_{\ab \eb \fb} \left( m^{\cb \eb } m^{\db \fb}- m^{\cb \db} m^{\eb \fb} \right) 
\end{equation}
 where $V_{gauged}$ is the known potential for the scalars in gauged supergravity given by \cite{Samtleben:2005bp}:
 \begin{equation}
 \label{Vgauged}
V_{gauged} = \frac{1}{64} \left( 2{m}^{\ab \bbr}  Y_{\bbr \cb} {m}^{\cb \db} Y_{\db \ab} -   ( {m}^{\ab \bbr} Y_{\ab \bbr})^2   \right) + Z^{\ab \bbr, \cb} Z^{\db \eb, \fb} \left( {m}_{\ab \db}  {m}_{\bbr \eb}  {m}_{\cb \fb} - {m}_{\ab \db}  {m}_{\bbr \cb}  {m}_{\eb \fb}     \right) \ . 
\end{equation}
 That is to say we have reproduced exactly the potential for the scalar fields expected for gauged supergravity up to the term 
\begin{equation}
\Lambda^{\ab}{}_{\bbr \cb \db} \Lambda^{\bbr}{}_{\ab \eb \fb} \left( m^{\cb \eb } m^{\db \fb}- m^{\cb \db} m^{\eb \fb} \right)  \  ,
\end{equation}
which, however, is a total derivative and after some algebra can be written as 
\begin{equation}
2 \pl_{kl} \left( m^{p k}  m^{\bar{q} \bar{l} } V_{\bar{q}}^{[q} \pl_{pq} V^{l] }_{\bar{l}} \right) \ . 
\end{equation}
It is worth remarking that the additional term in the Lagrangian $V_5$ was vital to achieve correct cancellations and contributions to this result.     
 
It is natural to ask whether the assumption that the trombone gauging vanishes is actually necessary; could one obtain an action principle for a trombone gauged supergravity?  From the above considerations it seem likely that an appropriate a scalar potential could be deduced. However, the trombone symmetry is only an on-shell symmetry of the full supergravity action and so to make such a conclusion it would be vital to include the other supergravity fields (i.e. the gauge and gravity sectors) in  a duality symmetric fashion.

Now we perform a variation of the action under a generalised diffeomorphism to find 
\begin{equation}
\delta V =  G_0 + \dots 
\end{equation} 
in which the ellipsis indicates total derivative terms and $G_0$ vanishes upon invoking the section condition. 

By substituting the Scherk--Schwarz ansatz to the action we obtain the action for the gauged supergravity \eqref{Vgauged}. This can be written in terms of $X_{mn,k}{}^l$ as follows
\begin{equation}
\begin{split}
V_{gauged}=&\fr{1}{64}\left(3X_{mn,r}{}^sX_{pq,s}{}^rm^{mp}m^{nq}-X_{mp,q}^{n}X_{nr,s}^{m}m^{pr}m^{qs}   \right)+\\
&+\fr{1}{96}\left(X_{mn,r}{}^sX_{pq,t}{}^um^{mp}m^{nq}m^{rt}m_{su} + X_{mp,q}{}^nX_{nr,s}{}^mm^{pq}m^{rs}    \right).
\end{split}
\end{equation}
Since we understand $X_{mn}$ as a generator of the algebra it does not transform under gauge variation while the transformation of the metric $m^{ab}$ reads
\begin{equation}
\d_\xi m^{ab} = (X_{kl,m}{}^am^{mb}+X_{kl,m}{}^bm^{am})\xi^{kl}.
\end{equation}
Thus we find the gauge transformation of the action to be
\begin{equation}
\delta_{\xi}V=\fr{1}{24}X_{a[b,c]}{}^dX_{dm,n}{}^{a}X_{kl,p}{}^bm^{cp}m^{mn}\xi^{kl}.
\end{equation}
The action of the gauged supergravity transforms as a scalar under generalised gauged transformation if $\q_{mn}=0$. Indeed, the expression above becomes
\begin{equation}
\begin{split}
\delta_{\xi}V=&-\fr{1}{12}\e_{abcpq}Z^{pq,d}\d^{a}_{[d}Y_{m]n}X_{kl,p}{}^bm^{cp}m^{mn}\xi^{kl}=\\
&=-\fr{1}{24}\left(\e_{abcpq}Z^{pq,a}Y_{mn}m^{mn} - \e_{abcpq}Z^{pq,d}Y_{dn}m^{an}\right)X_{kl,p}{}^bm^{cp}\xi^{kl}=0.
\end{split}
\end{equation}
The first term here is zero due to $Z^{[ab,c]}=0$ and the second is zero because of the quadratic constraint
\begin{equation}
Z^{mn,p}Y_{pq}=0.
\end{equation}

Hence, the generalised gauge transformation of the action is zero if one drops the trombone gauging. The trombone gauging does not leave the action invariant as it should do since it corresponds to the on--shell symmetry.

\subsection{$D=6$ supergravity}

The effective potential in six dimensions is given by \cite{Berman:2011jh}
\begin{equation}
\label{VSO}
\begin{split}
V_{eff}=&\fr{1}{16}M^{MN}\dt_M M^{KL}\dt_N M_{KL} -\fr12 M^{MN}\dt_N M^{KL} \dt_L M_{NK} +\\
  &+\fr{11}{1728} M^{MN} (M^{KL}\dt_MM_{KL})(M^{RS}\dt_NM_{RS}) +2 Y^{MN}_{KL}\dt_M E_\Q{}^K\dt_NE_\X{}^L\d^{\X\Q},
\end{split}
\end{equation}
where the extra term is included. Here the $16\times16$ matrix $M_{KL}$ is the generalised metric and it is written in terms of the metric $g_{\m\n}$ and the RR 3-form field $C_{\m\a\b}$
\begin{equation}
M=\fr{1}{\sqrt{g}}
\begin{pmatrix} g_{\mu \nu}+{\fr12} C_{\mu}{}^{\r\s} C_{\nu \r\s} +
{\fr{1}{16}}    X_{\mu}  X_{\nu}  &
{\fr{1}{\sqrt{2}}} C_{\mu}{}^{ \nu_1 \nu_2} + {\fr{1}{4 \sqrt
{2}}} X_{\mu} V^{\nu_1 \nu_2} & {\fr{1}{4}} {g}^{-{1 / 2}} X_{\mu} \\
{\fr{1}{\sqrt{2}}} C^{\mu_1 \mu_2}{}_{\nu} + {\fr{1}{4 \sqrt
{2}}}  V^{\mu_1 \mu_2} X_{\nu}  & g^{\mu_1 \mu_2,\nu_1 \nu_2}+ {\fr{1}
{2}} V^{\mu_1 \mu_2}V^{\nu_1 \nu_2} & {\fr{1}{\sqrt{2}}}  g^{-
{1 / 2}} V^{\mu_1 \mu_2} \\
{\fr{1}{4}} g^{-{1 / 2}} X_{\nu} & {\fr{1}{\sqrt{2}}} g^{-{1 /
2}} V^{\nu_1 \nu_2} & g^{-1}
\end{pmatrix},
\end{equation}
where the small Greek letters here run from 1 to 5 labelling 5 compact directions and
\begin{equation}
V^{\r\s}={\fr{1}{3!}} \epsilon^{\r\s\m\n\w}C_{\m\n\w}, \qquad \qquad X_{\m}=C_{\m\n\r} V^{\n\r}.
\end{equation}

The matrix $E_\X{}^K$ is the vielbein for $M^{MN}=E_\Q{}^ME_\X{}^N\d^{\Q\X}$ and the capital Greek indices run from 1 to 16 labelling flat spinorial indices.

For the convenience of notations we define the object
\begin{equation}
\label{f}
f_{\br{A}{B}}{}^{\bar{C}}=W_C{}^{\bar{C}}\dt_{\bar{A}}W^{C}{}_{\bar{B}},
\end{equation}
where $W^{C}{}_{\bar{B}}$ is the twist matrix introduced in \eqref{twist}. Then using the definition \eqref{Xgen} the structure constant can be written as
\begin{equation}
\label{X2f}
X_{MN}{}^K=f_{MN}{}^K-f_{NM}{}^K+Y^{KL}_{BN}f_{LM}{}^B,
\end{equation}
that us true for the extended geometry formalism in any dimension. 

From now on we assume that the trombone gauging vanishes and that the matrix $M_{MN}$ is unimodular. The latter can be always arranged by rescaling the generalised metric by $g=\mbox{det}(g_{\m\n})$. The only effect this has on the potential is change in the coefficients of the terms proportional to derivatives of the determinant. Summarising we have
\begin{equation}
\begin{split}
\q_M&=0, \quad\mbox{det} W=1,\\
f_{AB}{}^A&=0, \quad f_{A B}{}^B=0,\\
\dt_C W^C{}_{\bar{B}}&=0.
\end{split}
\end{equation}
In cases when it does not confuse the reader the bar notation is dropped to make expression less heavy. In all expressions which include terms with both barred and unbarred indices these are treated carefully. One should remember that such quantities like $X_{MN}{}^K$, $f_{MN}{}^K$ or gaugings always have flat barred indices and not be confused if they appear without bar. Taking this into account, the effective potential is given by
\begin{equation}
V_{eff}=V_1 + V_2 + V_3 + SC,
\end{equation}
where
\begin{equation}
\begin{split}
V_1 &= -\fr18M^{MN}f_{NP}{}^Lf_{ML}{}^P+M^{MN}f_{MP}{}^Lf_{LN}{}^P,\\
V_2 &= \fr12M^{MN}f_{PM}{}^Lf_{LN}{}^P,\\
V_3 &= M^{MN}M^{KL}M_{RS}\left(\fr18 f_{MK}{}^Rf_{NL}{}^S-\fr12f_{KM}{}^Rf_{NL}{}^S\right),\\
SC  &= \fr12 Y^{MN}_{KL}f_{MR}{}^Kf_{NS}{}^L M^{RS}.
\end{split}
\end{equation}
By integrating $\dt_P$ and $\dt_L$ by part in $V_2$ it can be shown that this term is zero up to a full derivative.
To proceed further and to be able to use gamma matrices algebra we need to define objects in the vector representation
\begin{equation}
\label{Gm}
\begin{split}
f_A{}_j{}^i&=\fr18(\G_j{}^i)_K{}^Lf_{AK}{}^L,\\
m_{\br{i}{j}}\G^{\bar{j}\br{A}{B}}&=\G_{\bar{i}\br{R}{S}}M^{\br{R}{A}}M^{\br{S}{B}},\\
X_M{}_i{}^j&=\fr18(\G_i{}^j)_K{}^LX_{ML}{}^K
\end{split}
\end{equation}
By making use of these definitions the part $V_3$ can be written as
\begin{equation}
\label{V3}
\begin{split}
V_3=&\fr14(\G_b\G^n)^N{}_Lf_K{}_i{}^jf_N{}_m{}^bm^{im}m_{jn}M^{KL}=\\
&\fr{1}{16}X_M{}_i{}^jX_N{}_k{}^lM^{MN}m^{ik}m_{jl}=\fr{1}{32} X_{MR}{}^KX_{NS}{}^LM^{MN}M^{RS}M_{KL}.
\end{split}
\end{equation}
Indeed, the first two lines of \eqref{Gm} imply that
\begin{equation}
\begin{aligned}
&f_{MK}{}^Rf_{NL}{}^SM_{RS}M^{MN}M^{KL}=2f_M{}_i{}^jf_N{}_m{}^nM^{MN}m_{nj}m^{mi},\\
&f_{KM}{}^Rf_{NL}{}^SM_{RS}M^{MN}M^{KL}=\fr12(\G^n{}_b)^N{}_Lf_K{}_i{}^jf_N{}_m{}^bm^{im}m_{jn}M^{KL}.
\end{aligned}
\end{equation}
These two equalities lead to the first line in \eqref{V3}. The definitions \eqref{f} and \eqref{Xgen} together with the condition $\q_M=0$ allow to write the structure constants $X_{MN}{}^K$ in terms of $f_{AB}{}^L$
\begin{equation}
\begin{aligned}
4Z^{iC}&=&&\G^{iAB}f_{AB}{}^C,\\
X_{MN}{}^K&=&&\fr14\G^{iAB}\G^j{}_{LM}(\G_{ij})_N{}^Kf_{AB}{}^L.
\end{aligned}
\end{equation}
Note, that this relation can not be inverted i.e. it is impossible to write $f_{AB}{}^C$ in terms of $X_{MN}{}^K$ and just substitute it into the potential. Basically, this follows from the first line of the equation above, that 
includes only symmetric part. Finally, substituting the last line of the equation \eqref{Gm} into the second line of \eqref{V3} and using the identities above one exactly recovers the first line in \eqref{V3}.

To obtain the term $V_1+SC$ one may use the following relations
\begin{equation}
\label{YXSO}
\begin{aligned}
Y_{SM}^{RL}X_{KR}{}^S&=-3X_{KM}{}^L,\\
Y_{SM}^{RL}f_{KR}{}^S&=-3f_{KM}{}^L,
\end{aligned}
\end{equation}
that follow from the explicit form of the structure constant \eqref{XSO}, relation between $X_{MN}{}^K$ and $f_{MN}{}^K$ \eqref{X2f}, identities \eqref{rel} involving the invariant tensor $Y_{MN}^{KL}$ and the condition $\q_M=0$. Then the term $V_1+SC$ of the effective potential can be written as
\begin{equation}
V_1+SC=-\fr{1}{8}X_{MK}{}^LX_{NL}{}^KM^{MN}.
\end{equation}
Indeed, substituting \eqref{X2f} into the expression above one encounters exactly $V_1+SC$ plus a term, proportional to $V_2$, that is a full derivative. 

Finally, the effective potential can be recast in the following form
\begin{equation}
\label{VSOt}
V_{eff}=-\fr{1}{8}X_{MK}{}^LX_{NL}{}^KM^{MN}+\fr{1}{32}X_{MR}{}^KX_{NS}{}^LM^{MN}M^{RS}M_{KL}.
\end{equation}
This expression reproduces exactly the scalar potential for maximal gauged supergravity in $D=6$ dimensions up to a prefactor
\begin{equation}
V_{eff}=6\Tr\left[T^{\hat{a}}\tilde{T}^{\hat{a}}-\fr12T\tilde{T}\right]=6V_{scalar}.
\end{equation}
The details of this calculation are provided in Appendix B.1.

The effective potential \eqref{VSOt} is invariant under transformations \eqref{G_transf} because of the quadratic constraint \eqref{closure2} (see Appendix B).

\subsection{$D=5$ supergravity}

The low energy effective potential for the $E_{6(6)}$ invariant M-theory has the same form as in the $SO(5,5)$ case up to coefficients \cite{Berman:2011jh}
\begin{equation}
\label{VE}
\begin{split}
V_{eff}=&\fr{1}{24}M^{MN}\dt_M M^{KL}\dt_N M_{KL} -\fr12 M^{MN}\dt_N M^{KL} \dt_L M_{NK} +\\
  &+\fr{19}{9720} M^{MN} (M^{KL}\dt_MM_{KL})(M^{RS}\dt_NM_{RS}) -\fr12Y^{MN}_{KL}\dt_M E_\Q{}^K\dt_NE_\X{}^L\d^{\X\Q}.
\end{split}
\end{equation}
We again add the term proportional to the section condition that includes the vielbein
\begin{equation}
E_{\Q}{}^{M} =  (\textup{det}e)^{-1/2}
\begin{pmatrix}
e_{\mu}{}^{i} & - {\fr{1}{\sqrt{2}}} e_{\mu}{}^{j} C_{j i_1 i_2} &
{\fr{1}{2}} e_{\mu}{}^{i_3} U + {\fr{1}{4}} e_{\nu}{}^{i_3} C_
{\mu j k} V^{\nu j k} \\
0 & e^{\mu_1}{}_{[i_1} e^{\mu_{2}}{}_{i_2]} & -{\fr{1}{\sqrt{2}}} e^
{\mu_1}{}_{j_1} e^{\mu_{2}}{}_{j_2} V^{j_1 j_2 i_3} \\
0 & 0 & (\textup{det}e)^{-1} e_{\mu_3}^{i_3}
\end{pmatrix},
\end{equation}
where the capital Greek letters now run from 1 to 27, the small Latin and Greek indices run from 1 to 6 labelling  curved and flat space respectively. The fields $U$ and $V^{ijk}$ are defined as
\begin{equation}
\begin{split}
U&=\fr{1}{6}\e^{ijklmn}C_{ijklmn},\\
V^{ikl}&=\fr{1}{3!}\e^{iklmnj}C_{mnj}.
\end{split}
\end{equation}
Here the 6-form field $C_{ijklmn}$ is a new field that was not present in the previous example because the dimension was lower than 6.

Using the same notations for $f_{MN}{}^K$ as in the previous subsection and setting $\det M=1$ and $\q_M=0$ we have for the twisted effective potential
\begin{equation}
V_{eff}=V_1 + V_2 + V_3 + SC,
\end{equation}
with
\begin{equation}
\begin{split}
V_1 &= -\fr{1}{12}M^{MN}f_{NP}{}^Lf_{ML}{}^P+M^{MN}f_{MP}{}^Lf_{LN}{}^P,\\
V_2 &= \fr12M^{MN}f_{PM}{}^Lf_{LN}{}^P,\\
V_3 &= M^{MN}M^{KL}M_{RS}\left(\fr{1}{12} f_{MK}{}^Rf_{NL}{}^S-\fr12f_{KM}{}^Rf_{NL}{}^S\right),\\
SC  &= \fr12Y^{MN}_{KL}f_{MR}{}^Kf_{NS}{}^L M^{RS}.
\end{split}
\end{equation}
Again the part $V_2$ is the full derivative and can be dropped.

It is straightforward to check the following identities
\begin{equation}
\label{YXE}
\begin{split}
Y^{RL}_{SM}X_{KR}{}^S&=-5X_{KM}{}^L,\\
Y^{RL}_{SM}f_{KR}{}^S&=-5f_{KM}{}^L,\\
Y^{KA}_{BL}X_{AN}{}^B&=X_{LN}{}^K+4X_{NL}{}^K,
\end{split}
\end{equation} 
that can be derived exactly in the same fashion as \eqref{YXSO}. The analogue of the second line of \eqref{Gm} is
\begin{equation}
\label{Md}
M_{\br{M}{N}}d^{\bar{N}\br{K}{L}}=d_{\bar{M}\br{N}{R}}M^{\br{N}{K}}M^{\br{R}{L}}
\end{equation}
and implies that the indices of the invariant tensor are raised and lowered by the generalised metric. This is in agreement with the definition of the unimodular matrix
\begin{equation}
M_{MN}=\mc{V}_M{}^{ij}\mc{V}_N{}^{kl}\W_{ik}\W_{jl}
\end{equation}
and the following representation of the invariant tensor \cite{Bergshoeff:2007ef}
\begin{equation}
d_{MNK}=\mc{V}_M{}^{ij}\mc{V}_N{}^{kl}\mc{V}_K{}^{mn}\W_{jk}\W_{lm}\W_{ni}
\end{equation}
if one takes into account the condition $\mc{V}_M{}^{ij}\W_{ij}=0$.

Using the identities \eqref{YXE}, the definition \eqref{Md} and the last line of \eqref{rel} we deduce for the effective potential
\begin{equation}
\label{VEt}
\begin{split}
V_{eff}=&-\fr{1}{12}X_{MK}{}^LX_{NL}{}^KM^{MN}+\fr{1}{12}X_{MR}{}^KX_{NS}{}^LM^{MN}M^{RS}M_{KL}+\\
&+\fr{1}{10}X_{RM}{}^KX_{NS}{}^LM^{MN}M^{RS}M_{KL}.
\end{split}
\end{equation}
The first term can be verified using the same technique as in the previous section. Namely, substituting the structure constant $X_{MN}{}^K$ from \eqref{X2f} and taking into account the identities \eqref{YXE} one obtains that the first term in the equation above is $V_1+SC$ plus a full derivative term.

The derivation of the second and the third term is longer but straightforward. Lets sketch the idea here on the example of the second term $X_{MR}{}^KX_{NS}{}^LM^{MN}M^{RS}M_{KL}$. Substituting here the expression \eqref{X2f} and expanding the brackets one obtains terms of the types
\begin{equation}
\label{termsE}
\begin{aligned}
& f_{MR}{}^Kf_{NS}{}^LM^{MN}M^{RS}M_{KL}, & & f_{RM}{}^Kf_{NS}{}^LM^{MN}M^{RS}M_{KL} \\
& f_{MR}{}^KY^{LA}_{BN}f_{AS}{}^BM^{MN}M^{RS}M_{KL}, & & f_{RM}{}^KY^{LA}_{BN}f_{AS}{}^BM^{MN}M^{RS}M_{KL},\\
& Y^{KQ}_{MP}f_{QR}{}^PY^{LA}_{BN}f_{AS}{}^BM^{MN}M^{RS}M_{KL}.
\end{aligned}
\end{equation}
Lets show that the third term in the second line is exactly proportional to the second term in the first line. Substituting the invariant tensor $Y^{MN}_{KL}=10d^{MNP}d_{PKL}$ as using the relation \eqref{Md} two times one can verify the following identities
\begin{equation}
\begin{aligned}
 &f_{RM}{}^KY^{LA}_{BN}f_{AS}{}^BM^{MN}M^{RS}M_{KL} = 10  f_{RM}{}^Kd^{CLA}d_{CBN}f_{AS}{}^BM^{MN}M^{RS}M_{KL}=\\
 &10  f_{RM}{}^Kd_{CBN}f_{AS}{}^B d_{KPQ}M^{CP}M^{AQ}M^{MN}M^{RS}=\\
 &10  f_{RM}{}^Kf_{AS}{}^B d^{PMJ}M_{JB}d_{KPQ}M^{AQ}M^{RS}=f_{RM}{}^Kf_{AS}{}^B Y^{MJ}_{KQ}M_{AB}M^{AQ}M^{RS}=\\
 &-5f_{RQ}{}^Jf_{AS}{}^BM_{JB}M^{AQ}M^{RS},
\end{aligned}
\end{equation}
where the identity \eqref{YXE} was used in the last line. Using the same idea one simplifies the last line in \eqref{termsE}. Finally, the contributions like the first term in the second line of \eqref{termsE} coming from two last terms in \eqref{VEt} precisely cancel each other.

After long algebraic calculations it can be derived that the expression \eqref{VEt} is up to a prefactor equal to the scalar potential of maximal gauged supergravity in 5 dimensions
\begin{equation}
V_{eff}=\fr92|A_1^{ij}|^2-\fr12|A_{2}^{i,jkl}|^2=\fr32V_{scalar},
\end{equation}
where the $|\;|^2$ stands for the contraction of all indices. To show this one expresses the potential in terms of the $T$-tensor by making use the relation \eqref{XT}. Finally, rewriting the $T$-tensor in terms of the $A$-tensor as \eqref{TA} and using the properties of the $A$-tensors one obtains the scalar potential of the maximal gauged supergravity.\footnote{These results were verified with the help of the computer algebra system Cadabra \cite{Cadabra, Peeters:2007wn}} We refer the reader to Appendix B for the proof that the potential \eqref{VEt} is invariant under gauge transformations \eqref{G_transf}.

\section{Summary}

The above results show that the idea of Scherk-Schwarz reduction works in detail for $D=5,6$ and 7. The most interesting point to mention here is that geometry of the extended space plays an important role in the picture presented above. It is not just a Kaluza-Klein reduction where fields does not depend on internal coordinates. The extended space should be an extended geometry analogue of a  parallelisable space so the dependence on the dual coordinates should be of a particular form. These constraints match the quadratic constraint on the embedding tensor of gauged supergravity. Although there are many papers \cite{Coimbra:2011ky, Hohm:2011si, Hohm:2012gk} considering geometry of the extended space it is not fully understood how to describe this object. In this work we investigate a very particular situation but we hope it may contribute to the full picture of the extended geometry.

\chapter{Boundary terms in extended geometry}
\label{bound}

\section{Gibbons--Hawking formalism}

The example of General Relativity where one first encounters a non-trivial contribution from boundary terms, teaches us that the action is not just a simple way of writing the equations of motion. In gravity the famous Gibbons-Hawking term that results from ambiguity in variations of the canonical fields in the action, encodes the thermodynamics of solutions.  As an introductory example the Gibbons-Hawking term of General Relativity that allows us to describe thermodynamics of black holes is taken. An important feature of General Relativity that differentiates it from say a vector field theory is that the boundary term cannot be written in terms of the canonical variables of the theory (bulk metric). Instead it is defined in terms of geometric properties of the boundary, i.e. the extrinsic curvature, which is related to the black hole entropy.

In this chapter we show that a similar situation takes place for the case of duality invariant formulations of string theory. Firstly, the corresponding boundary term can be written in terms of a normal to the boundary and the induced metric. Interestingly, by making use of the semi-covariant derivative this expression can be written exactly in the form of the (generalised) extrinsic curvature.

In the path integral approach to quantized fields one expresses the amplitude to go from the field with configuration
$\f_1$ at time $t_1$ to $\f_2$ at $t_2$ as
\begin{equation}
 \label{3.1}
 \langle\f_2,t_2|\f_1,t_1\rangle=\int_{\f(t_1)=\f_1}^{\f(t_2)=\f_2} \mc{D}\f e^{iI[\f]},
\end{equation}
where the integral is over all field configurations that take the values $\f_1$ at time $t_1$ and $\f_2$ at time $t_2$. On the other hand the same
quantity can be written in the following way using the Hamiltonian (operator of evolution) 
\begin{equation}
 \label{3.2}
 \langle\f_2,t_2|\f_1,t_1\rangle=\langle\f_2|e^{-iH(t_2-t_1)}|\f_1\rangle.
\end{equation}
After rotation to imaginary time $t_2-t_1=-i\b$ and taking the trace (sum over all $\f=\f_1=\f_2$) one obtains:
\begin{equation}
\label{3.3}
 \Tr\exp(-\b H)=\int\mc{D}\f e^{I[\f]},
\end{equation}
where the path integral is now taken over all fields that are periodic with period $\b$ in imaginary time. In a sense this integral describes
quantum field system in a space with one compact dimension. 

An important observation is that the left-hand side of \eqref{3.3} is just the partition function $Z$ for the canonical ensemble consisting of the
fields $\f$ at temperature $T=\b^{-1}$.  Thus, one can describe thermodynamics of field-theoretical systems and define such quantities as entropy and
free energy.

The object that connects classical and quantum gravity is the black hole. On the one hand it is a macroscopic object since it appears as a solution of the GR equations. On the other hand a black hole produces quantum effects, e.g. the Hawking radiation. According to Bekenstein and
Hawking, this object has entropy  proportional to the area of the black hole \cite{Hawking:1995fd}:
\begin{equation}
 S=\fr14A.
\end{equation}
This entropy was introduced to explain the phenomenon of Hawking radiation that is basically a flux of particles
emitted by a black hole. It is a pure quantum effect and a black hole evaporates during this process. The spectrum of the
radiated particles is described by the black body spectrum. In the Hawking description of this process  one assumes that the mass of the black hole slowly changes adiabatically
slowly i.e. there is no back-reaction. This means that at every moment of time the radiation and the black hole are (nearly) in thermodynamic
equilibrium and that the black hole should have well defined \emph{temperature}.

Thermodynamics defines temperature as a measure of how energy changes with the number of microstates corresponding to the given
macrostate. Applying this definition to black hole radiation, one encounters a paradox since according to the no-hair theorem any black hole has only one microstate. This implies zero entropy in contradiction to the Bekenstein formula.

String theory as a quantum theory of gravity has suggested a few ways of resolving such paradoxes of black holes. For example the contradiction stated above coul be resolved by the proposition made by Strominger and Vafa in \cite{Strominger:1996sh}. They have shown that one may count microscopic configurations of strings and branes to obtain the Bekenstein entropy. Another interesting approach that could explain black hole entropy is Loop Quantum Gravity. In this approach one assumes that the space-time is fundamentally triangulated, i.e. consists of the simplest objects, triangles. In certain sense this solves the information paradox by identifying one macroscopic configuration of the space-time with many different triangulations. These correspond to microstates in the thermodynamical state \cite{Rovelli:1996dv, Meissner:2004ju}. Finally, to resolve the information loss paradox in the framework of AdS/CFT correspondence one suggests that all information about an initial state of a collapsing object is returned by the Hawking radiation \cite{Lowe:1999pk}. For a review of the black hole thermodynamics see \cite{Bena:2007kg}.

Going back to General Relativity one could calculate the black hole temperature using \eqref{3.3}. However,
substituting the black hole solution to \eqref{3.3} one immediately acquires a problem: the usual
Einstein-Hilbert action for gravity
\begin{equation}
 I_{EH}=\int \sqrt{-g} R
\end{equation}
equals zero for empty-space solutions (including black hole). This is a reflection of the gauge nature of General Relativity and the same behaviour is observed in YM theories.

To resolve this difficulty one recalls that the action of general relativity contains second derivatives of the metric $g_{\m\n}$, that require us to set not only $\d g_{\m\n}=0$ on the boundary but also $\dt_\a\d g_{\m\n}=0$. To fit the extra conditions to the conventional Euler-Lagrange procedure a boundary term has to be added to the action.

It is useful to illustrate how this works in the simplest case of classical mechanics \cite{Winitzki}. Consider a particle moving in one dimensional space, parametrised by a coordinate $q$, whose action is given by
\begin{equation}
 \label{3.4}
I=-\int{}dtq\ddot{q}.
\end{equation}
Variation of the action leads to the following expression
\begin{equation}
 \label{3.5}
\d{}I=-\int{}dt\left(\ddot{q}\d{q}+q\d\ddot{q}\right).
\end{equation}
Usually at this step the second term is integrated by part two times and the boundary terms are neglected. However, the boundary condition $\d\dot q=0$
does not follow from $\d q=0$. More accurately the variation is written as:
\begin{equation}
\begin{aligned}
 \label{3.6}
  \d{I}&=-\int{dt\left[\ddot{q}\d{q}+\fr{d}{dt}\left(q\d\dot{q}\right)-\dot{q}\d\dot{q}\right]}\\
  &=-\int{dt\left[2\ddot{q}\d{q}+\fr{d}{dt}\left(q\d\dot{q}\right)-\fr{d}{dt}(\dot{q}\d{q})\right]}.
\end{aligned}
\end{equation}
To  end up with the ordinary equation of motion $\ddot{q}=0$, the second term should be somehow excluded. It can be compensated by introducing the following boundary term
\begin{equation}
 \label{3.7}
I_{tot}=-\int{dtq\ddot{q}}+\int{dt\fr{d}{dt}(q\dot{q})}.
\end{equation}
It is easy to check that now the variation of this action has the very familiar form:
\begin{equation}
 \label{3.8}
  \d{I}_{tot}=-2\int{dt\ddot{q}\d{q}}+2(\dot{q}\d{q})\left|^{t_2}_{t_1}\right.,
\end{equation}
which after fixing $\d{q}=0$ on the boundary gives $\ddot{q}=0$.

An important feature of theories of this kind is that the action \eqref{3.7} can be written in canonical variables 
\begin{equation}
 \label{3.9}
 I_{tot}=2\int{dt\dot{q}\dot{q}}
\end{equation}
with no second derivatives. On the contrary in the case of the Einstein-Hilbert action the Gibbons-Hawking boundary term cannot be written in terms of canonical variables in a covariant way. Instead it is written in terms of surface properties (first and
second fundamental form). And the total action for GR cannot be written nicely in the same way as \eqref{3.9}.

 Let us show that the full action for GR has the following form
\begin{equation}
\label{3.10}
 I[g]=I_{EH}[g]+I_{B}[g],
\end{equation}
where 
\begin{equation}
 \begin{split}
    I_{EH}=\int_{\mc M} &d^4x\sqrt g R;\\
    I_{B}=2 \oint_{\dt\mc M}&d^3 y \sqrt h K.
 \end{split}
\end{equation}
Here $h$ is the metric on the boundary $\dt\mc M$, $K$ is the second quadratic form (extrinsic curvature) of the boundary. Indeed, variation of the Einstein-Hilbert action gives (for details see \cite{Poisson:931451})
\begin{equation}
 \d I_{EH}=\int_\mc M \sqrt{-g} d^4x\left[R_{\m\n}-\fr12R g_{\m\n}\right]+\oint_{\dt \mc M}\sqrt{-h} d^3y \left(\bar\d v^\m n_\m\right),
\end{equation}
where $\bar\d v^\m$ is defined as:
\begin{equation}
 \bar\d v^\m=g^{\a\b}\d\G^\m_{\a\b}-g^{\a\m}\d\G^\b_{\a\b}
\end{equation}
and
\begin{equation}
 g^{\m\n}\d R_{\m\n}=\bar\d v^\m_{\ph{\m};\m}.
\end{equation}
The ``bar'' notation was introduced to emphasize that $\bar\d v^\m$ is not a variation of some quantity $v^\m$.
Finally, the vector $n^\m$ is the unit normal to the boundary $\dt \mc M$.

Varying the boundary term we obtain
\begin{equation}
 \d I_B=\oint_{\dt\mc M}h^{\a\b}\d g_{\a\b,\m}n^\m\sqrt{-h} d^3y.
\end{equation}
Finally, taking into account that  the following equality is true on the boundary
\begin{equation}
 n^\m\bar\d v_\m = -h^{\a\b}\d g_{\a\b,\m}n^\m
\end{equation}
we end up with the familiar Einstein equations:
\begin{equation}
 R_{\m\n}-\fr12 R g_{\m\n}=T_{\m\n}.
\end{equation}

Hence, we have shown that in general one has to add the Gibbons-Hawking boundary term  to the Einstein-Hilbert action for General Relativity. This term allows us to derive equations of motion for the metric $g_{\m\n}$ consistently by the conventional Euler-Lagrange procedure. An important implication of the boundary term is that the total action \eqref{3.10} does not vanish for empty-space solutions like the black hole. Instead, it implies that termodynamics of black holes is governed totally by the boundary term \cite{Hawking:1995fd}.

\section{Black holes thermodynamics}

Black holes are solutions of Einstein equations in the absence of matter $T_{\m\n}=0$. These objects are point-like and have a horizon, that is a boundary in the space-time of external observer that does not allow anything to get out from the black hole. The simplest example of such a solution is the Schwarzschild black hole described by the following metric
\begin{equation}
 \label{4.1.1}
ds^2=-\left(1-\fr{2M}{r}\right)dt^2+\fr{1}{1-\fr{2M}r}dr^2+r^2d\W^2.
\end{equation}
This solution is stationary and spherically symmetric and describes a space-time with a source at the point $r=0$. For an external observer the horizon is given by the surface $r=2M$.

According to the no-hair theorem, a black hole could only be characterized by three parameters: mass, angular momentum and charge. These correspond
to the Schwarzschild, Kerr (rotating) and Reissner--Nordstr\"om (charged) black holes \cite{Misner:1974qy}. The most general stationary metric for
rotating charged black hole is the Kerr-Newmann solution.

To follow the analogy with M-theory consider the Reissner--Nordstr\"om solution characterized by a charge $Q$
\begin{equation}
 \label{4.1.2}
ds^2=-\left(1-\fr{2M}{r}+\fr{Q^2}{r^2}\right)dt^2+\left(1-\fr{2M}{r}+\fr{Q^2}{r^2}\right)^{-1}dr^2+r^2d\W^2.
\end{equation}

In this case the action for the electromagnetic field should be added:
\begin{equation}
 I[A]=-\int \sqrt {-g} d^4xF_{\m\n}F^{\m\n}.
\end{equation}
For a solution of the Maxwell equations, $\dt_\m F^{\m\n}=0$ the integrand can be written as a divergence:
\begin{equation}
 F^2=(2F^{\m\n}A_{\m})_{;\n}
\end{equation}
and the action takes the same form as the Gibbons--Hawking term i.e. integral over the boundary
\begin{equation}
 -2\oint F^{\m\n}A_\m d\S_\n.
\end{equation}
 From the previous section we known that on empty space solutions the total action  becomes just a boundary term. Thus the
combined gravitational and electromagnetic actions can be written as \cite{Gibbons:1976ue}
\begin{equation}
 \label{3.11}
 \begin{aligned}
I&=I[g]+I[A]=2\oint K d\S-\int \sqrt{-g}d^4xF_{\m\n}F^{\m\n}\\
&=i16\p^2 k^{-1}(M-Q\F),
\end{aligned}
\end{equation}
where the gauge transformed electromagnetic vector potential is taken to be $A_a=(Qr^{-1}-\F)t_{;a}$ and $\F=Q r_g^{-1}$ is the scalar potential on the horizon.
The integral in \eqref{3.11} is taken over the surface near the horizon and $\k=(4M)^{-1}$ is the surface gravity of the black hole solution. The imaginary unit $i$ comes from the factor $\sqrt{-g}$ in the surface measure.

Now returning to \eqref{3.3} we can study the thermodynamics of our black hole. At first, let us mention the fact that the dominant contribution
to the
path integral comes from such configurations of the metric $g$ and the matter field $\ff$ which are close to classical solutions (background) $g_0$ and $\ff_0$. Expanding the action in  Taylor
series near the background $g_0$ and $\ff_0$ one obtains for the partition function
\begin{equation}
 \label{3.12}
\log Z = i I[g_0,\ff_0]+\log \int\mc D\tilde g \exp\left(iI_2[\tilde g]\lefteqn{\ph{\tilde\ff}}\right) +\log \int \mc D\tilde\ff\exp\left(iI_2[\tilde
\ff]\right),
\end{equation}
where $I_2[\tilde g]$ and $I_2[\tilde \ff]$ are quadratic in the fluctuations $\tilde g$ and $\tilde \ff$.

Thermodynamics teaches us that for the canonical ensemble $\log Z=-WT^{-1}$, where $W$ is the thermodynamic potential of a system. Thus one can
identify $-iI[g_0,\ff_0]T$ with background contribution to thermodynamic potential and the other terms with contributions of thermal gravitons and
matter quanta.

Thus from \eqref{3.11} it follows that $W=\fr12(M-\F Q)$ and the temperature $T=\k(2\p)^{-1}$. From the fact that $W=M-TS-\F Q$ one obtains that 
\begin{equation}
 \fr12M=TS+\fr12Q\F.
\end{equation}
And finally using the generalised Smarr formula $M=2\k A+Q\F$ we obtain:
\begin{equation}
 S=\fr14A.
\end{equation}
This famous Hawking formula introduces the notion of entropy and temperature for a black hole. This allows us to speak about \emph{black hole thermodynamics}.

These formulae connect gravity with thermodynamics in some strange way using \eqref{3.3}. But they give the correct answer that is used in construction of
``true'' black hole thermodynamics with counting states inside a black hole. Thus it can shed some light on the quantum gravity and string theory.

\section{Duality invariant topological terms}

\subsection{Double Field Theory}

In Double Field Theory which provides a duality covariant description of string theory backgrounds and was introduced in Section \ref{GG_metric} one encounters a constraint \eqref{closure} 
\begin{equation}
\h^{MN}\dt_M\bullet \dt_N\bullet =0,
\end{equation}
 that effectively restricts the extended space to some subspace if satisfied. Here the $2d\times 2d$ constant matrix $\h_{MN}$ is the flat $O(d,d)$ invariant metric. This closure condition can be solved by imposing two natural constraints on all fields defined on the extended space
\begin{equation}
\dt_{\tilde{x}} {} {} =0 \, \quad  {\rm{or \, \,  its \, \, dual}} \quad \dt_x {} {}= 0 \ ,
\end{equation}
that correspond to particular T-duality frames. Taking the first choice so that all fields are taken to be independent of the winding coordinates $\tilde{x}_i$ the duality covariant action \eqref{Odd_action} 
\begin{equation}
\label{Odd_action1}
\begin{aligned}
    S=\int dx d\tilde x e^{-2d}&\left(\fr18\mc{H}^{MN}\dt_M\mc{H}^{KL}\dt_{N}\mc{H}_{KL} -\fr12
\mc{H}^{KL}\dt_L\mc{H}^{MN}\dt_N\mc{H}_{KM} - \right.\\
      & \left.- 2\dt_M d\dt_N\mc{H}^{MN}+4\mc{H}^{MN}\dt_Md\dt_Nd\lefteqn{\ph{\fr12}}\right) \, .
      \end{aligned}
\end{equation}
reduces to the bosonic part of the supergravity action \eqref{S0} 
\begin{equation}
\label{S01}
\begin{aligned}
&&S=& \int d^dx\  e^{-2d}\left[-\fr14 g^{ij}\dt_i g^{kl}\dt_j g_{kl}+\fr12 g^{ij}\dt_i g^{kl}\dt_j g_{kl} +2\dt_i d\dt_jg^{ij} +4g^{ij}\dt_i d\dt_j d -\fr{1}{12}H^2 \right]\\
&&=&\int d^dx\  e^{-2\ff}\sqrt{-g}\left(R[g]+4(\dt\ff)^2-\fr{1}{12}H^2\right)+\mbox{boundary terms}.
\end{aligned}
\end{equation}
The boundary terms are usually dropped in this formalism, however imposing the strong constraint but keeping the boundary terms we can write the following
\begin{equation}
\label{2}
\begin{aligned}
 S&=\int d^dx\ \sqrt{g}e^{-2\ff}\left(R[g]+4(\dt\ff)^2-\fr1{12}H^2\right)-\\
 &-\int\dt_m\left[e^{-2\ff}\sqrt{g}g^{nb}g^{mc}\dt_ng_{bc}- e^{-2\ff
}\sqrt{g} g^ { mc } g^ { nb}\dt_cg_{nb} \right].
\end{aligned}
\end{equation}
It is then natural to combine the total derivative term in the above with the Gibbons-Hawking term (modified by
the dilaton). In the previous section it was shown that the Gibbons-Hawking boundary contribution can be  written in terms of the surface curvature \cite{Gibbons:1976ue}
\begin{equation}
\label{3}
\begin{aligned}
 S_{GH}&=2\oint\sqrt{h}e^{-2\ff}K=2\oint\sqrt{h}e^{-2\ff}h^{ab}\left(\dt_an_b-\G_{ab}^mn_m\right)\\
	&=2\oint\sqrt{h}e^{-2\ff}h^{ab}\dt_an_b-\oint\sqrt{h}e^{-2\ff}h^{ab}h^{mn}(2\dt_a h_{nb} - \dt_nh_{ab})n_m
\end{aligned}
\end{equation}
where $K=\nabla_i n^i$ is the second fundamental form for the boundary, $n_a$ and $h_{ab}$ are the normal and metric on the boundary respectively.

Comparing \eqref{2} and \eqref{3} (and with the replacement of $g$ by $h$) one obtains:
\begin{equation}
\label{4}
 \int\sqrt{g}e^{-2\ff}\left(R[g]+4(\dt\ff)^2-\fr1{12}H^2\right)+S_{GH}=S+\oint\sqrt{h}e^{-2\ff}(2h^{ab}n_{a,b}-n^ch^{ab}\dt_b
h_{ac}).
\end{equation}
It is well known that in gravity (in contrast to other field theories) it is impossible to write the action with the GH term in a
covariant form (without introducing a boundary). In other words, the additional term cannot be written as just a full derivative.

We now wish to write the boundary term on the right hand side of \eqref{4} in an $O(d,d)$ covariant form by recasting it in terms of the generalised metric.
This produces
\begin{equation}
 \label{2.4}
S_{tot}=S+\oint_\dt  e^{-2d} \left[2\mc{H}^{AB}\dt_AN_B+N_A\dt_B \mc{H}^{AB}\right]  \, .
\end{equation}
The normal $N_A$ is now the unit normal to the boundary in the doubled space. At the moment it is not clear how to define such a normal since the notion of the extended space itself is not well-defined. In the next section we show that the boundary term actually reflects topological properties of the internal manifold such as monodromy. For example this receives a contribution from exotic branes \cite{deBoer:2012ma}.
Finally, it is important to mention that the expression $\eqref{2.4}$ is $O(d,d)$ covariant and should be true in any duality frame. 

In order for the generalised boundary term \eqref{2.4} to match the boundary term in \eqref{4} (after a duality 
frame is chosen to give the usual bulk action) we require that the possible boundary in the doubled space is restricted to be of the form:
\begin{equation}
 N_A=
\begin{bmatrix}
 0\\ \\n_a
\end{bmatrix},\quad
N^A=
\begin{bmatrix}
 -b^i_{\ph{i}j}n_i \\ \\ n^a
\end{bmatrix}.
\end{equation}
This normal is such that the normalization condition does not imply any constraints to the dynamical fields $g_{ij}$ and
$b_{ij}$:
\begin{equation}
 N^{A}N^{B}\mc{H}_{AB}=1 \Longrightarrow n_a n^a=1.
\end{equation}

The fact that the normal is only allowed components along the $x^i$ directions is due to the fact that we chose the particular duality frame where the fields are independent of $\tilde{x}_i$. A direct consequence of this is that there
could be no boundary located in $\tilde{x}_i$ in the chosen duality frame as this would break $\tilde{x}_i$  translation invariance. Of course, if we chose the T-dual frame where fields are independent of $x^i$ then we would have to choose the opposite condition on the boundary normal. A natural conjecture is that the general restriction on the boundary normal follows from the constraint which has its origins in the level matching condition so that in general we require that
\begin{equation}
N^A \eta_{AB} N^B =0  \, 
\end{equation}
 and the normal vector has components in both ordinary and dual directions
\begin{equation}
\begin{aligned}
N_A=
\begin{bmatrix}
\tilde{n}^a\\
\n_a
\end{bmatrix}.
\end{aligned}
\end{equation}
In order to satisfy the normalisation condition $\mc{H}^{AB}N_AN_B=1$ without introducing extra constraints on the gauge field $b_{ij}$ the component ${\n}_a$ has to be defined as
\begin{equation}
\n_a=n_a+b_{ab}\tilde{n}^b.
\end{equation}
Then the normalisation constraint and section condition imply
\begin{equation}
\begin{aligned}
 \mc{H}^{AB}N_AN_B=1 &  & \Longrightarrow & & \tilde{n}^2+n^2=1,\\
 \h^{AB}N_AN_B=0 & & \Longrightarrow & & \tilde{n}_a n^a=0.
\end{aligned}
\end{equation}

\subsection{$SL(5)$ covariant geometry. }

In Section \ref{M} it was shown that low-energy effective dynamics of M-theory can be described in terms of U-duality covariant fields by extending the internal space in a particular way. In this section we consider the case of $SL(5)$ duality group that corresponds to 4 toroidal directions.  Recall the effective action \eqref{V_sl5}
\begin{equation}
 \begin{split}
\label{V1}
    V=&\sqrt{g}\left[\fr{1}{12}M^{MN}(\dt_M{}M_{KL})(\dt_N{M^{KL}})-\fr12{}M^{MN}(\dt_N{}M^{KL})(\dt_L M_{MK})\right.+\\
      &\left.+\fr{1}{12}M^{MN}(M^{KL}\dt_M M_{KL})(M^{RS}\dt_N M_{RS})-\fr14 (M^{RS}\dt_K M_{RS})(\dt_L M^{KL})\right],
 \end{split}
\end{equation}
where $\dt_M=(\fr\dt{\dt x^\a},\fr\dt{\dt y_{\a\b}})$ and the generalised metric $M_{MN}$ is given by \eqref{2.3}. The section condition $Y^{MN}{}_{KL}\dt_M\bullet\dt_N\bullet=0$, where $Y^{MN}{}_{KL}=\e^{iMN}\e_{iKL}$ is a duality invariant tensor \eqref{Y},  effectively restricts the extended space to its physical subspace. As before the capital Latin indices run from 1 to 10 labelling the $\mathbf{10}$ representation of $SL(5)$, small Greek indices are the ordinary tensor indices and run from 1 to 4, small Latin indices label the $\mathbf{5}$ representation of $SL(5)$.

The section condition is written in the form of a differential equation on all fields living on the extended space. It can be solved by restricting these fields in various ways with a natural solution being $\dt_y=0$. This solution implies that no fields depend on the dual coordinate $y_{\m\n}$ and turns the effective action $V$ to the ordinary supergravity action \eqref{sgr} modulo boundary terms.

In the natural duality frame given by the solution $\dt_y=0$ the effective potential $V$  with all boundary terms included takes the following form
\begin{multline}
\label{4.2.1}
 \int V=\int\sqrt{g}\left(R(g)-\fr{1}{48}F(C)^2\right)-\\-\int\dt_\m\left[\sqrt{g}g^{\n\b}g^{\m\a}\dt_\n g_{\b\a}-\sqrt{g}g^{\m\a}g^{\n\b}\dt_\a g_{\n\b}\right].
\end{multline}
Comparing this expression with the Gibbons-Hawking term \eqref{3} one obtains
\begin{equation}
\label{4.2.3}
 \int\sqrt{g}\left(R(g)-\fr{1}{48}F(C)^2\right)+S_{GH}=\int V+\oint\sqrt{h}(2h^{\a\b}n_{\a,\b}-n^\m h^{\a\b}\dt_\b h_{\a\m}).
\end{equation}
As before the extra term can not be written without referring to a boundary and introducing a normal explicitly. To write the expression above in a duality covariant form one needs to define a generalised normal, that in general should have the form:
\begin{equation}
 N^M=\left[\begin{array}{c}
            n^\m\\ \\
	    \n_{\r\s}-\dfr{1}{\sqrt{2}}C^\a{}_{\r\s}n_\a
           \end{array}
\right].
\end{equation}
As before the form of the normal is determined by the simple requirement that the normalization $M_{AB}N^AN^B=1$ should not imply any constraint either on $C$ or $g$.
Thus, we have for the norm
\begin{equation}
 M_{AB}N^AN^B=|n|^2+|\n|^2.
\end{equation}
Finally, repeating calculations of the previous section one finds that \eqref{4.2.3} can be written in the following form:
\begin{equation}
\label{4.2.4}
 \int V+\oint\sqrt{h}\left(2M^{AB}\dt_A N_B+N_A\dt_B M^{AB} \right).
\end{equation}
Hence, the extra term can be written in a duality covariant form by introducing a generalised normal.

\section{Summary}

In this chapter it was shown that for consistency one should add a boundary term to the known Hohm-Zwiebach or Berman-Perry actions. Then the full effective action successfully reproduces all the terms in the Einstein-Hilbert action without need of integrating by parts. Moreover, the Gibbons-Hawking term, which captures the thermodynamics of empty-space solutions, follows from the full effective action as well. An important remark is that the Gibbons-Hawking boundary term that is always present in General Relativity is related to thermodynamic properties of black branes and basically to the topology of the space. A similar idea stands behind the results shown in the sections above.

In all the expressions of this chapter we do not specify the boundary since the geometry of the extended space is still unclear. However in the following chapter we explicitly show that the derived boundary term actually feels the non-trivial topology of string theory backgrounds generated by exotic branes with non-zero non-geometric $Q$-flux. It is demonstrated, that two contributions from these fluxes are T-dual to each other providing the full boundary term is T-duality invariant as it should be.

\chapter{Non-geometry in Double Field Theory}

\section{Exotic branes and monodromy}
The Gibbons-Hawking term in general relativity is relevant for backgrounds generated by objects with a horizon, such as black holes or black branes, and the horizon is related to the boundary. In string or M-theory one may meet even more fascinating situations when a consistent background is not defined globally. Instead, local patches are glued by duality transformations leading to non-trivial cycles. Encircling these cycles results in transformation of the metric and gauge fields, that in general mixes these objects, hence the name of non-geometric background.

A example of such background is provided by the twisted torus already mentioned in section \ref{GG_metric}. This geometry appears in Type II string theory compactified on a torus $\TT^2$. Consider an NS5-brane extending along six dimensions not wrapping the internal 2-torus. T-duality along one of cycles of the 2-torus turns the NS5 brane into the Kaluza-Klein monopole (or the $5_2^1$-brane in another notations, for a review see \cite{deBoer:2012ma}). Further action of T-duality along the remained cycle of the internal torus results in a non-geometric background generated by the so-called $5_2^2$-brane that carries a non-zero Q-flux. This duality chain can be represented by the following table.
\begin{table}[http]
\centering
\begin{tabular}{|r|ccccccc|cc|}
\hline
 & 1 & 2 & 3 & 4 & 5 & 6 & 7 & 8 & 9 \\  
 \hline
NS5  &  &   & $\times$ & $\times$ & $\times$ & $\times$ & $\times$ & $\cdot$  & $\cdot$   \\
KKM &  &   & $\times$ & $\times$ & $\times$ & $\times$ & $\times$ &  $\odot$  & $\cdot$ \\
 $5^2_2$&   &   & $\times$ & $\times$ & $\times$ & $\times$ & $\times$ &  $\odot$  & $\odot$\\
 \hline 
\end{tabular}
\caption{Under T-dualities an NS5-brane stretched in directions marked by $\times$ turns into a Kaluza-Klein monopole and a $5_2^2$-brane. Dotted circles denote special cycles along which T-duality acts. }
\end{table}

In the supergravity description the metric for a $5_2^1$ brane (a KK monopole) wrapped on compact 3,4,5,6,7 directions, placed at  $\bf{x}=\bf{x}_p$ in the transverse space $\RR^3_{129}$ is given by \cite{deBoer:2012ma}
\begin{equation}
\label{KKM}
\begin{aligned}
ds^2&=ds_{034567}^2+Hds_{129}^2+H^{-1}(dx^8+\w)^2,\\
B^{(2)}&=0, \quad ds_{129}^2=dr^2+r^2d\q^2+(dx^9)^2,\\
H&=1+\sum_p{H_p}, \quad H_p=\fr{R_8}{2|\bf{x-x}_p|}
\end{aligned}
\end{equation}
where $B^{(2)}$ is the Kalb-Ramond 2-form field, $\w$ is a 1-form and $R_9$ is the radius of the $x^9$ direction.

In order to T-dualize this solution along the direction $x^9$ we consider a set of KK monopoles with centres arrayed along $x^9$ at intervals of $2\p\tilde{R}_9$. Hence, the function $H$ becomes divergent
\begin{equation}
H=1+\sum_{n\in\mathbb{Z}}\fr{R_8}{2\sqrt{r^2+(x^9-2\p \tilde{R}_9 n)^2}}\approx 1+\s\log\fr{\L+\sqrt{r^2+\L^2}}{r},
\end{equation}
where the sum was approximated by an integral and a cut-off $\L$ was introduced. The constant $\s$ is defined as $\s=R_8/2\p \tilde{R}_9=R_8R_9/2\p\a'$.

The log divergence of this kind is common for a co-dimension two object and imply that it is ill-defined as a stand-alone object. Instead, one considers a configuration where at long distances this divergence is compensated by contributions from another co-dimension two objects. Hence one considers a regularised form of the function
\begin{equation}
H(r)=h_0+\s \log\fr{\m}{r},
\end{equation}
where the radius $r\in [0,\L]$. To have asymptotically flat space, i.e. $H(r=\infty)=1$ we rewrite $H$ in the following form
\begin{equation}
H(r)=1-\s \log\fr{r}{\L}.
\end{equation}

For this choice of the function $H$ the 1-form $\w$ can be written as $\w=-\s \q dx^9$ and one immediately see that encircling the cycle $\q\to\q+2\p$ results in twisting the special 2-torus as
\begin{equation}
\label{f_twist}
\begin{aligned}
x^8&\to x^8-2\p \s x^9,\\ 
x^9&\to x^9.
\end{aligned}
\end{equation}
This transformation glues the tori $\TT^2$ at the points $\q$ and $\q+2\p$ defining a monodromy group around the cycle. Since the monodromy is a diffeomorphism the KKM background is geometric, i.e.~the metric do not mix with the B-field. This background carries a non-zero f-flux $f^{8}{}_{9\q}=-\s$

An example of a non-geometric background is provided by the $5_2^2$ brane. Its metric can be derived by T-dualizing the remained $x^9$ coordinate of the special 2-torus in \eqref{KKM}. Straightforward application of the Buscher rules \eqref{Buscher} gives the following expression
\begin{equation}
\label{522}
\begin{aligned}
ds^2&=H(dr^2+r^2 d\q^2)+\fr{H}{H^2+\s^2 \q^2}ds_{89}^2+ds_{034567}^2,\\
B^{(2)}&=\fr{\s \q}{H^2+\s^2\q^2}dx^8\wedge dx^9,\\
e^{-2\ff}&=\fr{H}{H^2+\s^2\q^2},
\end{aligned}
\end{equation}
where $\ff$ denotes the dilaton. 

For this configuration the monodromy around the circle $\q\to\q+2\p$ is not a diffeomorphism, it mixes the metric and the B-field acting as a T-duality transformation. Namely, the size of the special 2-torus does not come back to itself
\begin{equation}
\begin{aligned}
\q=0: & && G_{88}=G_{99}=H^{-1},\\
\q=2\p: & && G_{88}=G_{99}=\fr{H}{H^2+(2\p\s)^2}.
\end{aligned}
\end{equation}
The resulting transformation can be most clearly written in terms of Double Field Theory. Let us focus on the $(8,9)$ part of the metric that corresponds to the special torus. From the point of view of transverse space the corresponding generalised metric encodes scalar moduli and has the following form (see Section \ref{GG_metric})
\begin{equation}
\label{GG_metr}
\mc{H}_{MN}=
\begin{bmatrix}
G^{-1} & G^{-1}B \\
-BG^{-1} & G-BG^{-1}B
\end{bmatrix}.
\end{equation}
In this notation the monodromy $\q\to\q+2\p$ takes the form of an $O(2,2)$ rotation
\begin{equation}
\mc{H}(\q'=\q+2\p)=\mc{O}^{tr}\mc{H}(\q)\mc{O},
\end{equation}
where the matrix $\mc{O}$ encodes the non-geometric $\b$-transform
\begin{equation}
\label{beta_transf}
\mc{O}=
\begin{bmatrix}
\bf{1}_2 & 0 \\
\b(\q') & \bf{1}_2
\end{bmatrix}.
\end{equation}
We will see in further sections, that although the bivector $\b$ is usually understood as a sign of non-geometry, it is not the only source of Q-flux. In the DFT formulation the Q-flux becomes written in terms of derivatives of the vielbein with respect to dual coordinates and does not vanish even if $\b=0$.

To switch on the Q-flux with the section condition imposed, i.e. when there is no dependence on dual coordinates, one need the bivector to be non-zero. For the case of a $5_2^2$ exotic brane the bivector has only one component \cite{Geissbuhler:2013uka}
\begin{equation}
\begin{aligned}
\b&=\b^{89}\fr{\dt}{\dt x^8}\wedge\fr{\dt}{\dt x^9},& & \b^{89}=\s\q.
\end{aligned}
\end{equation}
The explicit form of the Q-flux is then given by one component $Q_{\q}{}^{89}=-\s$. 

It is more natural to write the metric for non-geometric backgrounds in the so-called non-geometric frame, where the generalised metric is just a beta-transform of \eqref{GG_metr} by the matrix \eqref{beta_transf}. In this frame the bivector $\b$ replaces the Kalb-Ramond field in the generalised metric $\tilde{\mc{H}}_{MN}$ 
\begin{equation}
\tilde{\mc{H}}_{MN}=
\begin{bmatrix}
G^{-1}-\b G\b & \b G \\
-\b G & G
\end{bmatrix}.
\end{equation}
In further sections we derive the explicit form of the corresponding generalised vielbein. in this frame the metric for a $5_2^2$ brane is written in a suggestive form
\begin{equation}
\label{522_beta}
\begin{aligned}
ds^2&=H(dr^2+r^2 d\q^2)+H^{-1}ds_{89}^2+ds_{034567}^2,\\
\b&=\b^{89}\fr{\dt}{\dt x^8}\wedge\fr{\dt}{\dt x^9}.
\end{aligned}
\end{equation}
Since the monodromy \eqref{beta_transf} glues the space at the point $\q$ and $\q+2\p$, the generalised metric appears to be written in different frames at these points. This observation is of crucial importance for further sections, where the topological contribution becomes proportional to the monodromy.

\section{Gauged Doubled Field Theory}

To introduce a setup for further sections we briefly repeat the calculation of \cite{Grana:2012rr} here with all necessary details included. For a more detailed description of the generalised Scherk-Schwarz reductions the reader is referred to the Chapter \ref{reduction}.

In the manifestly T-duality covariant low-energy formalism for string theory, there is an object called the generalised metric, which appears to be a metric on the so-called extended space. The background, given by the direct product of an external manifold $M$ and the internal torus $\TT^d$, is replaced by the direct product of the external manifold and the doubled torus $\TT^d\times\tilde{\TT}^d$. Coordinates $\YY^M$ parametrising the doubled torus unify the coordinates corresponding to all string charges
\begin{equation}
\YY^M=\begin{bmatrix}
y^a\\
\ty_a
\end{bmatrix},
\end{equation}
where the small Latin indices run from 1 to $d$, and the capital Latin indices run from 1 to $2d$. The $O(d,d)$ covariant dynamics is formulated in terms of the generalised metric by introducing the following effective action that is invariant with respect to \eqref{gen_Lie_1} up to the section condition \cite{Hohm:2010jy}
\begin{equation}
 \label{HZ}	
\begin{split}
    S=\int \sqrt{g}\;d\XX \, e^{-2d}&\left(\fr18\mc{H}^{\hat{M}\hat{N}}\dt_{\hat{M}}\mc{H}^{\hat{K}\hat{L}}\dt_{\hat{N}}\mc{H}_{\hat{K}\hat{L}} -\fr12
\mc{H}^{\hat{K}\hat{L}}\dt_{\hat{L}}\mc{H}^{\hat{M}\hat{N}}\dt_{\hat{N}}\mc{H}_{\hat{K}\hat{M}} - \right.\\
      & \left.- 2\dt_{\hat{M}} d\dt_{\hat{N}}\mc{H}^{\hat{M}\hat{N}}+4\mc{H}^{\hat{M}\hat{N}}\dt_{\hat{M}}d\dt_{\hat{N}}d\lefteqn{\ph{\fr12}}\right) \, .
\end{split}
\end{equation}
Here the internal coordinates $\YY^M=(y^a,\tilde{y}_a)$ on the doubled torus and the external coordinates $x^\m$ are collected into one object $\XX^{\hat{M}}=(x^\m,\YY^M)$, where the hatted Latin indices run from 1 to $2d+n$. Measure on the external space is given by $\sqrt{g}d^nx=\sqrt{\det||g_{\m\n}||}d^nx$ where the Greek indices run from 1 to $n$. Then the generalised metric $\mc{H}_{\hat{M}\hat{N}}$ can be represented in the block-diagonal form
\begin{equation}
\mc{H}_{\hat{M}\hat{N}} =
\begin{bmatrix}
g_{\m\n} & 0 \\
0 &\mc{H}_{{M}{N}}
\end{bmatrix},
\end{equation}
while the doubled dilaton $d$ is written in terms of the usual dilaton $\ff$ 
\begin{equation}
e^{-2d}=\sqrt{G}e^{-2\ff}  \, .
\end{equation}
An important aspect of the formalism is that explicit solutions of the section condition correspond to certain choices of T-duality frame. In each frame the effective action takes the form of the Type II supergravity action (bosonic part)
\begin{equation}
\label{sugra}
S \to \int \sqrt{\hat{G}} dz \, e^{-2\f} \left(R[g]+4(\dt\ff)^2-\fr{1}{12}H^2\right),
\end{equation}
where $\hat{G}$ is the metric on the whole $(d+n)$-dimensional space parametrised by the coordinates $z$ and $H=dB$ is the field strength for the Kalb-Ramond field. Explicit relationship of the coordinates $z$ to the coordinates $x$ and $\YY$ depends on the duality frame chosen. A natural choice is to drop all dependence on the dual coordinates $\ty_a$ and end up with $z=(x^\m,y^a)$. The choice $z=(x^\m,\ty_1,y^2,\ldots,y^d)$ is equivalent to performing a T-duality transformation along the $y^1$ coordinate.

It was shown in \cite{Grana:2012rr} that the Scherk-Schwarz dimensional reduction of the described $O(d,d)$ invariant formalism completely reproduces the structure of the scalar sector of gauged supergravity. The reduction is performed by introducing the twisting matrices $U^M{}_{\bM}(\YY)$ that encode all of the dependence on the internal extended space coordinates
\begin{equation}
\label{twist1}
\begin{aligned}
T^{A_1\ldots A_m}(x,\YY)&\equiv U^{A_1}{}_{\bA_1}(\YY)\cdots U^{A_m}{}_{\bA_m}(\YY)T^{\bA_1\ldots \bA_m}(x),\\
d(x,\YY)&\equiv \bd(x)+\l(\YY),
\end{aligned}
\end{equation}
where the barred indices denote twisted directions and $T^{A_1\ldots A_m}$ is a rank $m$ generalised tensor on the extended space.

Substituting this anzatz into the definition of the generalised Lie derivative \eqref{gen_Lie_1} one recovers the following expression
\begin{equation}
\label{F1}
\begin{aligned}
\mc{L}_{V_1}V_2^M&=U^M{}_{\bM}(\YY) F^{\bM}{}_{\bK\bL}V_1^{\bK}(x)V_2^{\bL}(x),\\
F^{\bA}{}_{\bB\bC}&=2U^{\bA}{}_M U_{[\bB}{}^N\dt_N U_{\bC]}{}^M-U^{\bA}{}_MY^{MN}{}_{KL}\dt_N U_{[\bB}{}^K U_{\bC]}{}^L,
\end{aligned}
\end{equation}
where the coefficients $F^{\bM}{}_{\bK\bL}$ are taken to be constants. This leads to tructure of an algebra
\begin{equation}
[X_{\bM},X_{\bN}]=F^{\bA}{}_{\bM\bN}X_{\bA}
\end{equation}
with generators $X_{\bA}$ defined by their matrix form in the adjoint representation $(X_{\bA})^{\bM}{}_{\bN}\equiv F^{\bM}{}_{\bA\bN}$. 

The structure constants can be split into irreducible representations of the corresponding duality group implying that the algebra is an $O(d,d)$ (or $E_{d}$)-graded algebra.  This is in accordance with the structure of gauged supergravities where $F^{\bM}{}_{\bK\bL}$ is called the embedding tensor. From the point of view of the external space the structure constants encode all geometric and non-geometric fluxes \cite{Grana:2008yw} (for details see Section \ref{Fluxes}).

An important but straightforward consequence of Scherk-Schwarz anzatz is that one does not need the section condition for closure of the algebra. Instead, the structure constants (the embedding tensor) should satisfy a set of constraints, quadratic and linear \cite{Samtleben:2008pe}. The anzatz \eqref{twist1} then allows to rewrite the action \eqref{HZ} in terms of the gaugings $F^{\bM}{}_{\bK\bL}$
\begin{equation} 
\label{GDFT} 
S_{G}= \int d^d \YY e^{-2 \l(\YY)} \int d^{n}x e^{-2 \bd}\left({\cal R} + {\cal R}_f\right)\ , 
\end{equation}
where ${\cal R}_f$ is the gauged part of the action
\begin{equation} 
\begin{aligned}
{\cal R}_f = & - \fr12
F^{\bA}{}_{\bB\bC} {\mc H}^{\bB\bD} {\mc H}^{\bC\bE}
\dt_{\bD} {\mc H}_{\bA\bE} - \fr{1}{12} F^{\bA}{}_{\bB\bC}F^{\bD}{}_{\bE\bF}{\mc H}_{\bA\bD}{\cal H}^{\bB\bE}{\cal H}^{\bC\bF}
\\ & - \fr14 F^{\bA}{}_{\bB\bC} F^{\bB}{}_{\bA\bD}{\mc H}^{\bC\bD} - 2 F_{\bA} \dt_{\bB}  {\cal H}^{\bA\bB} + 4 F_{\bA} {\mc H}^{\bA\bB}\dt_{\bB} \bd- F_{\bA} F_{\bB}  {\mc H}^{\bA\bB}\ .
\end{aligned}
\end{equation}
The twisted derivative is defined as $\dt_{\bM}\equiv U^M{}_{\bM}\dt_M$ and the additional gauging reads
\begin{equation} 
F_{\bA} = \dt_M (U^{-1})^M{}_{\bA} - 2 (U^{-1})^M{}_{\bA} \dt_M \l
\end{equation}
To write the gauged action consistently in terms of the gaugings $F^{\bA}{}_{\bM\bN}$ and $F_{\bM}$ one has to supply the action \eqref{HZ} by an additional full-derivative term and a term that is zero up to the section condition 
\begin{equation}
SC \sim \int dx d\YY \,Y^{MN}{}_{KL}\dt_M E^K_{\underline{K}}E^L_{\underline{L}}\mc{H}^{\underline{K}\underline{L}},
\end{equation}
 where we denote flat indices by underlined Latin letters. At the moment one should not confuse underlined and barred indices since they refer to different types of vielbeins $E$ and $U$. In the section \ref{Fluxes} these will be identified.

\section{Full Double Field Theory action}

The action \eqref{GDFT} that gives the scalar potential of supergravity differs from the effective potential \eqref{HZ} by a full derivative and a term that is zero up to the section condition. 
\begin{equation}
\label{Grana_HZ}
S_{G}=S+SC+\int_{\mc{M}} dxd\YY \dt_{\hat{M}}\left[e^{-2d}(4\mc{H}^{\hat{M}\hat{N}}\dt_{\hat{N}}d-\dt_{\hat{N}}\mc{H}^{\hat{M}\hat{N}})\right].
\end{equation}
The integration is taken over some region of the extended space $\mc{M}\simeq\RR^n\times \TT^d\times\tilde{\TT}^d$ that may have non-trivial topological properties (see discussion in the next section). The generalised metric $\mc{H}_{MN}$ and the extended coordinates $\XX^{\hat{M}}=(x^\m,\YY^M)$ are defined as
\begin{equation}
\label{gg_metric}
 \mc{H}^{MN}=
\begin{bmatrix}
   G_{ij}-B_i^{\ph{i}a}B_{aj}   & B_i^{\ph{i}k} \\
	&\\
   -B^l_{\ph{l}j}		& G^{kl}
\end{bmatrix}, \quad 
\YY^M= 
\begin{bmatrix}
 \tilde{y}_m \\
  y^m
\end{bmatrix}.
\end{equation}
Here the hatted indices label all coordinates, including flat coordinates $x^\m$ of the space that is not doubled, and $\hat{M}=1,\ldots,2d\ldots 2d+n$. Capital Latin indices without hat label only coordinates of the doubled space and run from 1 to $2d$.

On the other hand we have the action that gives the full SUGRA action with the Gibbons-Hawking term. It differs from the Hohm-Zwiebach action by a boundary term that can be written in the duality invariant form \cite{Berman:2011kg}
\begin{equation}
\label{full_HZ}
S_{Full}=S_{HZ}+\oint_\dt e^{-2d}\left[2\mc{H}^{\hat{A}\hat{B}}\dt_{\hat{A}}N_{\hat{B}}+N_{\hat{A}}\dt_{\hat{B}}\mc{H}^{\hat{A}\hat{B}}\right].
\end{equation}
In what follows this action is referred to as 'the full action'. 

Substituting the form of the action $S_{HZ}$ given by \eqref{Grana_HZ} to \eqref{full_HZ} the full action becomes equal to the action $S_{G}$ plus a term that is an integral over a boundary $\dt$ of a full derivative
\begin{equation}
\label{full}
S_{Full}=S_{G}+2\oint_\dt\dt_{\hat{A}}\left(e^{-2d}\mc{H}^{\hat{A}\hat{B}}N_{\hat{B}}\right).
\end{equation}
Although the integrand in the second term can be represented in a from of a total derivative the integral is in general non-zero. The extra term vanishes only if the expression in brackets in \eqref{full} can be defined globally meaning that the space $\mc{M}$ has trivial topology.

In the particular T-duality frame defined by the solution $\tilde{\dt}^i=0$ of the section condition when  $\tilde{n}_a=0$ reflecting the translational invariance along $\tilde{y}_m$, the  boundary term in \eqref{full} can be written in the following way
\begin{equation}
\begin{aligned}
2\oint_\dt\dt_A\left(e^{-2d}\mc{H}^{AB}N_B\right) & \longrightarrow & 2\oint _\dt dS \,K.
\end{aligned}
\end{equation}
Here $K=g^{ab}\nabla_a n_b$ is the extrinsic curvature of the boundary,$\nabla_a$ is the ordinary covariant derivative with Levi-Civita connection and $dS=\sqrt{-g}\,d^{d-1}x$ is the area element of the boundary.

The boundary term can be rewritten in the very same form in any T-duality frame using the semi-covariant formalism developed in \cite{Jeon:2011sq,Jeon:2011vx,Jeon:2012hp}.  The semi-covariant derivative is defined as
\begin{equation}
\nabla_C T_{A_1A_2\ldots A_N}=\dt_C T_{A_1A_2\ldots A_N}-\w_T \G^B{}_{BC}T_{A_1A_2\ldots A_N}+\sum_{i=1}^N \G_{CA_i}{}^BT_{A_1\ldots A_{i-1}BA_{i+1}\ldots A_N},
\end{equation}
where the weight $\w_T$ is non-zero only for the dilaton $d$, that is by definition covariantly constant $\nabla_C d=0$. Using these definitions, the boundary term readily takes the following form
\begin{equation}
\begin{aligned}
2\oint_\dt\dt_A\left(e^{-2d}\mc{H}^{AB}N_B\right)d\S=2\oint _\dt d\S \,e^{-2d}\mc{K},
\end{aligned}
\end{equation}
where $\mc{K}=\mc{H}^{AB}\mc{K}_{AB}$ and $\mc{K}_{AB}=\nabla_A N_B$. The quantity $\mc{K}$ can be identified with the extrinsic curvature of the generalised boundary. Although, the form of the boundary term looks very familiar it is not clear how the generalised area element $d\S$ is defined.

\section{Topology of extended space}

The form \eqref{full} of the boundary term does not manifestly specify the boundary $\dt$. To understand its geometry it is convenient to focus only on terms that can be represented as an integral of a full derivative
\begin{equation}
S_{full}=S_{Grana}+2\oint_\dt d\S\, \dt_{\hat{A}}N_{\hat{B}}\mH^{\hat{A}\hat{B}}e^{-2d}+2\int_{\mc{M}}\sqrt{-g} d^nx d\YY \dt_{\hat{A}}\left(\dt_{\hat{B}}\mH^{\hat{A}\hat{B}}e^{-2d}\right).
\end{equation} 
As was mentioned in the previous sections in the DFT formalism the space $\mc{M}$ can be represented as the flat manifold $\RR^n$ (parametrized by $x^\m$) times the doubled torus $\TT^d\times \tilde{\TT}^d$ parametrized by $\YY^M$.
Terms that involve boundaries in flat directions have the following form
\begin{equation}
\label{ext}
\int_\dt \dt_\mu N_\nu g^{\m\nu} d\s +\int d\YY \int \sqrt{-g}d^nx \dt_\a \left(\dt_\b g^{\a\b} e^{-2d}\right),
\end{equation}
where small Greek indices run from 1 to $n$ labelling the flat space coordinates, the area element has the transparent meaning $d\s=\sqrt{-h}d^{n-1}x$ and $h$ is an induced metric on the boundary. 

We choose the coordinates $x^\m$ to label the space $\RR^n$ that do not have any boundary or non-trivial topology. Then all terms that have derivatives with respect to $x^\m$ become zero. The second term in \eqref{ext} is zero according to the Poincar\'e lemma since it is a full derivative. The first term would not appear in our consideration from the very beginning since there is no boundary and the normal cannot be defined. Alternatively, one can cut the space $\RR^n$ introducing two boundaries $\dt_1$ and $\dt_2$ by hand. 
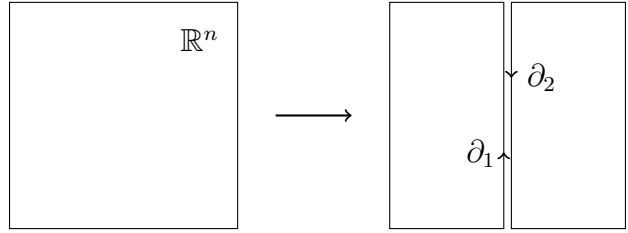
\begin{figure}[ht]
\centering
\begin{tikzpicture}[scale=1]
\draw [black] (0,0) rectangle (3,3);
\node at (2.5,2.5) {$\RR^n$};
\draw [->,thick] (3.5,1.5) -- (4.5,1.5);
\draw [black] (5,0) rectangle (6.5,3);
\draw [->,thick] (6.5,1) -- (6.5,1.01);
\draw [->,thick] (6.6,2.01) -- (6.6,2);
\draw [black] (6.6,0) rectangle (8.1,3);
\node at (6.2,1) {$\dt_1$};
\node at (7,2) {$\dt_2$};
\end{tikzpicture}
\caption{Contributions from the boundaries $\dt_1$ and $\dt_2$ in $\RR^n$ cancel each other}
\end{figure}
This results in two terms each involving an integral over the corresponding boundary. Since all fields are well defined when crossing this kind of boundaries in the flat directions, these contributions cancel each other:
\begin{equation}
\begin{aligned}
& & \int_\dt \dt_\mu N_\nu g^{\m\nu} d\s  &= \int_{\dt_1} \dt_\mu N^{(1)}_\nu g^{\m\nu} d\s + \int_{\dt_2} \dt_\mu N^{(2)}_\nu g^{\m\nu} d\s\\
& & &= \int_{\dt_1} \dt_\mu N^{(1)}_\nu g^{\m\nu} d\s -\int_{\dt_2} \dt_\mu N^{(1)}_\nu g^{\m\nu} d\s=0.
\end{aligned}
\end{equation}
In the second line it was used that the boundaries $\dt_1$ and $\dt_2$ are virtually the same surface with the normal $N^{(1)}_M=-N^{(2)}_M$.

The same reasoning does not work for the doubled torus $\TT^d\times\tilde{\TT}^d$ in the presence of fluxes, both geometric or non-geometric. In this case the doubled torus becomes a fibration with non-trivial monodromy properties that result in a non-zero contribution. Before we proceed in this direction it is suggestive to consider the classical monopole solution, which demonstrates the similar behaviour \cite{Nakahara:2003nw}.

The monopole appears as a topologically non-trivial  configuration of a gauge field given by a 1-form $A$. The flux of the corresponding field strength through a 2-sphere is then defined as
\begin{equation}
\int_{\mathbb{S}^2}F=\int_{\mathbb{S}^2}dA=\int_{U_N}dA_N+\int_{U_S}dA_S.
\end{equation}
Here the sphere is split into two charts each carrying gauge potentials $A_N$ and $A_S$ related by a gauge transformation $A_N=A_S+d\l$. 
\begin{figure}[ht]
\centering
\begin{tikzpicture}[scale=1.3]
\node at (-4,1.5) {$\SS^2$};
\node at (-0.3,1.5) {$U_N$};
\node at (-0.3,-0.5) {$U_S$};
\draw [black] (-3,0.5) ellipse (1 and 1);
\draw [black, thick] (-4,0.5) arc (180:360:1cm and 0.2cm);
\draw [black, thick, dashed] (-4,0.5) arc (180:0:1cm and 0.2cm);
\draw [->, thick] (-1.5,0.5) -- (-0.5,0.5);
\draw (0,1) arc (180:0:1cm and 1cm);
\draw [black, thick] (0,1) arc (180:360:1cm and 0.2cm);
\draw [black, thick, dashed] (0,1) arc (180:0:1cm and 0.2cm);
\draw [black, thick] (1,0) ellipse (1 and 0.2);
\node at (1,0.5) {$\SS^1$};
\draw (0,0) arc (180:360:1cm and 1cm);
\node at (2.7,1.5) (N) {$A_N$};
\node at (2.7,-0.5) (S) {$A_S$};
\draw [->, thick] (S) -- (N);
\node at (3.5,0.5) {$\l\in U(1)$};
\end{tikzpicture}
\caption{Gauge field $A$ is not defined globally on the sphere $\SS^2$. Two patches $U_{N}$ and $U_S$ carry the potentials $A_{N}$ and $A_S$ related by a gauge transformation. }
\end{figure}
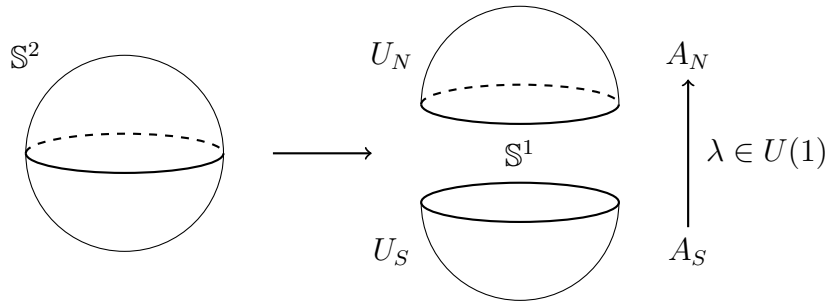
Using Stoke's theorem each term can be written as integration over boundaries of $U_{N,S}$ that  are virtually the same with the curve with topology of a circle $\SS^1$
\begin{equation}
\int_{U_N}dA_N+\int_{U_S}dA_S=\int_{\dt U_N}d\l=\int_{\mathbb{S}^1}d\l.
\end{equation}
The situation here is in certain sense similar to what we have had before. Naively, one could say that this integral should be zero since there is no boundary. However, the gauge parameter $\l$ is an element of $U(1)$ and thus has non-trivial monodromy when going around the circle. It acquires a shift when going around the circle.

Explicitly, one can cut the circle, introducing a coordinate $\q$ that runs from 0 to $2\p$. Then the integral becomes
\begin{equation}
\int_{\mathbb{S}^1}d\l=\int_0^{2\p} d\q \,\dt_\q\l(\q)=\l(\q)\bigg|_{\q=0}^{\q=2\p}=\l(2\p)-\l(0)\sim n\in \mathbb{Z}
\end{equation}
providing quantization of monopole charge. 

Going back to the boundary term in \eqref{full} consider only terms that involve derivatives along $\YY^M$ 
\begin{equation}
\int \sqrt{-g}d^nx \int d\YY \dt_A\left(e^{-2d}\dt_B \mH^{AB} \right).
\end{equation}
The internal torus $\mathbb{T}^d$ can be represented as a torus fibration over a circle $\mathbb{S}^1$ with a fibre $\mathbb{T}^{d-1}$. For backgrounds with non-zero geometric f-flux or non-geometric fluxes one acquires a non-zero holonomy around the circle $\q\rightarrow \q+2\p$. Fibres at $\q=0$ and $\q=2\p$ are glued by a T-duality transformation that in general mixes metric and gauge fields, hence the name of non-geometric background. An example of such a situation is the twisted torus that describes a background with f-flux \cite{Wecht:2007wu}.

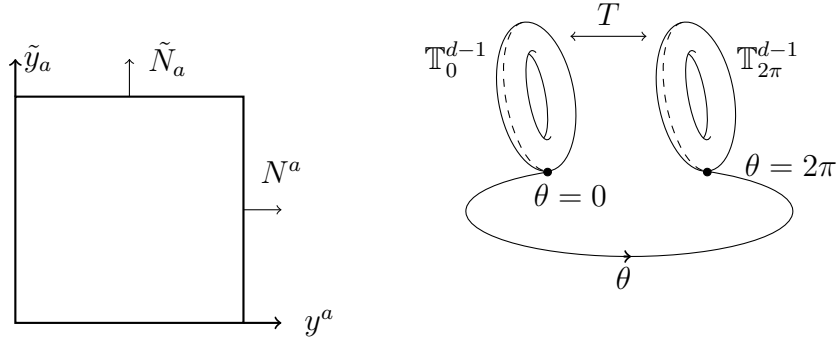
\begin{figure}[ht]
\centering
\begin{tikzpicture}
\draw [black, thick] (0,0) rectangle (3,3);
\draw [->, thick] (3,0) -- (3.5,0);
\node at (4,0) {$y^a$};
\draw [->, thick] (0,3) -- (0,3.5);
\node at (0.3,3.5) {$\tilde{y}_a$};
\draw [->] (1.5,3) -- (1.5,3.5);
\node at (2,3.5) {$\tilde{N}_a$};
\draw [->] (3,1.5) -- (3.5,1.5);
\node at (3.5,2) {${N}^a$};
\draw [black, rotate around={0:(8,2)}] (7,2) arc (-240:63:2.15 and 0.6);
\torus{8.95}{3}{-80}{0.5};
\torus{6.85}{3}{-80}{0.5};
\node at (5.8,3.5) {$\TT^{d-1}_{0}$};
\node at (9.9,3.5) {$\TT^{d-1}_{2\p}$};
\node at (8,0.6) {$\q$};
\draw [->, thick] (8,0.88) -- (8.1,0.88);
\node at (10.2,2.1) {$\q=2\p$};
\node at (7.3,1.7) {$\q=0$};
\draw [black, fill] (7,2) ellipse (0.05 and 0.05);
\draw [black, fill] (9.1,2) ellipse (0.05 and 0.05);
\draw [<->] (7.3,3.8) -- (8.3,3.8);
\node at (7.8,4.1) {$T$};
\end{tikzpicture}
\caption{The boundaries $\TT^{d-1}\times \tilde{\TT}^{d}$ and $\TT^{d}\times \tilde{\TT}^{d-1}$ are obtained by cutting the base $\SS^1$ of the corresponding torus fibrations $\TT^{d-1}\times \SS^1\times \tilde{\TT}^{d}$ and $\TT^{d}\times \tilde{\TT}^{d-1}\times \tilde{\SS}^1$.}
\end{figure}

We conjecture that the same is true for the dual torus meaning there is a non-trivial monodromy around the circle $\tilde{\SS}^1$ parametrised by  the dual coordinate $\tilde{\q}$. Taking this into account the expression above can be written as follows
\begin{multline}
\label{T_bound}
\int \sqrt{-g}d^nx \int_{\mathbb{T}^{d-1}_\q\times\tilde{\mathbb{T}}^{d}} {N}_a\left(e^{-2d}\dt_B \mH^{aB} \right)\bigg|^{\q=2\p}_{\q=0}+\\
+\int \sqrt{-g}d^nx\int_{\mathbb{T}^{d}\times\tilde{\mathbb{T}}^{d-1}_{\tilde{\q}}} \tilde{N}^a\left(e^{-2d}\dt_B \mH_{a}{}^{B} \right)\bigg|^{\tilde{\q}=2\p}_{\tilde{\q}=0}.
\end{multline}
Where we integrated out the cycle $\mathbb{\SS}^1$ in the first term and the  dual cycle  $\tilde{\SS}^1$ in the second term. The normal $N^a$ and the dual normal $\tilde{N}_a$ here have only components in the directions $\q$ and $\tilde{\q}$ respectively. 

The full action \eqref{full} is then written as a sum $S_{Full}=S_G+S_B$ where the boundary term has the following form
\begin{equation}
\begin{aligned}
\label{B_term}
S_B&=\int \sqrt{-g}d^nx \int_{\mathbb{T}^{d-1}_\q\times\tilde{\mathbb{T}}^{d}} \dt_B\left({N}_ae^{-2d} \mH^{aB} \right)\bigg|^{\q=2\p}_{\q=0}+\\
&+\int \sqrt{-g}d^nx\int_{\mathbb{T}^{d}\times\tilde{\mathbb{T}}^{d-1}_{\tilde{\q}}} \dt_B\left(\tilde{N}^ae^{-2d} \mH_{a}{}^{B} \right)\bigg|^{\tilde{\q}=2\p}_{\tilde{\q}=0}.
\end{aligned}
\end{equation}
It is important to mention that $S_B$ is T-duality invariant since the second term is exactly T-dual of the first term. As in the case of monopole discussed above, this expression is non-zero in general, depending on monodromy properties of the background. In this sense it is analogous to the term 
\begin{equation}
\int_{\mathbb{S}^1}d\l\neq 0.
\end{equation}
In the next section we show that the boundary term involves fluxes and thus probes global properties of the boundary, exactly as the gauge parameter $\l$.


\section{Duality invariant formulation of fluxes}
\label{Fluxes}

Dimensional reductions of effective theories that appear as low-energy limits of string theory are characterised by certain charges $H$, $f$, $Q$ and $R$, also known as fluxes. The geometric flux $H_{abc}$ is related to the 2-form Kalb-Ramond field $B=B_{mn}dy^m\wedge dy^n$ as $H=dB$. A T-duality transformation along one of cycles of the internal torus turns the geometric $H$-flux into the geometric $f$-flux. Under this transformations the  becomes twisted, with torsion defined by
\begin{equation}
de^{\bar{a}}=-\fr12 f^{\ba}{}_{\br{b}{c}}e^{\bar{b}}\wedge e^{\bar{c}},
\end{equation}
where barred indices are used for flat directions and $e^{\ba}=e^{\ba}_adx^a$ is a 1-form that defines vielbein. This equation can be written in an equivalent form by making use of the Lie bracket of two vector fields
\begin{equation}
\label{bracket_e}
[e_{\bar{a}},e_{\bar{b}}]=f^{\bar{a}}{}_{\br{b}{c}}e_{\bar{a}}.
\end{equation}
Here the inverse vielbein is a vector field $e_{\ba}=e_{\ba}^a\,\dt_a$ and $f^{\ba}{}_{\bm\bn}=2e^{\ba}_{r}e^b_{[\bm}\dt_be^r_{\bn]}$.

Further T-duality transformations along the remaining two 1-cycles of the 3-torus lead to the non-geometric fluxes $Q$ and $R$ respectively \cite{Wecht:2007wu}. Hence, one may think of the geometric $f$-flux as of structure constants defined by the algebra of vector fields \eqref{bracket_e}.

It was suggested in \cite{Grana:2008yw} to generalise the construction \eqref{bracket_e} to the case of Double Field Theory using the C-bracket \eqref{C_brckt} that is a natural multiplication of generalised vectors  in Double Field Theory
\begin{equation}
\label{bracket_E}
[E_{\bA},E_{\bB}]_C=F^{\bC}{}_{\bB\bC}E_{\bC},
\end{equation} 
where the barred indices denote flat directions and $E^B_{\bar{B}}$ is a generalised vielbein defined as
\begin{equation}
\mH^{MN}=E^M_{\bar{M}}E^N_{\bar{N}}\mH^{\br{M}{N}}.
\end{equation}
The diagonal form of the flat generalised metric $\mc{H}_{\br{A}{B}}=\mbox{diag}[h_{\br{a}{b}},h^{\br{a}{b}}]$ corresponds to two natural gauge choices for the generalised vielbein 
\begin{equation}
\label{vielbein}
\begin{aligned}
	& \hat{E}^M_{\bA}=
			\begin{bmatrix}
				e^m_{\ba} & & 0\\
					&\\
				e_{\ba}^kB_{nk} & & e^{\bb}_n
			\end{bmatrix}, &&
     \hat{E}_M^{\bA}=
			\begin{bmatrix}
				e_m^{\ba} & & 0\\
					&\\
				e_{\ba}^kB_{nk} & & e_{\bb}^n
			\end{bmatrix},\\
	& \tilde{E}^M_A=
				\begin{bmatrix}
					e^m_{\ba} & & e^{\bb}_k\b^{mn}\\
						&\\
					0 & & e^\bb_n
				\end{bmatrix},  &&
	 \tilde{E}^M_A=
			\begin{bmatrix}
				e_m^{\ba} & & e^{\bb}_k\b^{mn}\\
					&\\
				0 & & e_\bb^n
			\end{bmatrix}.
\end{aligned}
\end{equation}
Although usually the 2-vector $\b^{mn}$ is considered a sign of non-geometry it is not the only source of the non-geometric $Q$ and $R$ fluxes. As it will be shown further, the dual space parametrised by the coordinates $\tilde{y}_i$ itself generates non-geometric fluxes. In other words, even if  $\b^{mn}=0$ the flux $Q^{\ba}{}_{\bb\bc}$ is non-zero and is written in terms of dual derivatives $\tilde{\dt}^m$ of the fields.

One can think of the generalised vielbein as Scherk-Schwarz twist matrices \cite{Grana:2012rr} and the structure constants $F^{\bar{A}}{}_{\br{B}{C}}$ are thus gaugings of the corresponding supergravity \cite{Geissbuhler:2011mx}. 
Recall the explicit expression for gaugings \eqref{F1} written now in term of the generalised vielbein
\begin{equation}
F^{\bA}{}_{\bB\bC}=2E^{\bA}_M E_{[\bB}^N\dt_N E_{\bC]}^M-E^{\bA}_MY^{MN}{}_{KL}\dt_N E_{[\bB}^K E_{\bC]}^L
\end{equation}
The components of the generalised flux $F_{\bA \bB} ^{\bC}$ in the hatted frame when the generalised vielbein is chosen to be $\hat{E}^M_{\bA}$ have the following form
\begin{equation}
\begin{aligned}
F^{\ba}{}_{\bm \bn}&=f^{\ba}{}_{\bm \bn}+2B_{bk}e^{\ba}_{r}e^k_{[\bm}\tdt^be^r_{\bn]}+e^{\ba}_{r}e^k_{\bm}e^l_{\bn}\tdt^r B_{kl}\equiv\mc{F}^{\ba}{}_{\bm \bn},\\
F_{\ba \bm\bn}&=3e^b_{\bm}e^k_{\bn}e^r_{\ba}H_{brk}-3e^k_{[\ba}e^r_{\bm}e^l_{\bn]}B_{bk}\tdt^bB_{rl}\equiv\mc{H}_{\ba \bm\bn},\\
F^{\ba\bm}{}_{\bn}&=2e^{[\bm}_pe^{\ba]}_q\tdt^pe^q_{\bn}\equiv\mc{Q}^{\ba\bm}{}_{\bn},\\
F^{\ba\bm\bn}&=0,
\end{aligned}
\end{equation}
where $f^{\ba}{}_{\bm\bn}=2e^{\ba}_{r}e^b_{[\bm}\dt_be^r_{\bn]}$, $H_{brk}=\dt_{[b}B_{rk]}$ and all antisymmetrisations of $n$ indices include the factor $1/n!$.

The generalised vielbein \eqref{vielbein} can be written in compact notations by making use of the natural basis on the generalised tangent space $\{\dt_a,\tdt^a\}$
\begin{equation}
\begin{aligned}
&\hE^{\ba}=\hE^{\ba a}\dt_a+\hE^{\ba}_{a}\tdt^a=e^{\ba}_a\tdt^a=e^{\ba},\\
&\hE_{\ba}=\hE_{\ba}^{ a}\dt_a+\hE_{\ba a}\tdt^a=e^a_{\ba}\dt_a+e_{\ba}^b B_{ab}\tdt^a=e_{\ba}-i_{e_{\ba}}B,
\end{aligned}
\end{equation}
where $i_X\w$ denotes the intrinsic multiplication of a vector $X$ and a form $\w$. Written in this notation the expression \eqref{bracket_E} defines the fluxes $\mc{F}$, $\mc{H}$ and $\mc{Q}$ that contain the ordinary fluxes $f$, $H$ and $Q$ and terms with derivatives with respect to the dual coordinates $\tx_i$:
\begin{equation}
\begin{aligned}
&[\hE_{\ba},\hE_{\bb}]=\mc{F}^{\bc}{}_{\ba\bb}\hE_{\bc}+\mc{H}_{\ba\bb\bc}\hE^{\bc},\\
&[\hE^{\ba},\hE_{\bb}]=\mc{F}^{\ba}{}_{\bb\bc}\hE^{\bc}+\mc{Q}^{\bc\ba}{}_{\bb}\hE_{\bc},\\
&[\hE^{\ba},\hE^{\bb}]=\mc{Q}^{\ba\bb}{}_{\bc}\hE^{\bc}.
\end{aligned}
\end{equation}
It is useful to show the derivation of these expressions explicitly on an example. Consider the bracket $[E_{\ba},E_{\bb}]$ that can be written in the natural basis as
\begin{equation}
[E_{\ba},E_{\bb}]=[E_{\ba},E_{\bb}]^a\dt_a+[E_{\ba},E_{\bb}]_a\tdt^a.
\end{equation}
Taking into account \eqref{bracket_E} this generalised vector reads
\begin{equation}
\begin{aligned}
[\hE_{\ba},\hE_{\bb}]&= F^{\bc}{}_{\ba\bb}\hE_{\bc}^a\dt_a+F^{\bc}{}_{\ba\bb}\hE_{\bc a}\tdt^a+F_{\bc\ba\bb}\hE^{\bc}_a\tdt^a\\
&=F^{\bc}{}_{\ba\bb}\left(\hE_{\bc}^a\dt_a+\hE_{\bc a}\tdt^a\right)+F_{\bc\ba\bb}\hE^{\bc}_a\tdt^a\\
&=F^{\bc}{}_{\ba\bb}\hE_{\bc}+F_{\bc\ba\bb}\hE^{\bc}.
\end{aligned}
\end{equation}

In the tilde gauge where the Kalb-Ramond field is zero $B_{mn}=0$ but the two-vector $\b^{mn}$ is non-zero the situation is very similar. The components of the generalised flux then have the following form
\begin{eqnarray}
\begin{aligned}
F^{\ba}{}_{\bm\bn}&=f^{\ba}{}_{\bm\bn}\equiv\mc{F}^{\ba}{}_{\bm\bn},\\
F^{\ba\bm}{}_{\bn}&=2e^{[\ba}_pe^{bm}_q\tdt^qe^p_{\bn}+2e^{[\ba}_re^{\bm]}_m\dt_be^r_{\bn}\b^{bm}-e^{\ba}_re^{\bm}_me^b_{\bn}\dt_b\b^{rm}\equiv\mc{Q}^{\ba\bm}{}_{\bn},\\
F^{\ba\bm\bn}&=-3e^{\ba}_re^{\bm}_me^{\bn}_n\tdt^{[a}\b^{mn]}+3e^{\ba}_re^{\bm}_me^{\bn}_n\b^{bm}\dt_b\b^{rn}\equiv\mc{R}^{\ba\bm\bn},\\
F_{\ba\bm\bn}&=0.
\end{aligned}
\end{eqnarray}
Then the corresponding commutation relations involving the components of the generalised vielbein in the tilde gauge read
\begin{equation}
\begin{aligned}
&[\tE_{\ba},\tE_{\bb}]=\mc{F}^{\bc}{}_{\ba\bb}\tE_{\bc},\\
&[\tE^{\ba},\tE_{\bb}]=\mc{F}^{\ba}{}_{\bb\bc}\tE^{\bc}+\mc{Q}^{\bc\ba}{}_{\bb}\tE_{\bc},\\
&[\hE^{\ba},\hE^{\bb}]=\mc{Q}^{\ba\bb}{}_{\bc}\tE^{\bc}+\mc{R}^{\ba\bb\bc}\tE_{\bc}.
\end{aligned}
\end{equation}

The above expressions show that the natural frames correspond to backgrounds with various fluxes. It is worth mentioning that the bivector $\b^{mn}$ is not the only sign of non-geometry as the fluxes $\mc{Q}$ and $\mc{R}$ are non-zero even for vanishing $\b$. Dependence of the background on the dual coordinates itself leads to non-zero non-geometric fluxes.

\section{Towards the monodromy properties of Double Field Theory}

Consider the $5_2^2$ background written in the $\b$-frame \eqref{522_beta}. The metric does not have any non-trivial monodromy properties, though the background is still non-geometric due to the presence of the bivector field. This suggests that solutions that we normally call non-geometric are characterized by non-trivial monodromies only when written in terms of the ordinary coordinates $x^i$. It is natural then to expect that backgrounds purely geometric in terms of the ordinary coordinates acquire non-trivial monodromies in terms of the dual coordinates. 

In the extended geometry formalism and in particular in Double Field Theory one is allowed to glue patches by a generalised diffeomorphism. Action of these transition functions on fibers is represented by the corresponding duality group. Hence, one is allowed to glue say the KK monopole with the $5_2^2$ brane. In general one is able to describe this solution globally in terms of both conventional and dual coordinates. This is simply the idea of section condition that projects the extended space to a known supergravity solution. 

The T-duality transformation that turns f-flux into Q-flux can be represented as an $O(d,d)$ rotation that swaps an ordinary coordinate with the corresponding dual one. Indeed, consider a background with non-zero f-flux. This can be represented by the twisted torus 
\begin{equation}
\label{torus_f}
\begin{aligned}
ds_f^2&=(dx-N z dy)^2+dy^2+dz^2, \\
B&=0.
\end{aligned}
\end{equation}
Using Buscher rules one can show that a T-duality transformation in the $y$ direction brings us to a non-geometric background
\begin{equation}
\label{torus_Q}
\begin{aligned}
ds_Q^2&=\fr{1}{1-N^2 z^2}(dx^2+dy^2)+dz^2, \\
B&=\fr{N z}{1+N^2 z^2}dx \wedge dy.
\end{aligned}
\end{equation}
Since the metric is mixed with the gauge field as we shift $z\to z+1$ this background is said to be non-geometric. 
The transformation that turns the first background to the second one is given by an $O(2,2)$ rotation in the $(y,\tilde{y})$ plane
\begin{equation}
\label{Odd_rot}
\mc{O}=
\begin{bmatrix}
1 & 0 & 0 & 0 \\
0 & 0 & 0 & 1 \\
0 & 0 & 1 & 0 \\
0 & 1 & 0 & 0 \\
\end{bmatrix}.
\end{equation}
It is straightforward to check that the corresponding transformation of the generalised metric $\mc{H}\to \mc{O}^{tr}\mc{H}\mc{O}$ turns the background \eqref{torus_f} to \eqref{torus_Q}.

This simple example suggests to consider a solution of Double Field Theory equations that depends on both dual and ordinary coordinates. Section condition then defines a subspace of the extended space that is represented by backgrounds with different fluxes glued by a generalised diffeomorphism. On the way of realising these kind of solutions one has to understand how to write finite coordinate transformation properly \cite{Hohm:2012gk}.

One should not confuse the rotation \eqref{Odd_rot} that relates to different backgrounds with the transformation \eqref{beta_transf} that represents the monodromy $\q \to \q+2\p$ of the $5_2^2$-brane although these are related.

To investigate the monodromy properties of non-geometric backgrounds one should notice that an $O(d,d)$ transformation in general breaks the upper triangular form of the generalised vielbein $E^{\bA}_A$. One ends up with both the $B$-field and the $\b$-tensor that breaks the degrees of freedom count. To restore the counting one performs an $O(d)\times O(d)$ transformation of an appropriate form
\begin{equation}
E^{\bA}_{A}\to \mc{O}_A{}^B E^{\bB}_{B} g^{\bA}{}_{\bB}, \quad g \in O(d)\times O(d).
\end{equation}
Although the boundary term $S_B$ naively seems to be invariant under these transformation it is sensitive to its global properties. In the same sense the term $\int F = \int dA$ counts windings of the circle $\mathbb{S}^1$ around the $U(1)$. 

To see this one notes that an element $\L(x^i,\tx_i)$ of $O(d)\times O(d)$ defines a map
\begin{equation}
\L : \mathbb{T}^{m}\times \tilde{\mathbb{T}}^n \longrightarrow O(d)\times O(d),
\end{equation}
where the torus $\mathbb{T}^{m}\times \tilde{\mathbb{T}}^n$ represents the boundary $\mathbb{T}^{d-1}\times \tilde{\mathbb{T}}^d$ or $\mathbb{T}^{d}\times \tilde{\mathbb{T}}^{d-1}$ depending on the cycle chosen.

Since the first fundamental group of $O(d)$ is non-zero $\p_1[O(d)]=\mathbb{Z}_2$ the map above is classified by $\mathbb{Z}_2^{m+n}$. Hence, depending on the global properties of the $O(d)\times O(d)$ rotation the boundary term is non-zero \cite{Berman:2013ab}.

\chapter{Conclusion}

To conclude, we briefly review the results presented above and note possible directions of further research. The main topic of this thesis is the framework of extended geometry which has appeared recently, which provides a duality covariant description of the low energy dynamics of string and M-theory. The most important feature of this approach is the extension of space by dual coordinates that reflect the extended nature of fundamental objects in string and M-theory. In particular this is necessary to describe non-geometric backgrounds in geometric terms. 

It is shown above that dimensional reductions of extended space by Scherk-Schwarz U-duality valued twist matrices successfully reproduce structures of gauged supergravities. In particular, the embedding tensor which defines consistent deformations of toroidally reduced supergravities appears as a structure constant of an algebra generated by C-bracket. An intriguing feature of this approach is that the section condition which has the form of differential equations becomes replaced by algebraic constraints on the embedding tensor. These were known before as quadratic constraints of gauged supergravity. These algebraic equations restrict the form of the twist matrices and hence geometry of the extended space. It is a fascinating task to obtain the twist matrices in a manifest form by solving these constraints and to  compare the corresponding embedding tensor with the known solutions \cite{Pernici:1984xx,Pernici:1984zw, deWit:2004nw,Bergshoeff:2007ef,Samtleben:2005bp}.

We have demonstrated that in the case when the extended space is taken to be internal, the effective potential of extended geometry becomes precisely the scalar potential of gauged supergravity. To obtain this one has to add a term which is zero up to the section condition, but is crucial for the invariance of the effective potential. The generalised metric then collects the scalar moduli in a duality covariant way.

The tensor sector of gauged supergravities was not considered in this work and may be approached by further research on generalised Scherk-Schwarz reductions. One can straightforwardly include the tensor sector by taking a part of the extended space to be internal. Then the remaining directions will give space-time tensors or at least some of their components. The whole picture can always be restored using general covariance arguments. A fascinating task would be to investigate if the generalised reduction naturally provides a non-abelian structure of gauged supergravities \cite{deWit:2005hv, deWit:2008ta}. 

In Chapter \ref{bound} we presented the form of the actions for Double Field Theory and for the $SL(5)$ covariant formulation of M-theory that reduce to the known supergravity actions plus the Gibbons-Hawking term. The derived actions naturally contain all boundary terms and consistently reproduce the equations of motion of General Relativity. To write the boundary term in duality covariant notations we introduce a generalised normal, that is defined by consistency conditions with respect to the strong constraint and normalisation.

We showed manifestly that the boundary term is non-zero for backgrounds with non-trivial monodromies. These are typically generated by the so-called exotic branes that appear in string/M-theory compactifications to lower dimensions with non-geometry turned on. The boundary term takes the form of magnetic coupling of the Q and f fluxes to the B-field and the bivector $\b$ respectively. At the end of this chapter we calculate the boundary contribution for the background generated by the $5_2^2$-brane. For a review of exotic branes see \cite{deBoer:2012ma}.

It is suggestive to continue the results of Chapter \ref{bound} to M-theory exotic branes and explicitly calculate the boundary term for the corresponding backgrounds. One may in general assume that the topological contribution to the effective action can be written in a duality covariant way for all U-duality groups in a similar way as was done for the $SL(5)$ geometry.

\chapter{Appendix}

This Appendix consists mainly of calculations, that are necessary but too detailed to be presented in the main sections. In addition some interesting results, that do not fit organically to the main narrative thread, are collected here.


\section{$SO(5,5)$ gaugings}

While the trombone is obtained in a straightforward way from the gauge group generators $X_{MN}{}^K$ one has to do some algebra to get the remained gauging $\q^{iM}$. This section is to show how this gauging can be obtained by suitable projections of the gauge group generators.

The gauge group generators $X_{MN}{}^K$ evaluated in the representation $\mc{R}_V$ have the form
\begin{equation}
X_{MN}{}^K=\Q_M^\a (t_\a)_N{}^K+\left(\fr{16}{5} (t^\a)_M{}^P(t_\a)_N{}^K + \d_M^P\d_N^K\right)\q_{P},
\end{equation}
where $t_\a$ are the generators of the global duality group and are given by $(\G_{ij})_N{}^K$. The embedding tensor reads
\begin{equation}
\Q_M{}^{ij}=-\q^{L[i}\G^{j]}{}_{LM}.
\end{equation}
Thus the gaugings can be explicitly separated out by the following contractions
\begin{equation}
\label{XcontrSO}
X_{MN}{}^K(\G^i{}_j)_K{}^N\G^{jMR}=128\,\q^{iR}-\fr{144}{5}\G^{iRS}\q_S.
\end{equation}
By making use of the first line of the definitions \eqref{Gm} one can show that the generators \eqref{Xgen} contracted in the same way give exactly \eqref{XcontrSO} with gaugings defined as \eqref{gaugingsSO}. Indeed, lets rewrite the generators $X_{MN}{}^K$ using the second line of \eqref{rel}
\begin{equation}
X_{MN}{}^K=f_{MN}{}^K+\fr18(\G_{ij})_N{}^K(\G^{ij})_C{}^Bf_{BM}{}^C+\fr14\d_N^Kf_{BM}{}^B.
\end{equation}
Contracting with the generator and the gamma matrix as in \eqref{XcontrSO} we obtain
\begin{equation}
X_{MN}{}^K(\G^i{}_j)_K{}^N\G^{jMR}=(f_{MN}{}^K-4f_{NM}{}^K)(\G^i{}_j)_K{}^N\G^{jMR}
\end{equation}
To show that this is exactly \eqref{XcontrSO} one needs to do some simple algebra and use the following identities
\begin{equation}
\begin{split}
\q^{iM}&=-\fr14\G^{jMD}f_{Dj}{}^i-\fr25\G^{iMN}\q_N,\\
f_{AP}{}^R&=\fr14\G_{iPQ}\G^{jQR}f_{Aj}{}^i-\fr14f_{AB}{}^B\d_P^R,\\
f_{Aj}{}^i\G_{i}{}^{AB}\h^{jk}&=4\q^{kB}+\fr85\G^{kBC}\q_C+\fr14\G^{kAB}f_{AR}{}^R,\\
Y^{BK}_{CL}f_{AK}{}^L&=-3f_{AC}{}^B-2f_{KA}{}^K\d_C^B-8\d_C^B\q_A.
\end{split}
\end{equation}
Here the first line is just a rewriting of \eqref{gaugingsSO}, the second line is a consequence of the definition \eqref{Gm} and the last line here. Finally, the third and the last lines are obtained directly by making use of  properties of twist matrices.

\section{Effective potential for $SO(5,5)$ case}

Since the generalised metric $M_{MN}$ is a coset representative we identify it with the unimodular matrix of \cite{Bergshoeff:2007ef} that has the same meaning and is defined as
\begin{equation}
\label{M_V}
M_{MN}={V}_{M}{}^{\a\dot{\a}}{V}_{N}{}^{\b\dot{\b}}\W_{\a\b}\W_{\dot{\a}\dot{\b}},
\end{equation}
where $\W_{\a\b}$ and $\W_{\dot{\a}\dot{\b}}$ are the symplectic invariants of $Spin(4)$ corresponding to each $SO(5)$  in the coset. These matrices are antisymmetric $\W_{\a\b}=-\W_{\b\a}$ and are used to raise and lower spinor indices $\W_{\a\b}\W^{\b\m}=\d_\a{}^\m$. The matrices $V_M^{\a\dot{\a}}$ are coset representatives of 
\begin{equation}
\fr{SO(5,5)}{SO(5)\times SO(5)}.
\end{equation}

Recall the effective potential \eqref{VSOt} that comes from Scherk-Schwarz reduction of M-theory in the extended space formalism
\begin{equation}
\label{VSOt1}
V_{eff}=-\fr{1}{8}X_{MK}{}^LX_{NL}{}^KM^{MN}+\fr{1}{32}X_{MR}{}^KX_{NS}{}^LM^{MN}M^{RS}M_{KL}.
\end{equation}
To show that this expression exactly reproduces the scalar potential of $D=6$ gauged supergravity one needs the following relation
\begin{equation}
\label{Gamma_V}
\begin{aligned}
\G_{iAB}V^{B\b\dot{\b}}=\mc{V}_i^{\hat{a}}(\g^{\hat{a}})_\a{}^\b V_A^{\a\dot{\b}}-
\mc{V}_i^{\hat{\dot{a}}}(\g^{\hat{\dot{a}}})_{\dot{\a}}{}^{\dot{\b}} V_A^{\b\dot{\a}},
\end{aligned}
\end{equation}
that follows from the invariance of the $SO(5,5)$ gamma-matrices \cite{Bergshoeff:2007ef}.

Consider the first term of the potential since it is easier to proceed. The calculations for the second term are longer but the idea is the same. In the absence of the trombone gauging the structure constants read
\begin{equation}
X_{MN}{}^K=-\q^{iL}\G^j_{LM}(\G_{ij})_N{}^K.
\end{equation}
Taking into account the quadratic constraint $\q^{iM}\q^{jN}\h_{ij}=0$, where $\h_{ij}$ is 10-dimensional flat metric and simple gamma-matrix algebra, one can write
\begin{equation}
V_1=-\fr18X_{MK}{}^LX_{NL}{}^KM^{MN}=-2\,\q^{iA}\q^{kB}\G_{iBN}\G_{kAM}M^{MN}.
\end{equation}
The next step is to substitute the explicit expression of the generalised metric $M_{MN}$ in terms of the coset representatives \eqref{M_V} and use the identity \eqref{Gamma_V}. This gives
\begin{equation}
\begin{aligned}
V_1=-2\q^{iA}\q^{kB}&\left(\mc{V}_i^{\hat{a}}(\g^{\hat{a}})_\m{}^\a V_B^{\m\dot{\a}}-
\mc{V}_i^{\hat{\dot{a}}}(\g^{\hat{\dot{a}}})_{\dot{\m}}{}^{\dot{\a}} V_B^{\a\dot{\m}}\right)\times\\
&\left(\mc{V}_k^{\hat{b}}(\g^{\hat{b}})_\n{}^\b V_A^{\n\dot{\b}}-
\mc{V}_k^{\hat{\dot{b}}}(\g^{\hat{\dot{b}}})_{\dot{\n}}{}^{\dot{\b}} V_A^{\b\dot{\n}}\right)\W_{\a\b}\W_{\dot{\a}\dot{\b}}.
\end{aligned}
\end{equation}
Using the definition of the T-tensor \eqref{T_tensor} this expression can be written only in terms of $(T^{\hat{a}})_{\a\dot{\a}}$ and $(T^{\hat{\dot{a}}})_{\a\dot{\a}}$
\begin{equation}
\begin{aligned}
V_1=&2(T^{\hat{a}})_{\n\dot{\a}}(T^{\hat{b}})^{\m\dot{\a}}(\g^{\hat{a}})_{\m}{}^{{\a}}(\g^{\hat{b}})_{\a}{}^{{\n}}
-2(T^{\hat{\dot{a}}})_{\n\dot{\a}}(T^{\hat{b}})^{\a\dot{\m}}(\g^{\hat{\dot{a}}})_{\dot{\m}}{}^{\dot{\a}}(\g^{\hat{b}})_{\a}{}^{{\n}}-\\
&2(T^{\hat{{a}}})_{\a\dot{\n}}(T^{\hat{\dot{b}}})^{\m\dot{\a}}(\g^{\hat{{a}}})_{{\m}}{}^{{\a}} (\g^{\hat{\dot{b}}})_{\dot{\a}}{}^{\dot{\n}}+2
(T^{\hat{\dot{a}}})_{\a\dot{\n}}(T^{\hat{\dot{b}}})^{\a\dot{\m}}(\g^{\hat{\dot{a}}})_{\dot{\m}}{}^{\dot{\a}} (\g^{\hat{\dot{b}}})_{\dot{\a}}{}^{\dot{\n}},
\end{aligned}
\end{equation}
where one should note that the matrices $\g^{\hat{a}}$ and $\g^{\hat{\dot{a}}}$ are antisymmetric. Reversing the order of the gamma matrices in the first and the last terms one obtains
\begin{equation}
\begin{aligned}
V_1=&4(T^{\hat{a}})_{\a\dot{\a}}(T^{\hat{a}})^{\a\dot{\a}}-4T_{\a\dot{\a}}T^{\a\dot{\a}}+4(T^{\hat{a}})_{\a\dot{\a}}(T^{\hat{a}})^{\a\dot{\a}}\\
=&8\Tr\left[T^{\hat{a}}\tilde{T}^{\hat{a}}-\fr12T\tilde{T}\right],
\end{aligned}
\end{equation}
where the identities in the last line of \eqref{T_tensor} and \eqref{TT} were used and the tilde denotes transposition that implies $\Tr\left[T\tilde{T}\right]\equiv T_{\a\dot{\a}}T^{\a\dot{\a}}$.

The same but longer calculation shows that the second term in the potential $V_2$ gives the same expression up to  prefactor
\begin{equation}
V_2=-2\Tr\left[T^{\hat{a}}\tilde{T}^{\hat{a}}-\fr12T\tilde{T}\right].
\end{equation}
Together $V_1$ and $V_2$ result in
\begin{equation}
V_{eff}=6\Tr\left[T^{\hat{a}}\tilde{T}^{\hat{a}}-\fr12T\tilde{T}\right]=6V_{scalar}.
\end{equation}

\section{Invariance of the action}

In this section we show the details of the proof that the actions \eqref{VSOt} and \eqref{VEt} are invariant under the gauge transformations \eqref{G_transf}. In the dynamical picture of the extended geometry the action is invariant due to the section condition. In the Scherk-Schwarz reduction of the theory the invariance of the action is assured by the quadratic constraint \eqref{closure2}.

The terms that contribute to the effective potentials in $d=5,6$ 
\begin{equation}
\label{terms}
\begin{split}
&X_{MK}{}^LX_{NL}{}^KM^{MN},\\
&X_{MR}{}^KX_{NS}{}^LM^{MN}M^{RS}M_{KL},\\
&X_{RM}{}^KX_{NS}{}^LM^{MN}M^{RS}M_{KL}
\end{split}
\end{equation}
are invariant separately. Let us start with the first term whose transformation gives
\begin{equation}
\begin{split}
&\d_\S\left(X_{MK}{}^LX_{NL}{}^KM^{MN}\right)=2X_{MK}{}^LX_{NL}{}^KX_{RS}{}^M\S^RM^{SN}=\\
&=-2[X_R,X_S]_K{}^LX_{NL}{}^K\S^RM^{SN}=-4\Tr[X_R,X_S,X_N]\S^{[R}M^{S]N}=\\
&=-2X_{SN}{}^K\Tr[X_K,X_R]\S^RM^{SN}=-2X_{(SN)}{}^KX_{KP}{}^QX_{RQ}{}^{P}\S^RM^{SN}=0,
\end{split}
\end{equation}
where we used the closure constraint \eqref{closure2} in the first line and cyclic symmetry of the trace in the second line. The last step here exploits the condition $X_{(AB)}{}^CX_{CK}{}^L=0$. 

For the transformation of the second term we have
\begin{equation}
\label{Vtransf1}
\begin{split}
&\d_\S(X_{MR}{}^KX_{NS}{}^LM^{MN}M^{RS}M_{KL})=2X_{MR}{}^K X_{NS}{}^LX_{PQ}{}^M\S^P M^{QN}M^{RS}M_{KL}+\\
&+2X_{MR}{}^K X_{NS}{}^LX_{PQ}{}^M\S^P M^{MN}M^{QS}M_{KL}-2X_{MR}{}^K X_{NS}X_{PK}{}^Q\S^{P}M^{MN}M^{RS}M^{QL}.
\end{split}
\end{equation}
After relabelling the indices the last two terms can be recast in the following form
\begin{equation}
\begin{split}
&(X_{MR}{}^KX_{PQ}{}^R-X_{MQ}{}^RX_{PR}{}^K)X_{NS}{}^L\S^{P}M^{MN}M^{QS}M_{KL}=\\
&=\left(X_P X_M-X_MX_P\right)_Q{}^K\S^{P}M^{MN}M^{QS}M_{KL}=\\
&=-X_{PM}{}^RX_{RQ}{}^K\S^{P}M^{MN}M^{QS}M_{KL}.
\end{split}
\end{equation}
This is exactly the first term in \eqref{Vtransf1} but with the opposite sign. Thus the second term in \eqref{terms} is invariant under the gauge transformations. The proof of the invariance of the third term is exactly the same.

 
\bibliographystyle{utphys}
\bibliography{thesis}

\providecommand{\href}[2]{#2}\begingroup\raggedright\begin{thebibliography}{10%
0}

\bibitem{Berman:2011kg}
D.~S. Berman, E.~T. Musaev, and M.~J. Perry, ``{Boundary Terms in Generalized
  Geometry and doubled field theory},''
  \href{http://dx.doi.org/10.1016/j.physletb.2011.11.019}{{\em Phys.Lett.}
  {\bfseries B706} (2011) 228--231},
\href{http://arxiv.org/abs/1110.3097}{{\ttfamily arXiv:1110.3097 [hep-th]}}.

\bibitem{Berman:2012uy}
D.~S. Berman, E.~T. Musaev, D.~C. Thompson, and D.~C. Thompson, ``{Duality
  Invariant M-theory: Gauged supergravities and Scherk-Schwarz reductions},''
  \href{http://dx.doi.org/10.1007/JHEP10(2012)174}{{\em JHEP} {\bfseries 1210}
  (2012) 174},
\href{http://arxiv.org/abs/1208.0020}{{\ttfamily arXiv:1208.0020 [hep-th]}}.

\bibitem{Musaev:2013rq}
E.~T. Musaev, ``{Gauged supergravities in 5 and 6 dimensions from generalised
  Scherk-Schwarz reductions},''
  \href{http://dx.doi.org/10.1007/JHEP05(2013)161}{{\em JHEP} {\bfseries 1305}
  (2013) 161},
\href{http://arxiv.org/abs/1301.0467}{{\ttfamily arXiv:1301.0467 [hep-th]}}.

\bibitem{Berman:2013ab}
D.~S. Berman and E.~T. Musaev, ``{Boundary terms and non-geometry in Double
  Field Theory},'' \href{http://arxiv.org/abs/to appear}{{\ttfamily to
  appear}}.

\bibitem{Bolmatov:2013kpa}
D.~Bolmatov, E.~T. Musaev, and K.~Trachenko, ``{Symmetry breaking gives rise to
  three states of matter},'' \href{http://dx.doi.org/10.1038/srep02794}{{\em
  Sci. Rep.} {\bfseries 3} (2013) },
\href{http://arxiv.org/abs/1306.1892}{{\ttfamily arXiv:1306.1892
  [cond-mat.str-el]}}.

\bibitem{Green:1987sp}
M.~B. Green, J.~Schwarz, and E.~Witten, {\em {Superstring theory. Vol. 1:
  Introduction}}.
\newblock Cambridge University Press, Cambridge, England, 1987.

\bibitem{Fradkin:1984ai}
E.~Fradkin and A.~A. Tseytlin, ``{QUANTUM EQUIVALENCE OF DUAL FIELD
  THEORIES},''
\href{http://dx.doi.org/10.1016/0003-4916(85)90225-8}{{\em Annals Phys.}
  {\bfseries 162} (1985) 31}.

\bibitem{Kikkawa:1984cp}
K.~Kikkawa and M.~Yamasaki, ``{Casimir effects in superstring theories},''
  \href{http://dx.doi.org/10.1016/0370-2693(84)90423-4}{{\em Phys.Lett.}
  {\bfseries B149} (1984) 357}.

\bibitem{Sakai:1985cs}
N.~Sakai and I.~Senda, ``{Vacuum energies of string compactified on torus},''
  \href{http://dx.doi.org/10.1143/PTP.75.692, 10.1143/PTP.75.692}{{\em
  Prog.Theor.Phys.} {\bfseries 75} (1986) 692}.

\bibitem{Vafa:1997pm}
C.~Vafa, ``{Lectures on strings and dualities},''
  \href{http://arxiv.org/abs/hep-th/9702201}{{\ttfamily arXiv:hep-th/9702201
  [hep-th]}}.

\bibitem{Giveon:1994fu}
A.~Giveon, M.~Porrati, and E.~Rabinovici, ``{Target space duality in string
  theory},'' \href{http://dx.doi.org/10.1016/0370-1573(94)90070-1}{{\em
  Phys.Rept.} {\bfseries 244} (1994) 77--202},
  \href{http://arxiv.org/abs/hep-th/9401139}{{\ttfamily arXiv:hep-th/9401139
  [hep-th]}}.

\bibitem{Obers:1998fb}
N.~Obers and B.~Pioline, ``{U duality and M theory},''
  \href{http://dx.doi.org/10.1016/S0370-1573(99)00004-6}{{\em Phys.Rept.}
  {\bfseries 318} (1999) 113--225},
\href{http://arxiv.org/abs/hep-th/9809039}{{\ttfamily arXiv:hep-th/9809039
  [hep-th]}}.

\bibitem{Schwarz:1998fd}
J.~H. Schwarz, ``{Introduction to M theory and AdS / CFT duality},''
\href{http://arxiv.org/abs/hep-th/9812037}{{\ttfamily arXiv:hep-th/9812037
  [hep-th]}}.

\bibitem{delMoral:2012pr}
M.~G. del Moral, ``{Dualities as symmetries of the Supermembrane Theory},''
\href{http://arxiv.org/abs/1211.6265}{{\ttfamily arXiv:1211.6265 [hep-th]}}.

\bibitem{Hull:1994ys}
C.~Hull and P.~Townsend, ``{Unity of superstring dualities},''
  \href{http://dx.doi.org/10.1016/0550-3213(94)00559-W}{{\em Nucl.Phys.}
  {\bfseries B438} (1995) 109--137},
  \href{http://arxiv.org/abs/hep-th/9410167}{{\ttfamily arXiv:hep-th/9410167
  [hep-th]}}.

\bibitem{Grana:2008yw}
M.~Grana, R.~Minasian, M.~Petrini, and D.~Waldram, ``{T-duality, Generalized
  Geometry and Non-Geometric Backgrounds},''
  \href{http://dx.doi.org/10.1088/1126-6708/2009/04/075}{{\em JHEP} {\bfseries
  0904} (2009) 075},
\href{http://arxiv.org/abs/0807.4527}{{\ttfamily arXiv:0807.4527 [hep-th]}}.

\bibitem{Becker:2003yv}
K.~Becker, M.~Becker, K.~Dasgupta, and P.~S. Green, ``{Compactifications of
  heterotic theory on nonKahler complex manifolds. 1.},'' {\em JHEP} {\bfseries
  0304} (2003) 007,
\href{http://arxiv.org/abs/hep-th/0301161}{{\ttfamily arXiv:hep-th/0301161
  [hep-th]}}.

\bibitem{Becker:2003sh}
K.~Becker, M.~Becker, P.~S. Green, K.~Dasgupta, and E.~Sharpe,
  ``{Compactifications of heterotic strings on nonKahler complex manifolds.
  2.},'' \href{http://dx.doi.org/10.1016/j.nuclphysb.2003.11.029}{{\em
  Nucl.Phys.} {\bfseries B678} (2004) 19--100},
\href{http://arxiv.org/abs/hep-th/0310058}{{\ttfamily arXiv:hep-th/0310058
  [hep-th]}}.

\bibitem{Tseytlin:1990nb}
A.~A. Tseytlin, ``{Duality symmetric formulation of string world sheet
  dynamics},'' \href{http://dx.doi.org/10.1016/0370-2693(90)91454-J}{{\em
  Phys.Lett.} {\bfseries B242} (1990) 163--174}.

\bibitem{Tseytlin:1990ar}
A.~A. Tseytlin and P.~C. West, ``{TWO REMARKS ON CHIRAL SCALARS},''
\href{http://dx.doi.org/10.1103/PhysRevLett.65.541}{{\em Phys.Rev.Lett.}
  {\bfseries 65} (1990) 541--542}.

\bibitem{Duff:1989tf}
M.~Duff, ``{Duality rotations in string theory},''
  \href{http://dx.doi.org/10.1016/0550-3213(90)90520-N}{{\em Nucl.Phys.}
  {\bfseries B335} (1990) 610}.

\bibitem{Aldazabal:2013sca}
G.~Aldazabal, D.~Marques, and C.~Nunez, ``{Double Field Theory: A Pedagogical
  Review},''
\href{http://arxiv.org/abs/1305.1907}{{\ttfamily arXiv:1305.1907 [hep-th]}}.

\bibitem{Berman:2013eva}
D.~S. Berman and D.~C. Thompson, ``{Duality Symmetric String and M-Theory},''
\href{http://arxiv.org/abs/1306.2643}{{\ttfamily arXiv:1306.2643 [hep-th]}}.

\bibitem{Hohm:2013bwa}
O.~Hohm, D.~Lust, and B.~Zwiebach, ``{The Spacetime of Double Field Theory:
  Review, Remarks, and Outlook},''
\href{http://arxiv.org/abs/1309.2977}{{\ttfamily arXiv:1309.2977 [hep-th]}}.

\bibitem{Brink:1976sc}
L.~Brink, P.~Di~Vecchia, and P.~S. Howe, ``{A Locally Supersymmetric and
  Reparametrization Invariant Action for the Spinning String},''
\href{http://dx.doi.org/10.1016/0370-2693(76)90445-7}{{\em Phys.Lett.}
  {\bfseries B65} (1976) 471--474}.

\bibitem{Deser:1976rb}
S.~Deser and B.~Zumino, ``{A Complete Action for the Spinning String},''
\href{http://dx.doi.org/10.1016/0370-2693(76)90245-8}{{\em Phys.Lett.}
  {\bfseries B65} (1976) 369--373}.

\bibitem{Polchinski:1998rq}
J.~Polchinski, {\em {String theory. Vol. 1: An introduction to the bosonic
  string}}.
\newblock Cambridge University Press, Cambridge, England, 1998.

\bibitem{Tong:2009np}
D.~Tong, ``{String theory},'' \href{http://arxiv.org/abs/0908.0333}{{\ttfamily
  arXiv:0908.0333 [hep-th]}}.

\bibitem{Akhmedov:2009zz}
E.~Akhmedov, ``{Review of modern string theory},''
\href{http://dx.doi.org/10.1134/S106377880909021X}{{\em Phys.Atom.Nucl.}
  {\bfseries 72} (2009) 1574--1600}.

\bibitem{Buscher:1985kb}
T.~Buscher, ``{Quantum corrections and extended supersymmetry in new sigma
  models},'' \href{http://dx.doi.org/10.1016/0370-2693(85)90870-6}{{\em
  Phys.Lett.} {\bfseries B159} (1985) 127}.

\bibitem{Buscher:1987sk}
T.~Buscher, ``{A symmetry of the string background field equations},''
  \href{http://dx.doi.org/10.1016/0370-2693(87)90769-6}{{\em Phys.Lett.}
  {\bfseries B194} (1987) 59}.

\bibitem{Buscher:1987qj}
T.~Buscher, ``{Path integral derivation of quantum duality in nonlinear sigma
  models},'' \href{http://dx.doi.org/10.1016/0370-2693(88)90602-8}{{\em
  Phys.Lett.} {\bfseries B201} (1988) 466}.

\bibitem{Duff:1990hn}
M.~Duff and J.~Lu, ``{Duality rotations in membrane theory},''
  \href{http://dx.doi.org/10.1016/0550-3213(90)90565-U}{{\em Nucl.Phys.}
  {\bfseries B347} (1990) 394--419}.

\bibitem{Bakhmatov:2011ab}
I.~Bakhmatov, ``{Fermionic T-duality and U-duality in type II supergravity},''
\href{http://arxiv.org/abs/1112.1983}{{\ttfamily arXiv:1112.1983 [hep-th]}}.

\bibitem{Bakhmatov:2011be}
I.~Bakhmatov, D.~Berman, and E.~Musaev, ``{Generalised metric for a system of
  D1+D3 branes},'' \href{http://arxiv.org/abs/unpublished}{{\ttfamily
  unpublished}}.

\bibitem{Berman:2010is}
D.~S. Berman and M.~J. Perry, ``{Generalized geometry and M theory},''
  \href{http://arxiv.org/abs/1008.1763}{{\ttfamily arXiv:1008.1763 [hep-th]}}.

\bibitem{Giveon:1988tt}
A.~Giveon, E.~Rabinovici, and G.~Veneziano, ``{Duality in string background
  space},'' \href{http://dx.doi.org/10.1016/0550-3213(89)90489-6}{{\em
  Nucl.Phys.} {\bfseries B322} (1989) 167}.

\bibitem{Hitchin:2004ut}
N.~Hitchin, ``{Generalized Calabi-Yau manifolds},''
  \href{http://dx.doi.org/10.1093/qjmath/54.3.281}{{\em Quart.J.Math.Oxford
  Ser.} {\bfseries 54} (2003) 281--308},
  \href{http://arxiv.org/abs/math/0209099}{{\ttfamily arXiv:math/0209099
  [math-dg]}}.

\bibitem{Hitchin:2005in}
N.~Hitchin, ``{Brackets, forms and invariant functionals},''
  \href{http://arxiv.org/abs/math/0508618}{{\ttfamily arXiv:math/0508618
  [math-dg]}}. dedicated to the memory of Shiing-Shen Chern.

\bibitem{Gualtieri:2003dx}
M.~Gualtieri, ``{Generalized complex geometry},''
  \href{http://arxiv.org/abs/math/0401221}{{\ttfamily arXiv:math/0401221
  [math-dg]}}. Ph.D. Thesis (Advisor: Nigel Hitchin).

\bibitem{Cavalcanti:2011wu}
G.~R. Cavalcanti and M.~Gualtieri, ``{Generalized complex geometry and
  T-duality},''
\href{http://arxiv.org/abs/1106.1747}{{\ttfamily arXiv:1106.1747 [math.DG]}}.

\bibitem{courant1990dm}
T.~J. Courant, ``{Dirac manifolds},'' {\em Trans. Amer. Math. Soc} {\bfseries
  319} no.~2, (1990) 631--661.

\bibitem{Jeon:2010rw}
I.~Jeon, K.~Lee, and J.-H. Park, ``{Differential geometry with a projection:
  Application to double field theory},''
  \href{http://dx.doi.org/10.1007/JHEP04(2011)014}{{\em JHEP} {\bfseries 1104}
  (2011) 014},
\href{http://arxiv.org/abs/1011.1324}{{\ttfamily arXiv:1011.1324 [hep-th]}}.

\bibitem{Jeon:2011cn}
I.~Jeon, K.~Lee, and J.-H. Park, ``{Stringy differential geometry, beyond
  Riemann},'' \href{http://dx.doi.org/10.1103/PhysRevD.84.044022}{{\em
  Phys.Rev.} {\bfseries D84} (2011) 044022},
\href{http://arxiv.org/abs/1105.6294}{{\ttfamily arXiv:1105.6294 [hep-th]}}.

\bibitem{Jeon:2011sq}
I.~Jeon, K.~Lee, and J.-H. Park, ``{Supersymmetric Double Field Theory: Stringy
  Reformulation of Supergravity},''
  \href{http://dx.doi.org/10.1103/PhysRevD.86.089903,
  10.1103/PhysRevD.85.081501, 10.1103/PhysRevD.85.089908}{{\em Phys.Rev.}
  {\bfseries D85} (2012) 081501},
\href{http://arxiv.org/abs/1112.0069}{{\ttfamily arXiv:1112.0069 [hep-th]}}.

\bibitem{Hohm:2011si}
O.~Hohm and B.~Zwiebach, ``{On the Riemann Tensor in Double Field Theory},''
  \href{http://dx.doi.org/10.1007/JHEP05(2012)126}{{\em JHEP} {\bfseries 1205}
  (2012) 126},
\href{http://arxiv.org/abs/1112.5296}{{\ttfamily arXiv:1112.5296 [hep-th]}}.

\bibitem{Hohm:2012gk}
O.~Hohm and B.~Zwiebach, ``{Large Gauge Transformations in Double Field
  Theory},''
\href{http://arxiv.org/abs/1207.4198}{{\ttfamily arXiv:1207.4198 [hep-th]}}.

\bibitem{Park:2013gaj}
J.-H. Park and Y.~Suh, ``{U-geometry : SL(5)},''
\href{http://arxiv.org/abs/1302.1652}{{\ttfamily arXiv:1302.1652 [hep-th]}}.

\bibitem{Kugo:1992md}
T.~Kugo and B.~Zwiebach, ``{Target space duality as a symmetry of string field
  theory},'' {\em Prog.\ Theor.\ Phys.} {\bfseries {\bf 87}} (1992) 801,
\href{http://arxiv.org/abs/hep-th/9201040}{{\ttfamily arXiv:hep-th/9201040
  [hep-th]}}.

\bibitem{Hull:2009zb}
C.~Hull and B.~Zwiebach, ``{The Gauge algebra of double field theory and
  Courant brackets,},'' {\em JHEP} {\bfseries {\bf 0909}} (2009) 090,
\href{http://arxiv.org/abs/0908.1792}{{\ttfamily arXiv:0908.1792 [hep-th]}}.

\bibitem{Hull:2009mi}
C.~Hull and B.~Zwiebach, ``{Double field theory},''
  \href{http://dx.doi.org/10.1088/1126-6708/2009/09/099}{{\em JHEP} {\bfseries
  0909} (2009) 099}, \href{http://arxiv.org/abs/0904.4664}{{\ttfamily
  arXiv:0904.4664 [hep-th]}}.

\bibitem{Hohm:2010jy}
O.~Hohm, C.~Hull, and B.~Zwiebach, ``{Background independent action for double
  field theory},'' \href{http://dx.doi.org/10.1007/JHEP07(2010)016}{{\em JHEP}
  {\bfseries 1007} (2010) 016},
\href{http://arxiv.org/abs/1003.5027}{{\ttfamily arXiv:1003.5027 [hep-th]}}.

\bibitem{Hohm:2010pp}
O.~Hohm, C.~Hull, and B.~Zwiebach, ``{Generalized metric formulation of double
  field theory},'' \href{http://dx.doi.org/10.1007/JHEP08(2010)008}{{\em JHEP}
  {\bfseries 1008} (2010) 008},
  \href{http://arxiv.org/abs/1006.4823}{{\ttfamily arXiv:1006.4823 [hep-th]}}.

\bibitem{Siegel:1993th}
W.~Siegel, ``{Superspace duality in low-energy superstrings},''
  \href{http://dx.doi.org/10.1103/PhysRevD.48.2826}{{\em Phys.Rev.} {\bfseries
  D48} (1993) 2826--2837},
\href{http://arxiv.org/abs/hep-th/9305073}{{\ttfamily arXiv:hep-th/9305073
  [hep-th]}}.

\bibitem{Grana:2005jc}
M.~Grana, ``{Flux compactifications in string theory: A Comprehensive
  review},'' \href{http://dx.doi.org/10.1016/j.physrep.2005.10.008}{{\em
  Phys.Rept.} {\bfseries 423} (2006) 91--158},
\href{http://arxiv.org/abs/hep-th/0509003}{{\ttfamily arXiv:hep-th/0509003
  [hep-th]}}.

\bibitem{Shelton:2005cf}
J.~Shelton, W.~Taylor, and B.~Wecht, ``{Nongeometric flux compactifications},''
  \href{http://dx.doi.org/10.1088/1126-6708/2005/10/085}{{\em JHEP} {\bfseries
  0510} (2005) 085},
\href{http://arxiv.org/abs/hep-th/0508133}{{\ttfamily arXiv:hep-th/0508133
  [hep-th]}}.

\bibitem{Wecht:2007wu}
B.~Wecht, ``{Lectures on Nongeometric Flux Compactifications},''
  \href{http://dx.doi.org/10.1088/0264-9381/24/21/S03}{{\em Class.Quant.Grav.}
  {\bfseries 24} (2007) S773--S794},
\href{http://arxiv.org/abs/0708.3984}{{\ttfamily arXiv:0708.3984 [hep-th]}}.

\bibitem{Samtleben:2008pe}
H.~Samtleben, ``{Lectures on Gauged Supergravity and Flux Compactifications},''
  \href{http://dx.doi.org/10.1088/0264-9381/25/21/214002}{{\em
  Class.Quant.Grav.} {\bfseries 25} (2008) 214002},
\href{http://arxiv.org/abs/0808.4076}{{\ttfamily arXiv:0808.4076 [hep-th]}}.

\bibitem{Aldazabal:2010ef}
G.~Aldazabal, E.~Andres, P.~G. Camara, and M.~Grana, ``{U-dual fluxes and
  Generalized Geometry},''
  \href{http://dx.doi.org/10.1007/JHEP11(2010)083}{{\em JHEP} {\bfseries 1011}
  (2010) 083},
\href{http://arxiv.org/abs/1007.5509}{{\ttfamily arXiv:1007.5509 [hep-th]}}.

\bibitem{Grana:2006is}
M.~Grana, ``{Flux compactifications and generalized geometries},''
\href{http://dx.doi.org/10.1088/0264-9381/23/21/S02}{{\em Class.Quant.Grav.}
  {\bfseries 23} (2006) S883--S926}.

\bibitem{Andriot:2012an}
D.~Andriot, O.~Hohm, M.~Larfors, D.~Lust, and P.~Patalong, ``{Non-Geometric
  Fluxes in Supergravity and Double Field Theory},''
\href{http://arxiv.org/abs/1204.1979}{{\ttfamily arXiv:1204.1979 [hep-th]}}.

\bibitem{Andriot:2012wx}
D.~Andriot, O.~Hohm, M.~Larfors, D.~Lust, and P.~Patalong, ``{A geometric
  action for non-geometric fluxes},''
\href{http://arxiv.org/abs/1202.3060}{{\ttfamily arXiv:1202.3060 [hep-th]}}.

\bibitem{Dibitetto:2012ia}
G.~Dibitetto, A.~Guarino, and D.~Roest, ``{Exceptional Flux
  Compactifications},'' {\em JHEP} {\bfseries 1205} (2012) 056,
\href{http://arxiv.org/abs/1202.0770}{{\ttfamily arXiv:1202.0770 [hep-th]}}.

\bibitem{Dibitetto:2012rk}
G.~Dibitetto, J.~Fernandez-Melgarejo, D.~Marques, and D.~Roest, ``{Duality
  orbits of non-geometric fluxes},''
\href{http://arxiv.org/abs/1203.6562}{{\ttfamily arXiv:1203.6562 [hep-th]}}.

\bibitem{Witten:1995zh}
E.~Witten, ``{Some comments on string dynamics},''
\href{http://arxiv.org/abs/hep-th/9507121}{{\ttfamily arXiv:hep-th/9507121
  [hep-th]}}.

\bibitem{Townsend:1995gp}
P.~Townsend, ``{P-brane democracy},''
\href{http://arxiv.org/abs/hep-th/9507048}{{\ttfamily arXiv:hep-th/9507048
  [hep-th]}}.

\bibitem{Schwarz:1983wa}
J.~H. Schwarz and P.~C. West, ``{Symmetries and Transformations of Chiral N=2
  D=10 Supergravity},''
\href{http://dx.doi.org/10.1016/0370-2693(83)90168-5}{{\em Phys.Lett.}
  {\bfseries B126} (1983) 301}.

\bibitem{Cremmer:1978km}
E.~Cremmer, B.~Julia, and J.~Scherk, ``{Supergravity theory in
  eleven-dimensions},''
  \href{http://dx.doi.org/10.1016/0370-2693(78)90894-8}{{\em Phys.Lett.}
  {\bfseries B76} (1978) 409--412}.

\bibitem{Cremmer:1978ds}
E.~Cremmer and B.~Julia, ``{The N=8 supergravity theory. 1. The lagrangian},''
  \href{http://dx.doi.org/10.1016/0370-2693(78)90303-9}{{\em Phys.Lett.}
  {\bfseries B80} (1978) 48}.

\bibitem{Cremmer:1979up}
E.~Cremmer and B.~Julia, ``{The SO(8) supergravity},''
  \href{http://dx.doi.org/10.1016/0550-3213(79)90331-6}{{\em Nucl.Phys.}
  {\bfseries B159} (1979) 141}.

\bibitem{Campbell:1984zc}
I.~Campbell and P.~C. West, ``{N=2 D=10 Nonchiral Supergravity and Its
  Spontaneous Compactification},''
\href{http://dx.doi.org/10.1016/0550-3213(84)90388-2}{{\em Nucl.Phys.}
  {\bfseries B243} (1984) 112}.

\bibitem{Giani:1984wc}
F.~Giani and M.~Pernici, ``{N=2 SUPERGRAVITY IN TEN-DIMENSIONS},''
\href{http://dx.doi.org/10.1103/PhysRevD.30.325}{{\em Phys.Rev.} {\bfseries
  D30} (1984) 325--333}.

\bibitem{Schwarz:1995dk}
J.~H. Schwarz, ``{An SL(2,Z) multiplet of type IIB superstrings},''
  \href{http://dx.doi.org/10.1016/0370-2693(95)01138-G}{{\em Phys.Lett.}
  {\bfseries B360} (1995) 13--18},
\href{http://arxiv.org/abs/hep-th/9508143}{{\ttfamily arXiv:hep-th/9508143
  [hep-th]}}.

\bibitem{Aspinwall:1995fw}
P.~S. Aspinwall, ``{Some relationships between dualities in string theory},''
  \href{http://dx.doi.org/10.1016/0920-5632(96)00004-7}{{\em
  Nucl.Phys.Proc.Suppl.} {\bfseries 46} (1996) 30--38},
\href{http://arxiv.org/abs/hep-th/9508154}{{\ttfamily arXiv:hep-th/9508154
  [hep-th]}}.

\bibitem{Hohm:2012mf}
O.~Hohm and B.~Zwiebach, ``{Towards an invariant geometry of double field
  theory},''
\href{http://arxiv.org/abs/1212.1736}{{\ttfamily arXiv:1212.1736 [hep-th]}}.

\bibitem{Hohm:2010xe}
O.~Hohm and S.~K. Kwak, ``{Frame-like Geometry of Double Field Theory},''
  \href{http://dx.doi.org/10.1088/1751-8113/44/8/085404}{{\em J.Phys.}
  {\bfseries A44} (2011) 085404},
\href{http://arxiv.org/abs/1011.4101}{{\ttfamily arXiv:1011.4101 [hep-th]}}.

\bibitem{Isham:1971dv}
C.~Isham, A.~Salam, and J.~Strathdee, ``{Nonlinear realizations of space-time
  symmetries. Scalar and tensor gravity},''
\href{http://dx.doi.org/10.1016/0003-4916(71)90269-7}{{\em Annals Phys.}
  {\bfseries 62} (1971) 98--119}.

\bibitem{Ogievetsky:1973ik}
V.~Ogievetsky, ``{Infinite-dimensional algebra of general covariance group as
  the closure of finite-dimensional algebras of conformal and linear groups},''
{\em Lett.Nuovo Cim.} {\bfseries 8} (1973) 988--990.

\bibitem{Borisov:1974bn}
A.~Borisov and V.~Ogievetsky, ``{Theory of Dynamical Affine and Conformal
  Symmetries as Gravity Theory},''
\href{http://dx.doi.org/10.1007/BF01038096}{{\em Theor.Math.Phys.} {\bfseries
  21} (1975) 1179}.

\bibitem{West:1990in}
P.~West, {\em Introduction to Supersymmetry and Supergrativity}.
\newblock World Scientific Publishing Company, Incorporated, 1986.

\bibitem{West:2001as}
P.~C. West, ``{E(11) and M theory},''
  \href{http://dx.doi.org/10.1088/0264-9381/18/21/305}{{\em Class.Quant.Grav.}
  {\bfseries 18} (2001) 4443--4460},
\href{http://arxiv.org/abs/hep-th/0104081}{{\ttfamily arXiv:hep-th/0104081
  [hep-th]}}.

\bibitem{Riccioni:2007ni}
F.~Riccioni and P.~C. West, ``{E(11)-extended spacetime and gauged
  supergravities},''
  \href{http://dx.doi.org/10.1088/1126-6708/2008/02/039}{{\em JHEP} {\bfseries
  0802} (2008) 039},
\href{http://arxiv.org/abs/0712.1795}{{\ttfamily arXiv:0712.1795 [hep-th]}}.

\bibitem{Julia:1980gr}
B.~Julia, ``{Group disintegrations},'' in {\em Superspace and supergravity:
  proceedings of the Nuffield Workshop, Cambridge 1980}, S.~Hawking and
  M.~Rocek, eds., pp.~331--350.
\newblock Cambridge University Press, Cambridge, England, 1981.

\bibitem{Riccioni:2009xr}
F.~Riccioni, D.~Steele, and P.~West, ``{The E(11) origin of all maximal
  supergravities: The Hierarchy of field-strengths},''
  \href{http://dx.doi.org/10.1088/1126-6708/2009/09/095}{{\em JHEP} {\bfseries
  0909} (2009) 095},
\href{http://arxiv.org/abs/0906.1177}{{\ttfamily arXiv:0906.1177 [hep-th]}}.

\bibitem{West:2004iz}
P.~C. West, ``{Brane dynamics, central charges and E(11)},''
  \href{http://dx.doi.org/10.1088/1126-6708/2005/03/077}{{\em JHEP} {\bfseries
  0503} (2005) 077},
\href{http://arxiv.org/abs/hep-th/0412336}{{\ttfamily arXiv:hep-th/0412336
  [hep-th]}}.

\bibitem{Hull:2007zu}
C.~Hull, ``{Generalised geometry for M-theory},''
  \href{http://dx.doi.org/10.1088/1126-6708/2007/07/079}{{\em JHEP} {\bfseries
  0707} (2007) 079}, \href{http://arxiv.org/abs/hep-th/0701203}{{\ttfamily
  arXiv:hep-th/0701203 [hep-th]}}.

\bibitem{Berman:2011jh}
D.~S. Berman, H.~Godazgar, M.~J. Perry, and P.~West, ``{Duality Invariant
  Actions and Generalised Geometry},''
  \href{http://dx.doi.org/10.1007/JHEP02(2012)108}{{\em JHEP} {\bfseries 1202}
  (2012) 108},
\href{http://arxiv.org/abs/1111.0459}{{\ttfamily arXiv:1111.0459 [hep-th]}}.

\bibitem{Berman:2011cg}
D.~S. Berman, H.~Godazgar, M.~Godazgar, and M.~J. Perry, ``{The Local
  symmetries of M-theory and their formulation in generalised geometry},''
  \href{http://dx.doi.org/10.1007/JHEP01(2012)012}{{\em JHEP} {\bfseries 1201}
  (2012) 012},
\href{http://arxiv.org/abs/1110.3930}{{\ttfamily arXiv:1110.3930 [hep-th]}}.

\bibitem{Coimbra:2011ky}
A.~Coimbra, C.~Strickland-Constable, and D.~Waldram, ``{$E_{d(d)} \times
  \mathbb{R}^+$ Generalised Geometry, Connections and M Theory},''
\href{http://arxiv.org/abs/1112.3989}{{\ttfamily arXiv:1112.3989 [hep-th]}}.

\bibitem{Berman:2012vc}
D.~S. Berman, M.~Cederwall, A.~Kleinschmidt, and D.~C. Thompson, ``{The gauge
  structure of generalised diffeomorphisms},''
\href{http://arxiv.org/abs/1208.5884}{{\ttfamily arXiv:1208.5884 [hep-th]}}.

\bibitem{Berman:2011pe}
D.~S. Berman, H.~Godazgar, and M.~J. Perry, ``{SO(5,5) duality in M-theory and
  generalized geometry},''
  \href{http://dx.doi.org/10.1016/j.physletb.2011.04.046}{{\em Phys.Lett.}
  {\bfseries B700} (2011) 65--67},
\href{http://arxiv.org/abs/1103.5733}{{\ttfamily arXiv:1103.5733 [hep-th]}}.

\bibitem{Thompson:2011uw}
D.~C. Thompson, ``{Duality Invariance: From M-theory to Double Field Theory},''
  \href{http://dx.doi.org/10.1007/JHEP08(2011)125}{{\em JHEP} {\bfseries 1108}
  (2011) 125},
\href{http://arxiv.org/abs/1106.4036}{{\ttfamily arXiv:1106.4036 [hep-th]}}.

\bibitem{Scherk:1979zr}
J.~Scherk and J.~H. Schwarz, ``{How to Get Masses from Extra Dimensions},''
{\em Nucl.Phys.} {\bfseries B153} (1979) 61--88.

\bibitem{deWit:1981eq}
B.~de~Wit and H.~Nicolai, ``{N=8 Supergravity with Local SO(8) x SU(8)
  Invariance},''
\href{http://dx.doi.org/10.1016/0370-2693(82)91194-7}{{\em Phys.Lett.}
  {\bfseries B108} (1982) 285}.

\bibitem{Gunaydin:1985cu}
M.~Gunaydin, L.~Romans, and N.~Warner, ``{Compact and Noncompact Gauged
  Supergravity Theories in Five-Dimensions},''
\href{http://dx.doi.org/10.1016/0550-3213(86)90237-3}{{\em Nucl.Phys.}
  {\bfseries B272} (1986) 598}.

\bibitem{Pernici:1984xx}
M.~Pernici, K.~Pilch, and P.~van Nieuwenhuizen, ``{GAUGED MAXIMALLY EXTENDED
  SUPERGRAVITY IN SEVEN-DIMENSIONS},''
\href{http://dx.doi.org/10.1016/0370-2693(84)90813-X}{{\em Phys.Lett.}
  {\bfseries B143} (1984) 103}.

\bibitem{Hull:1984vg}
C.~Hull, ``{NONCOMPACT GAUGINGS OF N=8 SUPERGRAVITY},''
\href{http://dx.doi.org/10.1016/0370-2693(84)91131-6}{{\em Phys.Lett.}
  {\bfseries B142} (1984) 39}.

\bibitem{Hull:1984qz}
C.~Hull, ``{MORE GAUGINGS OF N=8 SUPERGRAVITY},''
\href{http://dx.doi.org/10.1016/0370-2693(84)90091-1}{{\em Phys.Lett.}
  {\bfseries B148} (1984) 297--300}.

\bibitem{Blumenhagen:2006ci}
R.~Blumenhagen, B.~Kors, D.~Lust, and S.~Stieberger, ``{Four-dimensional String
  Compactifications with D-Branes, Orientifolds and Fluxes},''
  \href{http://dx.doi.org/10.1016/j.physrep.2007.04.003}{{\em Phys.Rept.}
  {\bfseries 445} (2007) 1--193},
\href{http://arxiv.org/abs/hep-th/0610327}{{\ttfamily arXiv:hep-th/0610327
  [hep-th]}}.

\bibitem{Geissbuhler:2013uka}
D.~Geissbuhler, D.~Marques, C.~Nunez, and V.~Penas, ``{Exploring Double Field
  Theory},''
\href{http://arxiv.org/abs/1304.1472}{{\ttfamily arXiv:1304.1472 [hep-th]}}.

\bibitem{LeDiffon:2008sh}
A.~Le~Diffon and H.~Samtleben, ``{Supergravities without an Action: Gauging the
  Trombone},'' \href{http://dx.doi.org/10.1016/j.nuclphysb.2008.11.010}{{\em
  Nucl.Phys.} {\bfseries B811} (2009) 1--35},
\href{http://arxiv.org/abs/0809.5180}{{\ttfamily arXiv:0809.5180 [hep-th]}}.

\bibitem{deWit:2004nw}
B.~de~Wit, H.~Samtleben, and M.~Trigiante, ``{The Maximal D=5
  supergravities},''
  \href{http://dx.doi.org/10.1016/j.nuclphysb.2005.03.032}{{\em Nucl.Phys.}
  {\bfseries B716} (2005) 215--247},
\href{http://arxiv.org/abs/hep-th/0412173}{{\ttfamily arXiv:hep-th/0412173
  [hep-th]}}.

\bibitem{Bergshoeff:2007ef}
E.~Bergshoeff, H.~Samtleben, and E.~Sezgin, ``{The Gaugings of Maximal D=6
  Supergravity},'' \href{http://dx.doi.org/10.1088/1126-6708/2008/03/068}{{\em
  JHEP} {\bfseries 0803} (2008) 068},
\href{http://arxiv.org/abs/0712.4277}{{\ttfamily arXiv:0712.4277 [hep-th]}}.

\bibitem{Samtleben:2005bp}
H.~Samtleben and M.~Weidner, ``{The Maximal D=7 supergravities},''
  \href{http://dx.doi.org/10.1016/j.nuclphysb.2005.07.028}{{\em Nucl.Phys.}
  {\bfseries B725} (2005) 383--419},
\href{http://arxiv.org/abs/hep-th/0506237}{{\ttfamily arXiv:hep-th/0506237
  [hep-th]}}.

\bibitem{Tanii:1984zk}
Y.~Tanii, ``{N=8 Supergravity in Six Dimensions},''
\href{http://dx.doi.org/10.1016/0370-2693(84)90337-X}{{\em Phys.Lett.}
  {\bfseries B145} (1984) 197--200}.

\bibitem{Aldazabal:2011nj}
G.~Aldazabal, W.~Baron, D.~Marques, and C.~Nunez, ``{The effective action of
  Double Field Theory},'' {\em JHEP} {\bfseries 1111} (2011) 052,
\href{http://arxiv.org/abs/1109.0290}{{\ttfamily arXiv:1109.0290 [hep-th]}}.

\bibitem{Grana:2012rr}
M.~Grana and D.~Marques, ``{Gauged Double Field Theory},''
  \href{http://dx.doi.org/10.1007/JHEP04(2012)020}{{\em JHEP} {\bfseries 1204}
  (2012) 020},
\href{http://arxiv.org/abs/1201.2924}{{\ttfamily arXiv:1201.2924 [hep-th]}}.

\bibitem{Peeters:2007wn}
K.~Peeters, ``Introducing Cadabra: A Symbolic computer algebra system for field
  theory problems,''
\href{http://arxiv.org/abs/hep-th/0701238}{{\ttfamily arXiv:hep-th/0701238
  [hep-th]}}.

\bibitem{Cadabra}
K.~Peeters, ``A field-theory motivated approach to symbolic computer algebra,''
  {\em Comp. Phys. Comm.} {\bfseries 176} (2007) 550--558.

\bibitem{Hawking:1995fd}
S.~Hawking and G.~T. Horowitz, ``{The Gravitational Hamiltonian, action,
  entropy and surface terms},''
  \href{http://dx.doi.org/10.1088/0264-9381/13/6/017}{{\em Class.Quant.Grav.}
  {\bfseries 13} (1996) 1487--1498},
\href{http://arxiv.org/abs/gr-qc/9501014}{{\ttfamily arXiv:gr-qc/9501014
  [gr-qc]}}.

\bibitem{Strominger:1996sh}
A.~Strominger and C.~Vafa, ``{Microscopic origin of the Bekenstein-Hawking
  entropy},'' \href{http://dx.doi.org/10.1016/0370-2693(96)00345-0}{{\em
  Phys.Lett.} {\bfseries B379} (1996) 99--104},
\href{http://arxiv.org/abs/hep-th/9601029}{{\ttfamily arXiv:hep-th/9601029
  [hep-th]}}.

\bibitem{Rovelli:1996dv}
C.~Rovelli, ``{Black hole entropy from loop quantum gravity},''
  \href{http://dx.doi.org/10.1103/PhysRevLett.77.3288}{{\em Phys.Rev.Lett.}
  {\bfseries 77} (1996) 3288--3291},
\href{http://arxiv.org/abs/gr-qc/9603063}{{\ttfamily arXiv:gr-qc/9603063
  [gr-qc]}}.

\bibitem{Meissner:2004ju}
K.~A. Meissner, ``{Black hole entropy in loop quantum gravity},''
  \href{http://dx.doi.org/10.1088/0264-9381/21/22/015}{{\em Class.Quant.Grav.}
  {\bfseries 21} (2004) 5245--5252},
\href{http://arxiv.org/abs/gr-qc/0407052}{{\ttfamily arXiv:gr-qc/0407052
  [gr-qc]}}.

\bibitem{Lowe:1999pk}
D.~A. Lowe and L.~Thorlacius, ``{AdS / CFT and the information paradox},''
  \href{http://dx.doi.org/10.1103/PhysRevD.60.104012}{{\em Phys.Rev.}
  {\bfseries D60} (1999) 104012},
\href{http://arxiv.org/abs/hep-th/9903237}{{\ttfamily arXiv:hep-th/9903237
  [hep-th]}}.

\bibitem{Bena:2007kg}
I.~Bena and N.~P. Warner, ``{Black holes, black rings and their microstates},''
  \href{http://dx.doi.org/10.1007/978-3-540-79523-0_1}{{\em Lect.Notes Phys.}
  {\bfseries 755} (2008) 1--92},
\href{http://arxiv.org/abs/hep-th/0701216}{{\ttfamily arXiv:hep-th/0701216
  [hep-th]}}.

\bibitem{Winitzki}
S.~Winitzki, {\em {Advanced general relativity}}.
\newblock lecture notes,
2007.
\newblock

\bibitem{Poisson:931451}
E.~Poisson, {\em A Relativist's Toolkit: The Mathematics of Black-Hole
  Mechanics}.
\newblock Cambridge Univ. Press, Cambridge, 2004.

\bibitem{Misner:1974qy}
C.~W. Misner, K.~Thorne, and J.~Wheeler,
``{Gravitation},''.

\bibitem{Gibbons:1976ue}
G.~Gibbons and S.~Hawking, ``{Action Integrals and Partition Functions in
  Quantum Gravity},''
\href{http://dx.doi.org/10.1103/PhysRevD.15.2752}{{\em Phys.Rev.} {\bfseries
  D15} (1977) 2752--2756}.

\bibitem{deBoer:2012ma}
J.~de~Boer and M.~Shigemori, ``{Exotic Branes in String Theory},''
\href{http://arxiv.org/abs/1209.6056}{{\ttfamily arXiv:1209.6056 [hep-th]}}.

\bibitem{Jeon:2011vx}
I.~Jeon, K.~Lee, and J.-H. Park, ``{Incorporation of fermions into double field
  theory},'' \href{http://dx.doi.org/10.1007/JHEP11(2011)025}{{\em JHEP}
  {\bfseries 1111} (2011) 025},
\href{http://arxiv.org/abs/1109.2035}{{\ttfamily arXiv:1109.2035 [hep-th]}}.

\bibitem{Jeon:2012hp}
I.~Jeon, K.~Lee, J.-H. Park, and Y.~Suh, ``{Stringy Unification of Type IIA and
  IIB Supergravities under N=2 D=10 Supersymmetric Double Field Theory},''
\href{http://arxiv.org/abs/1210.5078}{{\ttfamily arXiv:1210.5078 [hep-th]}}.

\bibitem{Nakahara:2003nw}
M.~Nakahara, {\em {Geometry, topology and physics}}.
\newblock Boca Raton, USA: Taylor \& Francis,
2003.
\newblock

\bibitem{Geissbuhler:2011mx}
D.~Geissbuhler, ``{Double Field Theory and N=4 Gauged Supergravity},'' {\em
  JHEP} {\bfseries 1111} (2011) 116,
\href{http://arxiv.org/abs/1109.4280}{{\ttfamily arXiv:1109.4280 [hep-th]}}.

\bibitem{Pernici:1984zw}
M.~Pernici, K.~Pilch, P.~van Nieuwenhuizen, and N.~Warner, ``{NONCOMPACT
  GAUGINGS AND CRITICAL POINTS OF MAXIMAL SUPERGRAVITY IN SEVEN-DIMENSIONS},''
\href{http://dx.doi.org/10.1016/0550-3213(85)90046-X}{{\em Nucl.Phys.}
  {\bfseries B249} (1985) 381}.

\bibitem{deWit:2005hv}
B.~de~Wit and H.~Samtleben, ``{Gauged maximal supergravities and hierarchies of
  nonAbelian vector-tensor systems},''
  \href{http://dx.doi.org/10.1002/prop.200510202}{{\em Fortsch.Phys.}
  {\bfseries 53} (2005) 442--449},
\href{http://arxiv.org/abs/hep-th/0501243}{{\ttfamily arXiv:hep-th/0501243
  [hep-th]}}.

\bibitem{deWit:2008ta}
B.~de~Wit, H.~Nicolai, and H.~Samtleben, ``{Gauged Supergravities, Tensor
  Hierarchies, and M-Theory},''
  \href{http://dx.doi.org/10.1088/1126-6708/2008/02/044}{{\em JHEP} {\bfseries
  0802} (2008) 044},
\href{http://arxiv.org/abs/0801.1294}{{\ttfamily arXiv:0801.1294 [hep-th]}}.

\end{thebibliography}\endgroup

\end{document}